\documentclass{article}
\usepackage[utf8]{inputenc}
\usepackage{authblk}
\usepackage[T1]{fontenc}
\usepackage{graphicx}
\usepackage{amsmath, amssymb, graphics, setspace}
\newcommand{\mathsym}[1]{{}}
\newcommand{\unicode}[1]{{}}
\usepackage[font=small,labelfont=it]{caption}
\usepackage{float}
\usepackage{subfig}
\usepackage{url}
\usepackage{hyperref}
\usepackage{multicol}
\usepackage{tensor}

\usepackage[a4paper, portrait, margin=1in]{geometry}
\title{Visual Appearance of Extended objects in Special Relativity}
\author[1,2]{Utkarsh Bajaj}
\affil[1]{Junior Research Affiliate, Wolfram Physics Project}
\affil[2]{DPS International school, New Delhi}
\newcommand{\levicivita}{}
\def\levicivita#1#{\tensor#1{\epsilon}}
\begin{document}
\maketitle
\begin{abstract}
\noindent
The Lorentz transformation is a spontaneous measurement. We first highlight the difference between “measuring” and “seeing”, where the latter considers the time light rays (emitted by each point on the object) take to reach the observer. We compute the apparent position of a point given its velocity, initial position, and observation time. The apparent speed of a point is calculated, and we obtain that it exceeds the speed of light when approaching the observer, similar to superluminal motion. For parameterizable surfaces, we analyze properties (such as curvature and torsion) of apparent shapes. The observation that a sphere retains its circular silhouette when transformed to its apparent shape, independent of the initial conditions, is proved mathematically. Plots describing the apparent speed and length of objects are made, and the metric tensor for a distorted sphere is calculated. A generalized equation for the Doppler effect and relativistic aberration is derived to analyze regions of redshift and blueshift. Using the Born-rigidity conditions, we compute the hyperbolic trajectories of each point on an extended object given an initial velocity, position, and proper acceleration for any reference point. The claim that a rigid body, accelerating in Special Relativity, cannot exceed a given length in certain circumstances is justified. We obtain many non-trivial results, which are proved algebraically and using light cones, that are tested by taking the limit of acceleration approaching 0 to retrieve results in the constant velocity scenario. In conclusion, these visualizations may be used by teachers to explain SR intuitively. Finally, we provide an overview of extending the same problem to curved spacetime and explain the potential applications of this project.
\end{abstract}
\section{Introduction}
\subsection{Background}
The strange visual appearance of objects is one of the puzzling predictions of Einstein's relativity. This is mainly due to the distinction between “measuring” and “seeing”. The first successful exploration of simulating the visual appearances of objects moving close to the speed of light was conducted by Terrell \cite{terrel} and Penrose \cite{penrose} in 1959. The purpose of visualizing the appearance of objects in Special Relativity is to create an intuitive understanding of the same, as well as for educational purposes described in \cite{6}. However, visualising accelerating objects in Special Relativity has been widely ignored, as mentioned by D. Weiskopf et al. in \cite{illumination}. They implement acceleration by proposing a virtual environment in which the user can control the camera's movement by changing its speed. This is simply based on the standard appearance of objects in moving with constant velocity except in co-moving reference systems for a small time steps. In this paper, however, we try to address the opposite. We position the observer at $(0, 0, 0)$, and analyse the apparent shapes of different objects moving on different accelerated trajectories. We will address optical effects in the presence of an external light source but will rather focus on analysing the shapes of objects produced travelling at different velocities with different initial conditions.
\subsection{The Problem for a Point}
In order to simulate the visual appearance of objects travelling at high speeds, we must first understand the difference between measuring and seeing.
Consider a point moving with a constant velocity of \(0.9c,\) passing \(x = 0\) on the stationary observer{'}s \textit{x}-axis at a time \(t =0\)
on the stationary observer{'}s clock. The stationary observer the decides to look at the point at time \textit{\(t_{\text{obs}}\)} on
his watch. He will \textit{measure} the position of the point at $0.9ct_{obs}$ metres
along his \textit{x}-axis, which is indeed the \textit{actual} position of the point on the stationary observer{'}s \textit{x}-axis. However,
he will \textit{see} the point at a distance \textit{less} than $0.9ct_{obs}$ metres on his \textit{x}-axis - simply because he sees the point at time \textit{\(t_{\text{obs}}\)} when a light ray leaves the point at
a time earlier than \(t_{\text{obs}}\) to ultimately reach him at a time \(\text{\textit{$t_{\text{obs}}$}}\) on his watch. This distinction
would have been rendered redundant if the speed of light, \textit{c}, was infinite. Let's call the time of emission \(t_{\text{em}}\), the time at which light must have been emitted to reach the observer at time \(\text{\textit{$t_{\text{obs}}$}}\) on his clock. We can compute \(t_{\text{em}}\) using the following equation: 
\begin{equation}\label{1}
c\left(t_{\text{obs}} -t_{\text{em}}\right) =\left|x\left(t_{\text{em}}\right)\right|, t_{\text{em}}<t_{\text{obs}}
\end{equation}
\noindent
In more technical terms, the apparent position of the point i.e. the location of the point as seen by the observer, is found by intersecting its
world-line with the backward light cone of the observer. It is obvious that the apparent position of the point along the \textit{x}-axis of the
observer is given by \textup{\(x\left(t_{\text{em}}\right)\)}, which can be computed once we calculate \textup{\(t_{\text{em}}\)} from equation
(\ref{1}). \(x(t)\) is the \textit{actual} i.e. \textit{measured} trajectory of the particle. The absolute value sign implies the necessary condition that \(t_{\text{em}}\) $<$ \(t_{\text{obs}}\), because the light must have been emitted at a time earlier than the time at which the point is seen. The
difference between \(t_{\text{em}}\) and \(t_{\text{obs}}\) is termed as the \textit{time delay} of light, arising from the fact that light travels
at a finite speed.
\subsection{The Problem for an Object}\label{prob_object}
Assuming that we have evaluated the required transformation law for a point, which gives us the {`}apparent{'} position of a point at a time \textit{
\(t_{\text{obs}}\)} on the observer{'}s watch, the question of extending the same to a 3-D object still remains. A straightforward approach would
be to divide the object into \(n\) points, evaluate the transformation for every point, and then join the points together to form the apparent view
of the object. However, the clear issue with this computation is its lack of accuracy$-$100 percent accuracy is only achieved when we to take infinitely
many points on the surface of the object. However, for parametrizable curves, we can determine the exact vector valued function for the transformed
object with 100 percent accuracy. To understand why, let's first set up the mathematical formalism for evaluating the apparent position of a point. Let \(f\):\(\overset{\to}{p}\) $\in $ \(\mathbb{R}^3\to \mathbb{R}^3\) be the function returning the apparent position of a point \(\overset{\to}{p}\)\ moving with a 3-velocity $\Vec{\beta}$. Let{'}s evaluate the explicit form of the function \(f\):\(\overset{\to}{p}\) $\in $ \(\mathbb{R}^3\to \mathbb{R}^3\). Let the four vector $a = (0, \Vec{a})$ represent the initial position of the moving frame of reference \(S'\) and
let three vector $\Vec{\beta}$ represent its velocity as measured with respect to \(S\). The particle is still with respect to \(S'\)
i.e. it has a velocity of $\Vec{\beta}$ as well. Let \(\overset{\to }{x}'\) represent the displacement of the particle from \(S'\),
which will be a constant 3 vector. Let \(\left(x^0\right.\), \(\left.x^1(x^0),x^2(x^0),x^3(x^0)\right)\) denote the world-line of the point with respect
to \(S\). Note that we will be using Greek indices that run from 0 to 3 and Latin indices that run from 1 to 3. We will also set \(c\) = 1.
\par\noindent
One might question the reason for inventing the need of \(\overset{\to }{x}'\), posing that it is enough to set \(\overset{\to }{x}'\) as 0 while
varying $\Vec{a}$ to change the initial position of the particle. While the initial position of a particle at any point in $S$ involves only 1 degree of freedom, which is the initial position vector itself, it is important to keep in mind that we are trying to evaluate the apparent shape of 3 dimensional objects. For example, when trying to find the apparent shape of a fast moving sphere, we should first state the actual shape of the sphere before applying Lorentz transformation equations and time delay. The actual shape i.e. the {`}proper
shape{'} of the object can only be defined with respect to the moving frame of reference, thus validating the need for \(\overset{\to }{x}'\)
rather than just using $\Vec{a}$. In other words, the radius \(r\) of the sphere will be defined with respect to the proper
(moving) frame of reference. Note that setting \(\overset{\to }{x}'\) =  $\Vec{a}$ and $\Vec{a}$ = 0 is \textbf{not} equivalent to keeping $\Vec{a}$ as it is and \(\overset{\to }{x}'\) = 0 instead. This is because the spatial axis of $S'$ is stretched with respect to that of $S$, i.e. the situations will be equivalent if \(\overset{\to }{x}'\) =$\gamma \Vec{a}$. A stationary observer can use the following strategy to describe the 
proper shape of the sphere without using \(\overset{\to }{x}'\). Infinite observers can be placed on the surface of a sphere, one at each point, implying that \(\overset{\to }{x}'\) = 0 and $\Vec{a}$ $= \Vec{x} + (r\sin\theta\sin\phi, r\sin\theta\cos\phi,r\cos\theta)$. The observer must realise he is describing the Lorentz contracted version of the sphere. In fact, the proper shape of the sphere if $\Vec{\beta} = (\beta, 0, 0)$ will be an ellipsoid, defined by: 
\begin{equation}\label{2}
    (\gamma r\sin\theta\sin\phi, r\sin\theta\cos\phi,r\cos\theta)
\end{equation}
\noindent
Clearly, the stationary observer is unable to correctly define the shape, validating the need for 2 position vectors in describing the spacetime coordinates of a point on an object.
\par\noindent
To evaluate the apparent position vector \(\overset{\to
}{p}'\) measured with respect to \(S\) for an \textit{arbitrary} point on surface of sphere of radius \(r\) initially centred at $\Vec{a}$ and moving with a velocity $\Vec{\beta}$ at a time \(x_{\text{obs}}^0\) on the stationary observer{'}s watch, we must apply our
function \(f(\)$\Vec{\beta}$, $\Vec{a}$, \(\overset{\to }{x}'\), \(x_{\text{obs}}^0\)) in the following way:
\begin{equation}\label{3}
\overset{\to }{p}' = f\left(\Vec{\beta},\Vec{a},\overset{\to }{x}',x_{\text{obs}}^0\right) = f \left(\Vec{\beta},\Vec{a},(r\sin\theta\sin\phi, r\sin\theta\cos\phi,r\cos\theta), x_{\text{obs}}^0\right)  0\leq \theta
\leq \pi, 0\leq \phi \leq 2\pi
\end{equation}
Note that \(f\) is not defined with the arguments \(\overset{\to }{\beta}, \Vec{a},\overset{\to }{x}',x_{\text{obs}}^0\)$-$they
are mentioned explicitly for clarity. Since \(f\):\(\overset{\to }{p}\) $\in $ \(\mathbb{R}^3\to  \mathbb{R}^3\), we should rewrite the above
equation as:
\begin{equation}\label{4}
    \overset{\to }{p}' = f \left(\Vec{a} +\Vec{\beta}x_{\text{obs}}^0\right) \text{at} \overset{\to }{x}' = (r\sin\theta\sin\phi, r\sin\theta\cos\phi,r\cos\theta) \text{ at time } x_{\text{obs}}^0
\end{equation}
It{'}s now straightforward to see that \(\overset{\to }{p}'\) will depend on only 2 variables: $\theta $ and $\phi$, where $0\leq\theta\leq\pi$ and $0\leq\phi\leq2\pi$. Thus, we can plot the transformed sphere. However, for non-parametrizable surfaces or those which can{'}t be
defined as \(f(x,y,z) = c\), the only reasonable approach is to evaluate the apparent positions of \(n\) points and join them together while maintaining
the original mesh.
\section{Constant Velocity}\label{section2}
\subsection{Finding \textit{f}} 
\label{ff}
To find \textit{f}, we first must find \(x_{\text{em}}^0\), the time at which the light was emitted by the point to reach the observer at \(x_{\text{obs}}^0\). We can rewrite (\ref{1}) as:
\begin{equation}\label{5}
\left(x_{\text{obs}}^0 - x_{\text{em}}^0\right)^2 = \sum _{ i=1}^3\left( x^{ i}\left(x_{\text{em}}^0\right)\right)^2
\end{equation}
The above equation is another way of saying that the spacetime interval between the event of emission and the event of observation is 0. Let{'}s
say that the word line \(x^{ i}\left(x^0\right)\) describes the trajectory of a particle positioned at \(\overset{\to }{x}'\) from the origin of
\(S'\), which is itself positioned at $\Vec{a}$ from the origin of \(S.\) We can rewrite equation 3 in terms of the variables \(x^{'\mu
}\) as measured from \(S'\). Therefore, we will use the Poincar\'e transformation \cite{Muller}: 
\begin{equation}\label{6}
x^{\mu } = \Lambda ^{\mu }_{\nu } x^{'\nu } + a^{\mu }
\end{equation}
where\(\)\textrm{ \(\Lambda ^0_0\)} = \textrm{ \(\gamma\)}, \textrm{ \(\Lambda ^0_i=\)}\textrm{ \(\gamma \beta _i\)}, \textrm{ \(\Lambda ^i_0\)}
= \textrm{ \(\gamma \beta ^i\)}, \textrm{ \(\Lambda ^i_j = \delta _j^i+ \frac{\gamma ^2}{1+\gamma }\beta ^i\beta _j\)}, \textrm{ \(\gamma =\)}\textrm{
\(\frac{1}{\sqrt{1-\beta ^i \beta _i}}\)}, and \(a^{\mu }\) are the components of $\Vec{a}$. Note that $\beta ^i$= $\beta_{i}$. Substituting in equation (\ref{5}), we obtain:
\begin{equation}\label{7}
\left(x_{\text{obs}}^0 -\Lambda ^0_{\nu } x^{'\nu }\right)^2 = \sum _{ i=1}^3\left(\Lambda ^{i}_{\nu} x^{'\nu}+a^i \right)^2
\end{equation}where we have used the summation convention ($\nu $ ranges from 0 to 3). Simplifying, we obtain:
\begin{equation}\label{8}
\left(x_{\text{obs}}^0 -\gamma  x^{'0}-\gamma  \beta _j x^{'j}\right)^2 = \sum _{ i=1}^3\left( \gamma \beta ^{i} x^{'0}+\left(\delta _k^i+ \frac{\gamma ^2}{1+\gamma }\beta ^i\beta _{k}\right)x^{'k}+a^i
\right)^2
\end{equation}
Simplifying further, we obtain the following quadratic equation in \(x^{'0}\). A derivation of \(x^{'0}\) is also given in \cite{Muller}. We
solve for \(x^{'0}\), i.e. the time on the moving observer{'}s clock at which the light ray is emitted, and obtain 2 solutions. We first declare the following variables:
\begin{equation}\label{9}
\Vec{\sigma} = \overset{\to }{x}'+\frac{\Vec{\beta}\gamma ^2\left( \Vec{\beta}\cdot \overset{\to
}{x}'\right)}{1+\gamma } + \Vec{a}, \Omega  = x_{\text{obs}}^0-\gamma \left( \Vec{\beta}\cdot \overset{\to }{x}'\right)
\end{equation}
Then \(x^{'0}\) is given by:
\begin{equation}\label{extra1}
    x^{'0} = \gamma (\Omega + \Vec{\beta}\cdot\Vec{\sigma}) \pm \gamma\sqrt{2\Omega(\Vec{\beta}\cdot\Vec{\sigma})+ (\beta\cdot\Vec{\sigma})^2 + (\Vec{\beta}\cdot\Vec{\beta}) (\Omega^2 - \Vec{\sigma}\cdot\Vec{\sigma}) + \Vec{\sigma}\cdot\Vec{\sigma}}
\end{equation}
We know that time at which light is emitted must be less than the time of observation. This is shown in figures (\ref{fig2}) and (\ref{fig3}), where the world lines can be seen intersecting both the past and future light cones. Since we need to consider the intersection with the past light cone, we choose the smaller solution (it can be checked algebraically that the solution corresponding to the positive square root exceeds \(x_{\text{em}}^{'0}\)). Therefore, we can find the apparent position of the particle using the Poincar\'e transformation again:
\begin{equation}\label{extra2}
\Vec{x} = \overset{\to }{x}' + \gamma\Vec{\beta}x^{'0} + \Vec{\beta}\frac{\gamma^2}{1+\gamma}(\Vec{\beta}\cdot\overset{\to }{x}') + \Vec{a} 
\end{equation}
Substituting $x^{'0}$ given in equation (\ref{extra1}), 
\begin{equation}\label{10}
f\left(\Vec{a} +\Vec{\beta}x_{\text{obs}}^0\right) =
\Vec{\sigma} + \Vec{\beta} \gamma^2 \left( \Omega + \Vec{\beta}\cdot\Vec{\sigma} - \sqrt{2\Omega(\Vec{\beta}\cdot\Vec{\sigma})+ (\beta\cdot\Vec{\sigma})^2 + (\Vec{\beta}\cdot\Vec{\beta}) (\Omega^2 - \Vec{\sigma}\cdot\Vec{\sigma}) + \Vec{\sigma}\cdot\Vec{\sigma} } \right)
\end{equation}
In summary, we have obtained the apparent position of a point moving with a velocity of $\Vec{\beta}$ at a time \(x_{\text{obs}}^0\)
on the observer{'}s watch, having a displacement \(\overset{\to }{x}'\) from \(S'\), which itself has a displacement of $\Vec{a}$ from
the origin. For clarity, we will now write \(f\)(\(\Vec{a} +\Vec{\beta}x_{\text{obs}}^0\)) as \(f\)(\(\Vec{a} +\Vec{\beta}x_{\text{obs}}^0\), \(\overset{\to }{x}'\)). We have used the formalism established in \cite{Muller} by M{\" u}ller and
Boblest.
\par\noindent
Let's explain the equations above diagrammatically. Assume that a point, initially positioned at the origin of \(S\), moves with a velocity of
(0.5, 0, 0) with respect to the stationary observer. It is easy to see that to find the time \(x_{\text{em}}^0\) (the time on the observer{'}s clock
when the light was emitted) and the apparent position of the point at time \(x_{\text{obs}}^0\) on the observer{'}s watch, we must intersect the world-line
of the point with the observer{'}s past light cone at time \(x_{\text{obs}}^0\). Let \(x_{\text{obs}}^0\) = 0:
\begin{figure}[H]
    \centering 
    \includegraphics[scale = 0.35]{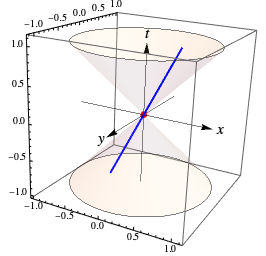}
    \caption{The blue word-line is the word-line of the point travelling with $\Vec{\beta}$ = (0.5, 0), and the red point represents the observer.
The light cone{'}s intersection with world line occurs at the spacetime coordinates (\(t\), \(x\), \(y\)) = (0, 0, 0), indicating that the apparent
position of the point at \(x_{\text{obs}}^0\) = 0 is indeed its true position as the observer is located at the origin.}
\label{fig1}
\end{figure} 
\noindent
Now, let{'}s find the apparent
position of the point at time \(x_{\text{obs}}^0\) = 1 on the observer{'}s watch:
\begin{figure}[H]
    \centering 
    \includegraphics[scale = 0.28]{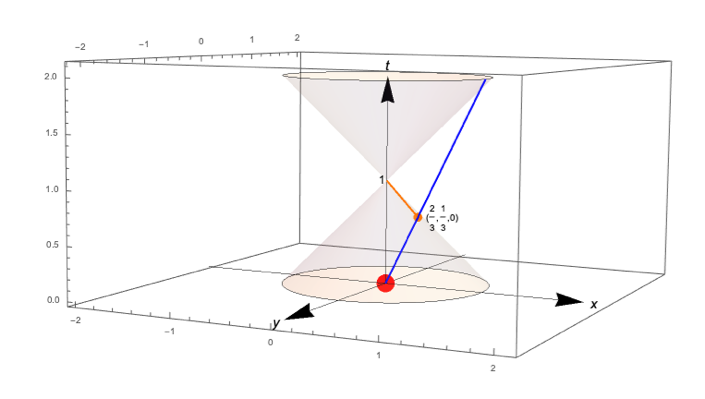}
    \caption{The observer{'}s past light cone at \(x_{\text{obs}}^0\)= 1 on his world-line intersects with the point{'}s trajectory at the spacetime coordinates (\(t,x,y\)) = \(\left(\frac{2}{3},\frac{1}{3},0\right.\)). These spacetime coordinates represent the emission of a light ray, which travels across the surface of the light cone to reach the observer at the event (1, 0, 0).}
    \label{fig2}
\end{figure}
\noindent Note that the actual position of the point at time \(x_{\text{obs}}^0\)= 1 is \(\frac{1}{2}\) and the apparent position of the point is \(\frac{1}{3}\) along the \textit{x}-axis. We can verify this result from the equation (\ref{10}). Furthermore, we can use these diagrammatic representations to intuitively explain the apparent shape of a vertical line travelling in a straight line in figure (\ref{fig3}). Let'ss describe a vertical line using endpoints (\(x',y'\)) = (0, -1) and (0, 1) with respect to the moving frame of reference. Assume that the vertical line is initially positioned at \(x= 0\) at \(x_{\text{obs}}^0\)= 0. Let{'}s try to find the apparent
shape of this vertical line travelling at $\Vec{\beta}$ = (0.5, 0) by analysing the intersections of word-lines of points regularly spaced along the vertical line with the observer{'}s light cone at (\(t,x,y\)) = \((1,0,0\)).
\begin{figure}
    \centering 
    \includegraphics[scale = 0.17]{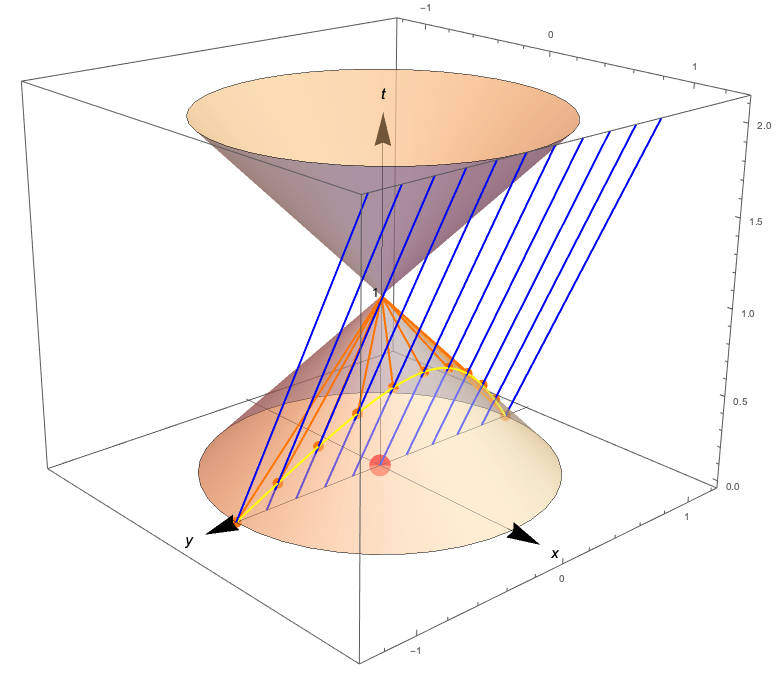}
    \caption{The blue world-lines are the trajectories of points along the vertical line. As expected, the world-lines further from the origin intersect the light cone at an earlier time. The above figure illustrates that the apparent shape of a vertical line is a hyperbola, which is indeed the case (refer to section (\ref{other_constant}))}
    \label{fig3}
\end{figure}
\noindent Note that the apparent shape of the hyperbola is found by projecting the yellow line{'}s shadow on the\textit{ \(x-y\)}
plane. The orange lines represent flashes of light travelling from the respective points to the observer at (\(t,x,y\)) = \((1,0,0\)). 
\par
\begin{figure}
    \centering 
    \includegraphics[scale = 0.16]{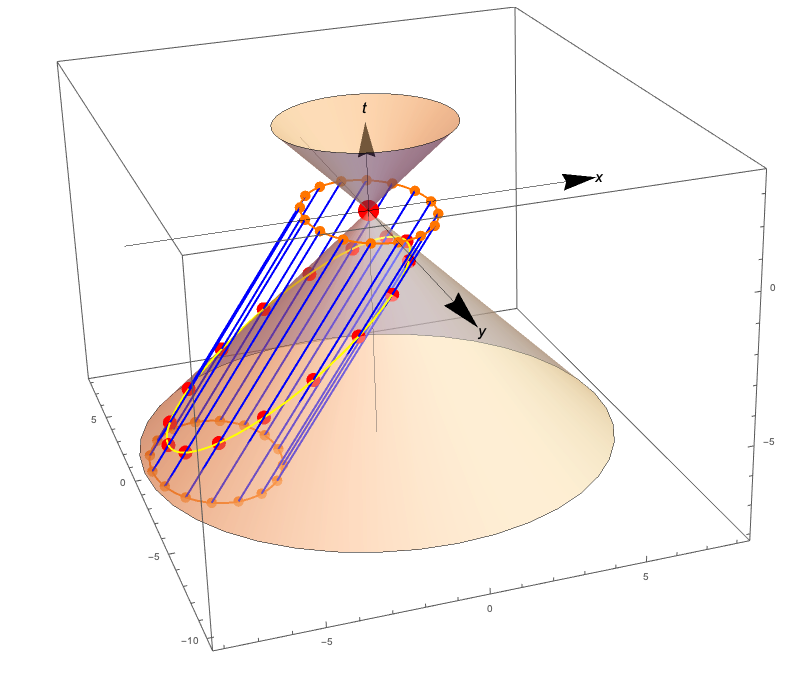}
    \caption{The apparent shape of the circle can be found by projecting the slanted figure on the light cone onto the \(x-y\) plane. }
    \label{fig4}
\end{figure}

\noindent
We can also find the apparent shape of a 1-sphere (circle) travelling in the positive \(x\)-direction assuming that the origin of \(S'\) is the center
of the circle. To repeat the analysis shown in the light cone above, we will first plot the trajectories of regularly spaced points on the circle{'}s
boundary. However, since the circle extends in the \(x\) direction as well, the trajectories of the points in \(S\) will represent the Lorentz contracted
circle. Therefore, we will plot the trajectories after applying the Lorentz contraction and then see the intersection with the light cone as shown
previously. Assume that \(r=1\), \(\Vec{\beta} = (0.9, 0, 0)\), and \(x_{\text{obs}}^0\)= 0. The circle{'}s center is at the origin $(0,0,0)$ of \(S\)
at time \(x_{\text{obs}}^0\) = 0. In this case, the trajectory of the rightmost point (\(x\),\(y\)) = (1, 0) will begin from the spacetime coordinates
(0, \(\frac{1}{\gamma }\), \(0\)) and that of the leftmost point will begin from \(\left(0,\frac{-1}{\gamma },0\right)\). The starting spacetime points for an arbitrary point on the circle (cos$\theta $, sin$\theta $) can written as \(\left(0,\frac{\text{cos$\theta$}}{\gamma },\text{sin$\theta $}\right)\) for 0 $\leq $ $\theta $ $\leq $ 2$\pi$, as shown in figure (\ref{fig4}).
\noindent
To test our equations, we can substitute specific values for the parameters. By reducing equation (\ref{10}) to a simpler form by substituting \(\overset{\to
}{x}'\) = $\Vec{a}$ = \(\overset{\to }{0}\) and assuming that $\Vec{\beta}$ = $\beta$ $\hat{i}$, we obtain:
\begin{equation}\label{11}
\left(x_{\text{obs}}^0\beta  \gamma ^2-x_{\text{obs}}^0\beta  \gamma  \sqrt{-1+\gamma ^2},0,0\right)=\left(\frac{\beta  x_{\text{obs}}^0}{1+\beta
},0,0\right)
\end{equation}
Carrying out the computation simply using (\ref{1}), we find that our solution in (\ref{3}) satisfies this simple case:
\begin{equation}\label{12}
x_{\text{obs}}^0 - x_{\text{em}}^0 = \beta x_{\text{em}}^0, \Vec{x} = x_{\text{em}}^0 \hat{i}
= \left(\frac{\beta  x_{\text{obs}}^0}{1+\beta },0,0\right)
\end{equation}
We also expect that our equations give correct results for when \(\overset{\to }{\beta}\) = \(\overset{\to }{0}\). Assuming $\Vec{a}$ = (\(a\), \(b\), \(c\)) and \(\overset{\to }{x}'\) = (\(x, y, z\)), the apparent position
\(\overset{\to }{x}\) becomes:
\begin{equation}\label{13}
    \overset{\to }{x} = (a+x,b+y,c+z)
\end{equation} which is indeed the correct result. 
\subsection{Apparent Shapes of fast-moving Spheres}\label{spheres}
The parametric form representing the surface of the transformed sphere i.e. the apparent shape of the sphere should be of the form \(\overset{\to }{x}_s\): \(U\subset\mathbb{R}^2\to\mathbb{R}^3\), where the parameters are the spherical polar coordinates $\theta$ and $\phi$. Using equation (\ref{10}), we can find the explicit form of $ \Vec{x}_{s}(\theta,\phi)$. If a sphere of radius $r$ is travelling with a velocity \(\overset{\to }{\beta}\):
\begin{equation}\label{14}
   \Vec{x}_{s}(\theta,\phi) = f\left(\Vec{a} +\Vec{\beta}x_{\text{obs}}^0, r \sin(\theta) \sin(\phi), r \sin(\theta) \cos(\phi), r \cos(\theta)\right)
\end{equation}
If we assume that the sphere's centre is initially positioned at the origin of \(S\) i.e. $\Vec{a}$ = 0 and that \(\overset{\to}{\beta}\) = ($\beta $, 0, 0), then we obtain an expression for the $x$ component of $\Vec{x}_{s}(\theta,\phi)$:
\begin{equation}\label{15}
\frac{1}{1-\beta ^2}\left(x_{\text{obs}}^0\beta  +\sqrt{1-\beta^2} r\sin\theta\sin\phi -\beta \sqrt{\left(1-\beta ^2\right)r^2+ \beta  x_{\text{obs}}^0\left(2 r \sin\theta\sin\phi \sqrt{1-\beta ^2}+\beta  x_{\text{obs}}^0\right)}\right)
\end{equation}
The $y$ and $z$ components remain unchanged. At time \(x_{\text{obs}}^0\) = 0, it{'}s easy to see that the parametric equation of the sphere takes the form:
\begin{equation}\label{16}
\Vec{x}_{s}(\theta,\phi)=\left(-\frac{r\beta}{\sqrt{1-\beta^2}}+\frac{r\sin\theta\sin\phi}{\sqrt{1-\beta^2}},r\cos\phi\sin\theta,r\cos\theta \right)
\end{equation}
As expected, the parametric form above reduces to the ordinary spherical polar coordinates as $\beta $ \(\to\) 0. The difference in the apparent $x$ coordinates between the sphere seen at \(x_{\text{obs}}^0\) = 0 with $\Vec{a}$= \((a,0,0)\) and the apparent $x$ coordinate given in equation (\ref{10}) can be computed after some algebra:
\begin{equation}\label{17}
    \frac{a-\beta  \left(\sqrt{a^2+2 a \sqrt{1-\beta ^2} r \sin (\theta ) \sin (\phi )-\beta ^2 r^2+r^2}+\sqrt{1-\beta ^2} r\right)}{1-\beta ^2}
\end{equation}
We can produce a series of snapshots showing the apparent shape of a sphere, which is located at the origin of \(S\) at time \(x_{\text{obs}}^0\) = 0 moving at a velocity \(\overset{\to}{\beta}\) = (0.5, 0, 0) with a radius \(r\) = 2, at times \(x_{\text{obs}}^0\)= -10, \(x_{\text{obs}}^0\) = -5, \(x_{\text{obs}}^0\) = 0, \(x_{\text{obs}}^0\) = 5, and \(x_{\text{obs}}^0\)= 10:
\begin{figure}[H]
    \centering
    \includegraphics[scale = 0.4]{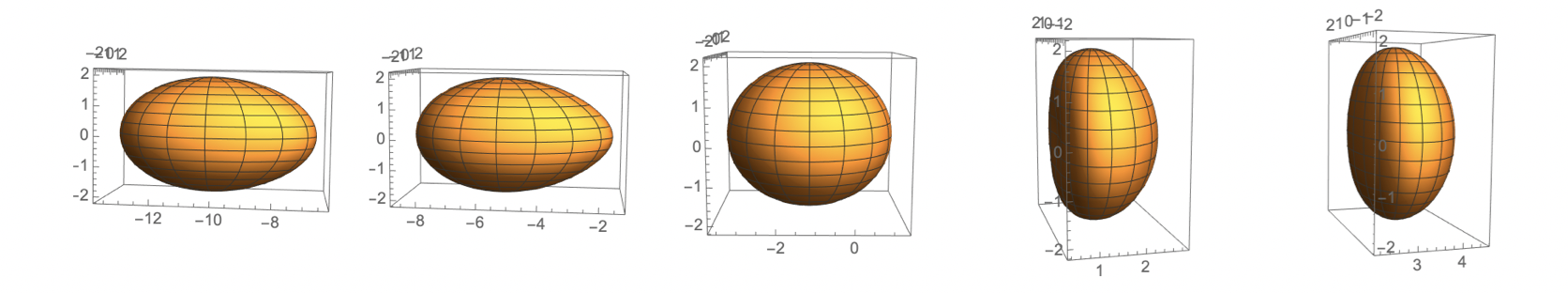}
    \caption{Note that these images do \pmb{NOT} represent what the sphere would \textit{look} like. To find the visual image of the sphere as would be seen
by an observer, we need to position a camera at the point $(0,0,0)$. If we do so, the deformation will create an illusion of the the
sphere being rotated as it moves across the positive \text{x}-axis.}
    \label{fig5}
\end{figure}
\noindent
below is a series of snapshots of the same sphere as in figure (\ref{fig5}) except that with a velocity of $\Vec{\beta}$ = $(0.7, 0, 0)$:
\begin{figure}[H]
    \centering
    \includegraphics[scale=0.4]{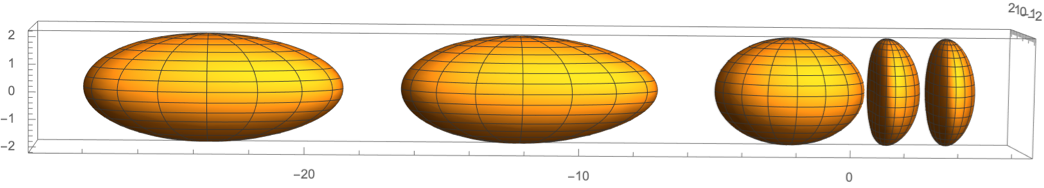}
    \caption{By comparing figure (\ref{fig6}) with figure (\ref{fig5}) we can easily see the difference in curvature when the velocity is increased.}
     \label{fig6}
\end{figure}
\noindent
Below is a series of snapshots of the same sphere at $\Vec{a}$ = $(0, 0, 0)$ moving with a velocity of $\Vec{\beta}$=(0.45, -0.45, \(\frac{0.9}{\sqrt{2}}\)): 
\begin{figure}[H]
    \centering
    \includegraphics[scale = 0.4]{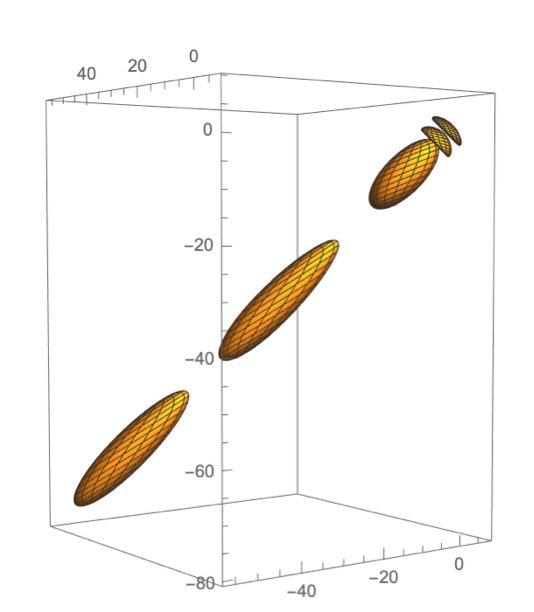}
    \caption{The apparent shape of a sphere moving in the north-east direction}
     \label{fig7}
\end{figure}
\noindent
As stated in \cite{penrose} and \cite{boas}, the silhouette of the apparent shape of a moving sphere will always be circular regardless of its direction of motion, speed, and actual initial position. The camera views of the sphere in figure (\ref{fig6}) from the point $(0,0,0)$ at times \(x_{\text{obs}}^0\) = $-5$ and $5$ are shown in figure (\ref{fig8}).
\begin{figure}
    \centering
    \includegraphics[scale = 0.25]{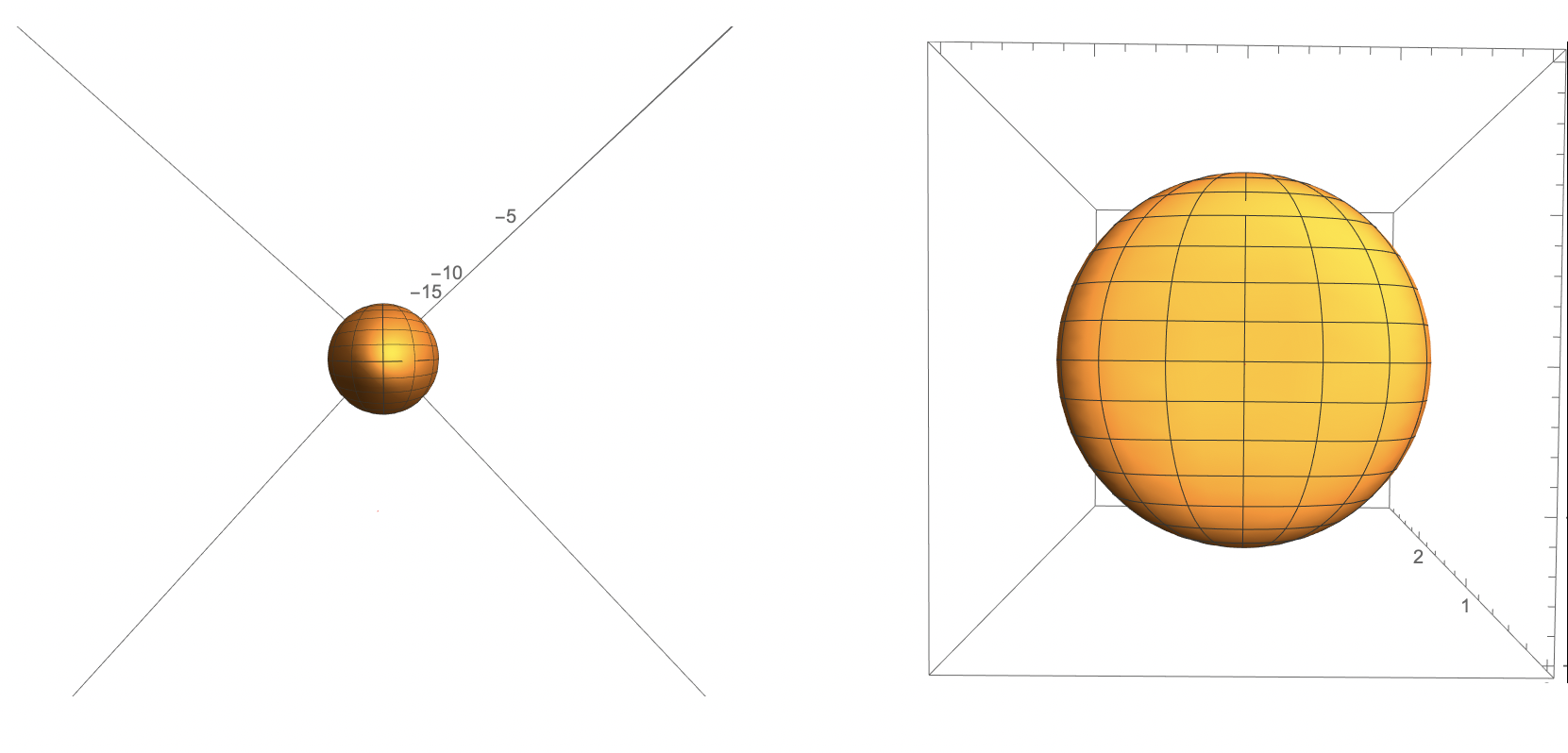}
    \caption{The silhouette of both these spheres are circular. The sphere on the left (at time \(x_{\text{obs}}^0\) = -5) looks smaller than the sphere on the right as it's farther away.} 
    \label{fig8}
\end{figure}
\noindent
When $\Vec{a} = k\Vec{\beta}$, where $k$ is a constant, and \(\overset{\to
}{x}'\) is the center of the sphere, then the observer will lie on the sphere's path of movement. In this case, it is obvious that the silhouette of a sphere will be circular. This is simply because the observer will not be able to notice any deformation in the sphere as the deformations take place along his line of sight. This is shown in figure (\ref{fig8_new}), where $\Vec{a} = 0$. However, for a non-zero vector $\Vec{a}$ and \(\overset{\to
}{x}'\) as the center of the sphere, it is not obvious why the silhouette of the apparent sphere should be a circle. This is similar to the case when $\Vec{a} = 0$ and \(\overset{\to}{x}'\) is not parallel to $\Vec{\beta}$. In short, we have to prove that the silhouette of sphere whose motion is not parallel to the observer's line of sight is also circular. For example, the following images show the apparent shape of a sphere travelling with a velocity of $\Vec{\beta} = (0.9,0,0)$ initially positioned at $\Vec{a} = (0,4,0)$ with respect to the observer located at the origin:
\begin{figure}[H]
    \centering
    \includegraphics[scale = 0.3]{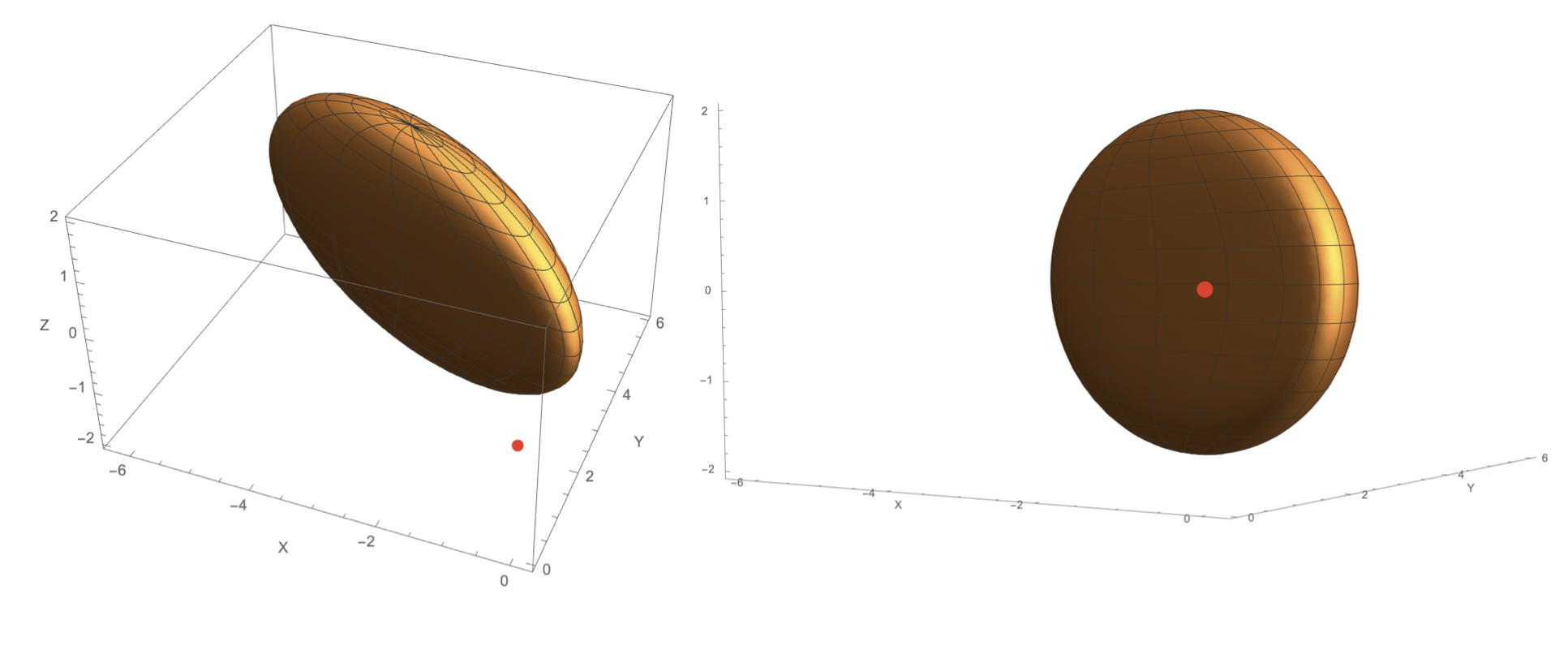}
    \caption{The red dot represents the observer. The figure to the left shows the apparent shape of the sphere, and the image on the right shows that an observer will indeed see a circular silhouette.}
    \label{fig8_new}
\end{figure}
\noindent 
Mathematically, a circular silhouette implies that the sphere must be tangent to a right circular cone whose apex is at the observer's position. In the rest frame of a sphere i.e. in the frame $S'$, a right circular cone with an apex angle $\omega'$ and a unit vector \(\overset{\to}{\varphi}'\) along the cone axis can be constructed. The silhouette of the sphere is defined by the intersection of the cone with a plane passing though the points that are tangent to the sphere. The cone can be thought as the region through which all light rays propagate from the surface of the sphere to the observer, and the silhouette is the circular outline of the sphere. Any position vector $\overset{\to}{x}'$ on the surface of the right circular cone is at angle $\omega'$ from the cone axis. Therefore, the implicit equation of the cone is given by:
\begin{equation}\label{18}
    \overset{\to}{x}'\cdot \overset{\to}{\varphi}' = \mid\overset{\to}{x}'\mid \cos\omega'
\end{equation}
\begin{figure}[]
    \centering
    \includegraphics[scale = 0.34]{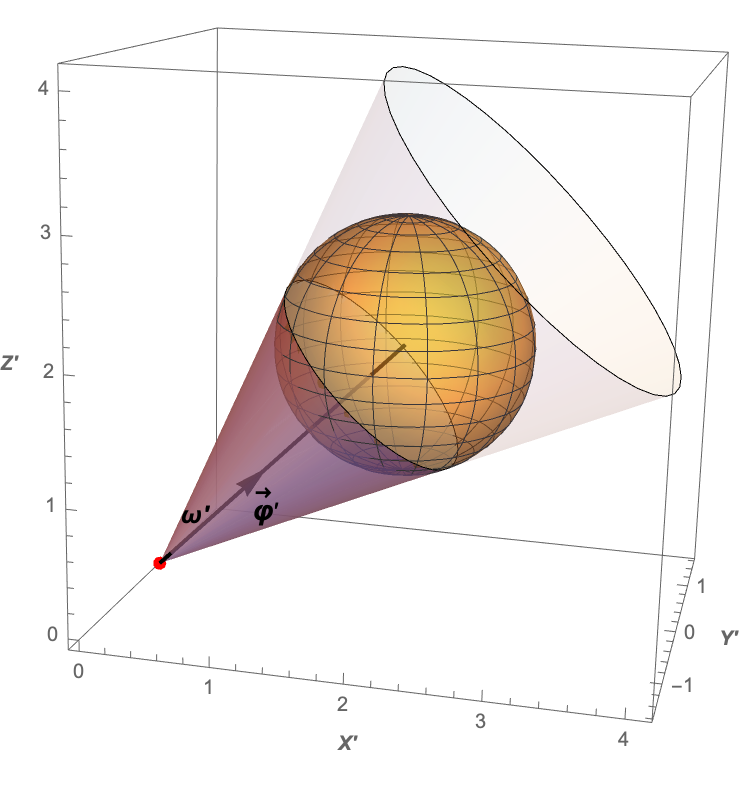}
    \caption{The rest observer $S'$ looking at a sphere with proper radius 1 centered at $(x',y',z') = (2,0,2)$, with $\cos\omega' = \frac{\sqrt{7}}{2\sqrt{2}}$. For a right-circular cone, the apex angle $\omega'$ is constant for every point on the cone.}
    \label{fig9}
\end{figure}
\noindent
The apparent shape of the sphere in $S'$ is the actual i.e. proper shape, as shown in figure (\ref{fig10}) on the next page. Let $x_{\text{em}}^{'0}$ be the time at which light is emitted from a point on the sphere $\overset{\to}{x}'$ to reach the moving observer at time $x_{\text{obs}}^{'0}$ according to the moving observer's watch. We can find $x_{\text{em}}^{'0}$ using equation \ref{1}, namely that $c^2(x_{\text{obs}}^{'0} - x_{\text{em}}^{'0})^2 = \mid\overset{\to}{x}'\mid^2$. Without loss of generality, we may assume that the stationary observer $S$ sees the sphere at $x_{\text{obs}}^{0} = 0$, implying that $x_{\text{obs}}^{'0} = 0$ as well (this was assumed in the Poincar\'e transformation before). Since we only want to consider points that lie on the right circular cone in $S'$, we substitute $\mid\overset{\to}{x}'\mid$ in equation (\ref{18}) to find the parametric form of the cone:
\begin{equation}\label{19}
     \overset{\to}{x}'\cdot \overset{\to}{\varphi}' = -c x_{\text{em}}^{'0} \cos\omega', \mid\overset{\to}{x}'\mid^2 = c^2 (x_{\text{em}}^{'0})^2
\end{equation}
These equations compute the apparent shape of the sphere in $S'$, which is obviously the actual shape. Now, we have to prove that the apparent shape of the sphere in $S$ is also tangential to a right circular cone defined by an apex angle $\omega$ and unit vector $\Vec{\varphi}$. Let's coincide the origin of $S'$ with the origin of $S$ at time $x_{\text{obs}}^{0}$, implying that both $S$ and $S'$ both receive same light rays at $x_{\text{obs}}^{0}$. To determine what observer $S$ will see at another time $x_{\text{obs}}^{0}$ using the same method, we must coincide the origin $S'$ with $S$ at that moment of time and vary $\Vec{x}'$ accordingly. Note that, due to relativistic aberration (see \cite{abberr}), $S$ and $S'$ will not agree on the direction from which light originates. Since $x_{\text{obs}}^{0} = 0$, we can set $\Vec{a}$ as 0. In this case, to ensure that the line of sight of the observer is not parallel to $\Vec{\beta}$, we must make sure that $\overset{\to}{x}'$ is arbitrary and lies outside the sphere in $S'$, as shown in figure (\ref{fig10}) and equations (\ref{19}).
\par\noindent
We can use the Lorentz transformation to compute $\Vec{x}$, the position of the points on the sphere that emitted the light ray at time $x_{\text{em}}^{0}$ to reach the observer at time $x_{\text{obs}}^{0}$ = 0. Writing the components of $\Vec{x}'$ in terms of the components of $\Vec{x}$ using the inverse Lorentz transformation $x'^{i} = \Bar{\Lambda}^{i}_{\alpha}x^\alpha$, we can rewrite equations (\ref{19}) in the following form:
\begin{equation}\label{20}
    \left(\Bar{\Lambda}^{i}_{j}x^j + \Bar{\Lambda}^{i}_{0}x_{\text{em}}^{0}\right)\varphi'_{i} = -c\left(\Bar{\Lambda}^{0}_{0} x_{\text{em}}^{0} + \Bar{\Lambda}^{0}_{j} x^j \right)\cos \omega', |\Vec{x}|^2 = c^2 \left(x_{\text{em}}^{0}\right)^2 
\end{equation}
\noindent
The second equation is true because the speed of light is same in both reference frames. If we can bring the first equation of equations (\ref{20}) to the form of $ x^i \varphi_{i} = -c x_{\text{em}}^{0} \cos \omega $, then we can prove that the transformed sphere in $S$ is tangent to a right circular cone defined by a vector $\Vec{\varphi}$ and apex angle $\omega$. Rearranging and after switching the indices $i$ and $j$, we obtain: 
\begin{equation}\label{21}
    \varphi_{i} = \Bar{\Lambda}^{j}_{i} \varphi_{j}' - \gamma \beta_{i}\cos \omega', \cos \omega = \gamma \cos\omega' - \gamma \beta^{j} \varphi_{j}' 
\end{equation}
\noindent
Note that $\varphi_{i}$ are the components of $\Vec{\varphi}$. Note that $\Vec{\varphi}$ can be normalised. As mentioned in \cite{boas}, the fact that the apex angle $\omega$ is different from $\omega'$ indicates that the cone tangent to the sphere is of a different size (see figure (\ref{fig9})) i.e. the apparent shape of a sphere is different from its actual shape. The fact that the unit vector $\Vec{\varphi}$ defining the sphere is different indicates that the direction from which light arrives is different (aberration). Note that $\varphi_{i}$ and $\omega$ approach their respective values in $S'$ as the velocity approaches 0.
\par\noindent
If $\Vec{x}(u,v)$ is the parametric equation for any regular surface is euclidean space, then the normal vector to the tangent space at any point on the surface can be written as:
\begin{equation}
    \Vec{n} = \frac{\partial \Vec{x}}{\partial u} \times \frac{\partial \Vec{x}}{\partial v} 
\end{equation}
The equation of a silhouette of a parametric surface viewed from the origin can be found by solving the following equation:
\begin{equation}\label{constraint}
   \levicivita{_{ijk}}x^{i}\frac{\partial x^{j}}{\partial \theta} \frac{\partial x^{k}}{\partial \phi} = 0
\end{equation}
i.e. the normal vector on any point on the silhouette is perpendicular to the position vector. Note that $\levicivita{_{ijk}}$ is the \textit{levi-civita} symbol. On any surface, equation (\ref{constraint}) provides a constraint on the type of points that must lie on the silhouette. To test whether these points $\Vec{x}$ constitute a circular silhouette, there must exist a \textit{unique} unit vector $\hat{\varphi}$ such that the points described by equation (\ref{constraint}) can lie on a right circular cone with apex angle $\omega$. In other words, the points $\Vec{x}$ that constitute the silhouette of the surface must also satisfy the following equation:
\begin{equation}\label{cone}
   \frac{ \Vec{x} \cdot \hat{\varphi}}{ \mid \Vec{x} \mid} = \cos{\omega}
\end{equation}
\noindent Note that for this equation to be a right circular cone, $\omega$ must be constant. However, we must note that the vector $\hat{\varphi}$, which is essentially the `viewing vector' from the origin, can be arbitrary. In other words, the statement of a sphere having `a circular silhouette' is only valid for one $\hat{\varphi}$. Therefore, we must prove there there exists a unique $\hat{\varphi}$ for which $\omega$ is constant. Since we want to discern which group of transformations can preserve the circular silhouette, we may assume that $\cos{\omega}$ is constant for all $\theta$ and $\phi$ for a surface $\Vec{x}$. This implies that the partial derivative of $\cos{\omega}$ with respect to the parameters are $0$ everywhere. Differentiating the LHS of equation (\ref{cone}) with respect to the parameters yield the following equations, written in index form:
\begin{equation}\label{theta}
    \varphi^{i}\frac{\partial x_{i}}{\partial \theta} \left(x^{j}x_{j}\right) - x^{i}\frac{\partial x_{i}}{\partial \theta}\left(\varphi^{j}x_{j}\right) = 0
\end{equation}
\begin{equation}\label{phi}
    \varphi^{i}\frac{\partial x_{i}}{\partial \phi} \left(x^{j}x_{j}\right) - x^{i}\frac{\partial x_{i}}{\partial \phi}\left(\varphi^{j}x_{j}\right) = 0
\end{equation}
In summary, we have assumed that equation (\ref{constraint}) and the equations above are satisfied for a unique $\hat{\varphi}$. Now, let $x^{'i} =x^{'i}\left(x^{j}\right)$ be a coordinate transformation. To prove that a certain active transformation of a surface preserves the circular silhouette, one must write equations ($\ref{theta}$) and ($\ref{phi}$) in terms of the new coordinates $x^{'i}$. If the equations can be rearranged in the form of equations ($\ref{theta}$) and ($\ref{phi}$) with the primed coordinates and a primed vector $\hat{\varphi}$, then the transformed will also have a circular silhouette. In doing so, we also substitute the primed coordinates $x^{'i}$ in equation (\ref{constraint}) to obtain the new constraints. We can understand this in the case of a simple scaling i.e. when $x^{'i} = k x^{i}$. We expect the unit vector to remain same. One can imagine this as, first, looking at a sphere and seeing a circular silhouette and, second, scaling the sphere in equally in all directions. In this case, the viewing vector for which the enlarged sphere should have a circular silhouette should stay the same, as the only variable that changes is $\omega$. Substituting  $x^{i} = \frac{1}{k}x^{'i}$ in the equations above and equation (\ref{constraint}), we obtain a factor of $\frac{1}{k^3}$ that can be eliminated as the RHS is 0. We easily obtain that the equations can indeed be rearranged in the form of equations ($\ref{theta}$) and ($\ref{phi}$) with $\varphi^{'i} = \varphi^{i}$, which is what we anticipated. 
\par\noindent
For a general linear transformation $x^{'i} = k x^{i} + a^{i}$ for a vector $\Vec{a}$, we need to carry out more algebraic computations. After substituting $x^{i}$ in terms of the primed coordinates in equations ($\ref{theta}$) and ($\ref{phi}$), we obtain:
\begin{equation}
    \varphi^{'i} = \frac{\hat{\varphi}^{i} + \hat{a}^{i}}{\cos \alpha}
\end{equation}
where $\alpha$ is the angle between $\hat{\varphi}$ and $\Vec{a}$. $\hat{a}$ is the unit vector pointing in the direction of $\Vec{a}$. This is also an expected result.  Although the formalism works for very simple cases, the set of silhouette-preserving maps may be too general as many shapes can have circular silhouettes when viewed from a certain angle. Nevertheless, the fact that the speed of light in both reference frames is constant made the proof much simpler.
\subsection{Apparent Shapes of other Objects}\label{other_constant}
\noindent
As explained in \cite{Muller}, a similar approach can be considered to evaluate the apparent shape of a fast moving line. A straight line can be determined by considering one end point at the origin of \(S'\) and considering the other end point as \(\overset{\to }{x}'\). Then, the line may be parametrized as $\lambda $\(\overset{\to }{x}'\) in \(S'\), with $\lambda $ ranging from 0 to 1. Let{'}s find the apparent shape of a line is traveling at a velocity
\(\Vec{\beta}= (\beta ,0,0)\) at time \(x_{\text{obs}}^0\) when \(S'\) is initially positioned at $\Vec{a}$ = 0. Let{'}s consider a horizontal line given by \(\overset{\to }{x}'\) = (\(x'\), 0, 0) in \(S'\). We need to evaluate $f$(\(\Vec{a} +\Vec{\beta}x_{\text{obs}}^0\), \(\lambda \overset{\to }{x}'\)):
\begin{equation}\label{22}
\overset{\to }{x}=\left(\frac{x_{\text{obs}}^0 \beta +x'\sqrt{1-\beta ^2} \lambda }{1+\beta },0,0\right), \lambda \in [0,1]
\end{equation}
\noindent
The above formula shows that the apparent shape of the line is also horizontal$-$this is obvious because the direction of velocity is in the positive \textit{x}-axis. Using equation (\ref{22}), we can plot a series of snapshots of a line with proper length 4 moving with a velocity \(\overset{\to}{\beta }=(0.7,0, 0)\) at times \(x_{\text{obs}}^0\) = $-$10, \(x_{\text{obs}}^0\) = $-$5, \(x_{\text{obs}}^0\)= 0, \(x_{\text{obs}}^0\) = 5, and \(x_{\text{obs}}^0\)
= 10:
\begin{figure}[H]
    \centering
    \includegraphics[scale= 0.5]{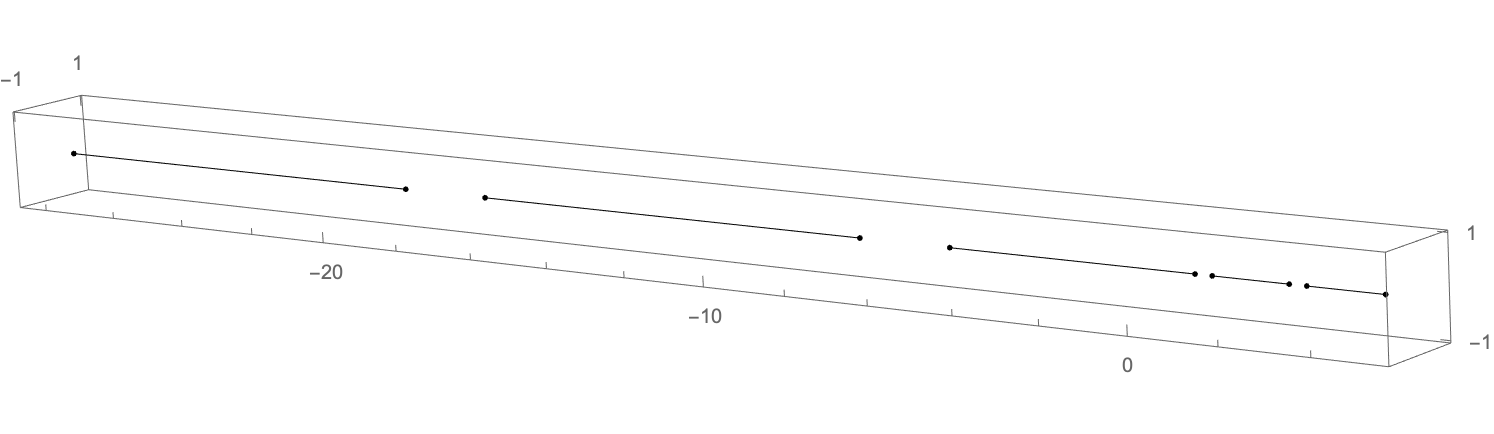}
    \caption{As seen in the above figure, the line is \emph{extended} before crossing the origin and \emph{contracts} afterwards.}
    \label{fig10}
\end{figure}
\noindent
As presented in the section above and throughout the rest of this paper, each snapshot represents the apparent shape of the object at that corresponding time. We can produce a plot of the apparent length of a line with proper length 1 as a function of \(x_{\text{obs}}^0\) for different velocities $\Vec{\beta}$ =\(\beta , 0, 0)\):
\begin{figure}[H]
    \centering
    \includegraphics[scale = 0.5]{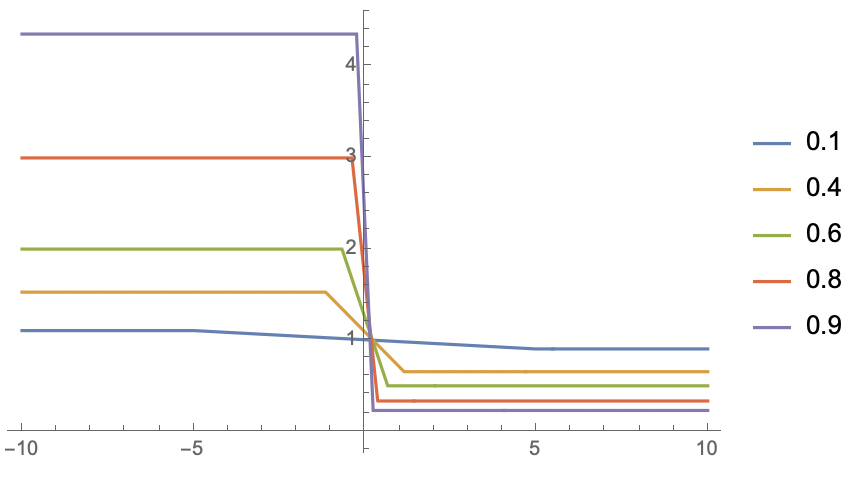}
    \caption{Apparent Length on the \textit{y}-axis and time \(x_{\text{obs}}^0\) on the \textit{x}-axis.}
    \label{fig11}
\end{figure}
\noindent
Let{'}s consider a vertical line given by \(\overset{\to }{x}'\) = $(0,0,x')$ in \(S'\) and $\Vec{a}$ = (0, 0, 0) moving in the positive \textit{x}-axis with \(\Vec{\beta} = (\beta , 0, 0)\). We need to evaluate $f$(\(\overset{\to}{a} +\Vec{\beta}x_{\text{obs}}^0\), \(\lambda \overset{\to }{x}'\)) in the same way as in equation (\ref{22}). To keep
the line symmetrical about the \textit{x}-axis, let $\lambda $ $\in $ [-1,1] instead.
\begin{equation}\label{23}
\left(\frac{\beta x_{\text{obs}}^0- \beta \sqrt{\left(x_{\text{obs}}^0\right)^2 \beta^2+ \left(x'\right)^2 \lambda ^2\left(1-\beta
^2\right)}}{1-\beta ^2}, 0, \text{\textit{$x'$}} \lambda \right), \lambda \in [-1,1]
\end{equation}
\noindent
Figure (\ref{fig12}) shows a series of snapshots of a vertical line with proper length 20 i.e. \(x'\) = 10 initially at $\Vec{a}$ = $(0, 0, 0)$ moving with a velocity of \(\Vec{\beta}= (0.9, 0, 0)\) observed at \(x_{\text{obs}}^0\) = -6, \(x_{\text{obs}}^0\)
= -3, \(x_{\text{obs}}^0\) - 0, \(x_{\text{obs}}^0\) = 3, \(x_{\text{obs}}^0\) = 6, and \(x_{\text{obs}}^0\) = 10.
\begin{figure}[H]
    \centering
    \includegraphics[scale = 0.4]{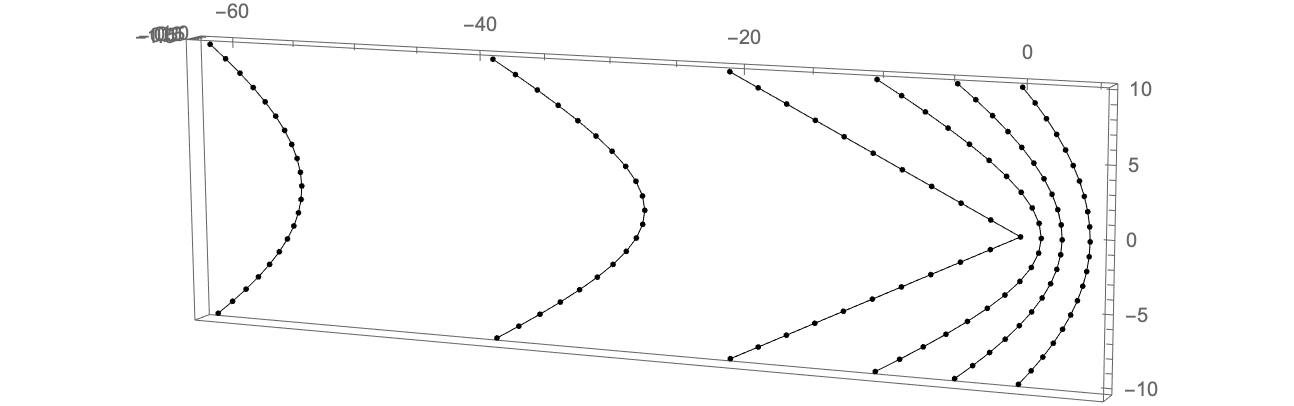}
    \caption{The hyperbola changes shape as it moves along the $x-$axis, degenerating to its asymptotes at  \(x_{\text{obs}}^0 = 0\).}
    \label{fig12}
\end{figure}
\noindent
From figure (\ref{fig12}), we can notice that the vertical line looks like a horizontal hyperbola which degenerates to its asymptotes at time \(x_{\text{obs}}^0\) = 0. From equation (\ref{23}), we can prove that the apparent shape of the line is indeed an hyperbola, and the equation of the asymptotes are \(\textrm{
\(x(y)=\pm \frac{\beta  }{\sqrt{1-\beta ^2}}y\)}\). We don{'}t need to translate the hyperbola near the origin to make the computation easier, as the curvature $\kappa $ and torsion $\tau $ are invariant under a change of orientation \cite{carmo}. Analysis of local properties is given in section (\ref{observe}). Figure (\ref{fig13}) shows a series of snapshots of a line with $\overset{\to }{x}' = (0, 0, 10)$ with $\lambda$ $\in$ [-1,1] moving with a velocity of $\Vec{\beta}$ = ($0.45$, $-0.45$, \(\frac{0.9}{\sqrt{2}}\)) (speed = 0.9 in the north east direction) at times \(x_{\text{obs}}^0\) = $-10$, \(x_{\text{obs}}^0\) = $-5$, \(x_{\text{obs}}^0\) = 0, \(x_{\text{obs}}^0\) = 5, \(x_{\text{obs}}^0\)= 10, and \(x_{\text{obs}}^0\)=15.
\begin{figure}
    \centering
    \includegraphics[scale = 0.3]{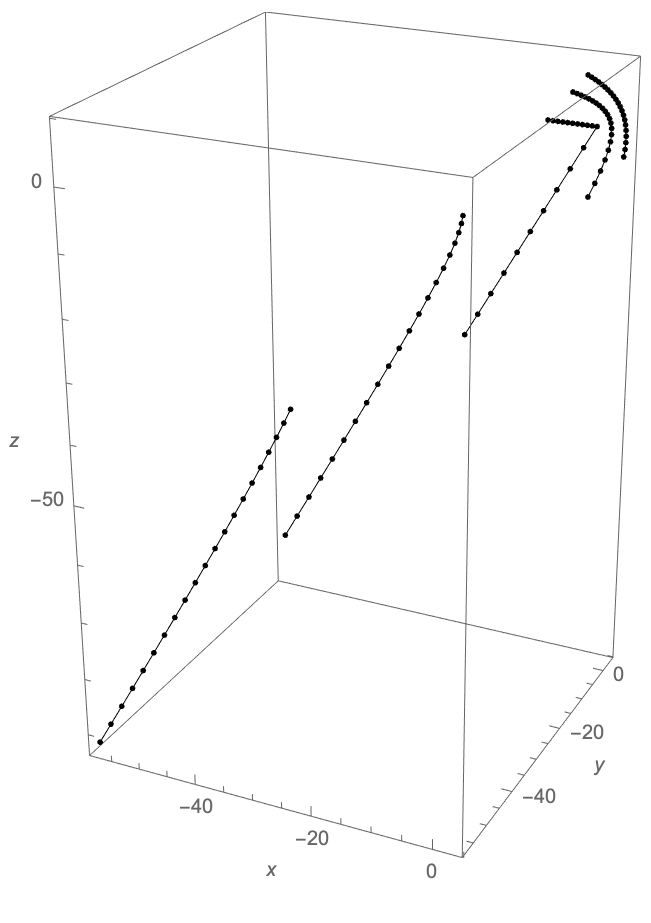}
    \caption{A vertical line moving diagonally up and to the right when $\Vec{a} = 0$. We can see that the line {`}bends{'} in the direction of motion.}
    \label{fig13}
\end{figure}
\noindent
We can also find the apparent shape of fast-moving arbitrary functions. Since any point can be written as \((x,y,f(x,y))\) in a given domain on the $x-y$ plane, we can use easily the methods described above. The
proper shape of the function will be defined in \(S'\), and \(\overset{\to }{x}'\) = (\(x'\), \(y'\), \textup{ \(f(x',y')\)}). The
4-vector $\Vec{a}$, as usual, will point from the origin of \(S\) to the origin of \(S'\). Let \(f(x',y')\) = \(\sin (x')\) for \(-\pi<x'<\pi\) and \(-1<y'<1\). Let $\Vec{\beta}$ = (\(\beta , 0, 0\)) and \(\Vec{a} =
(a,0,0)\). Below is a series of snapshots of the function above travelling at $\Vec{\beta}$ = \((0.9, 0, 0)\) with \textit{$\Vec{a}$} = $(0,0,0)$ seen at times \(x_{\text{obs}}^0 = -1\), \(x_{\text{obs}}^0 = 0\), \(x_{\text{obs}}^0 = 1\), \(x_{\text{obs}}^0 = 2\), and \(x_{\text{obs}}^0 = 5\):
\begin{figure}[H]
    \centering
    \includegraphics[scale = 0.37]{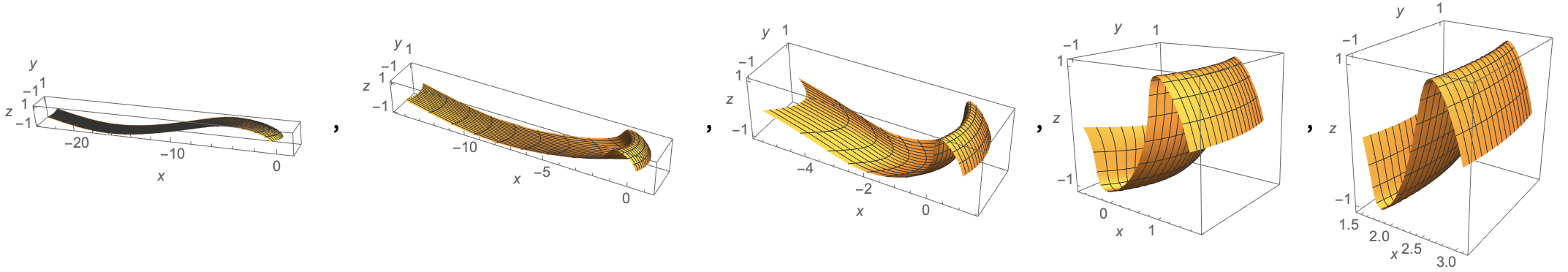}
    \caption{}
    \label{fig14}
\end{figure}
\noindent 
Below is a series of snapshots of the function above travelling with a speed of $|\Vec{\beta}| = 0.9$ in the direction defined by spherical polar coordinates $\theta = \frac{\pi}{6}$ and $\phi = \frac{\pi}{4}$ with $\Vec{a} = (1,-3,-0.6)$, seen at times \(x_{\text{obs}}^0 = -1\), \(x_{\text{obs}}^0 = 0\), \(x_{\text{obs}}^0 = 1\), \(x_{\text{obs}}^0
= 2\), and \(x_{\text{obs}}^0 = 5\):
\begin{figure}[H]
    \centering
    \includegraphics[scale = 0.37]{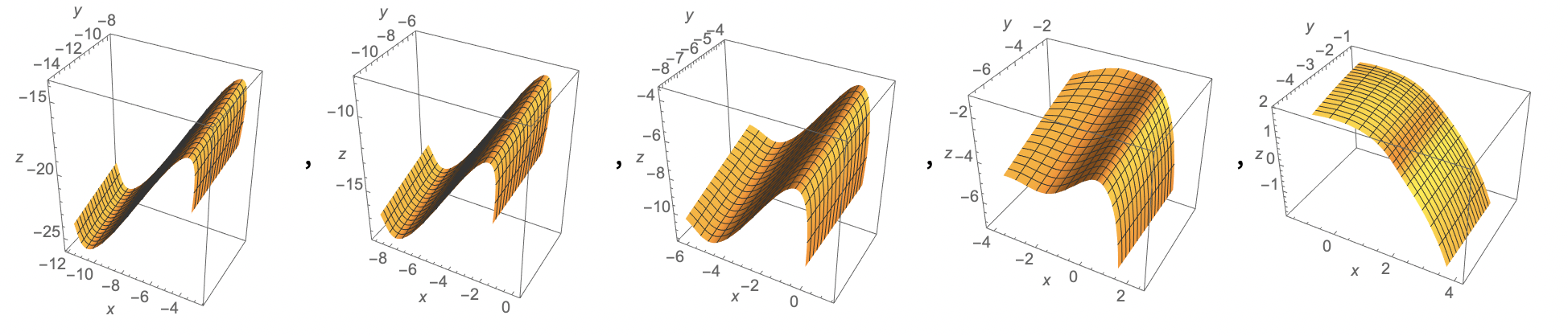}
    \caption{}
    \label{fig16}
\end{figure}
\noindent
We notice the same trend of the object contracting in direction of velocity. Below is a series of snapshots of the function $f(x',y') = x'^2 + y'^2$ for \(-1<x'<1\) and \(-1<y'<1\) travelling at \(\overset{\to }{\beta}\) = \((0,
0, 0.8)\) with $\Vec{a} = (0,0,0)$ for times \(x_{\text{obs}}^0 = -1\), \(x_{\text{obs}}^0 = 0\), \(x_{\text{obs}}^0 = 1\), \(x_{\text{obs}}^0 = 2\), and \(x_{\text{obs}}^0 = 5\):
\begin{figure}[H]
    \centering
    \includegraphics[scale = 0.35]{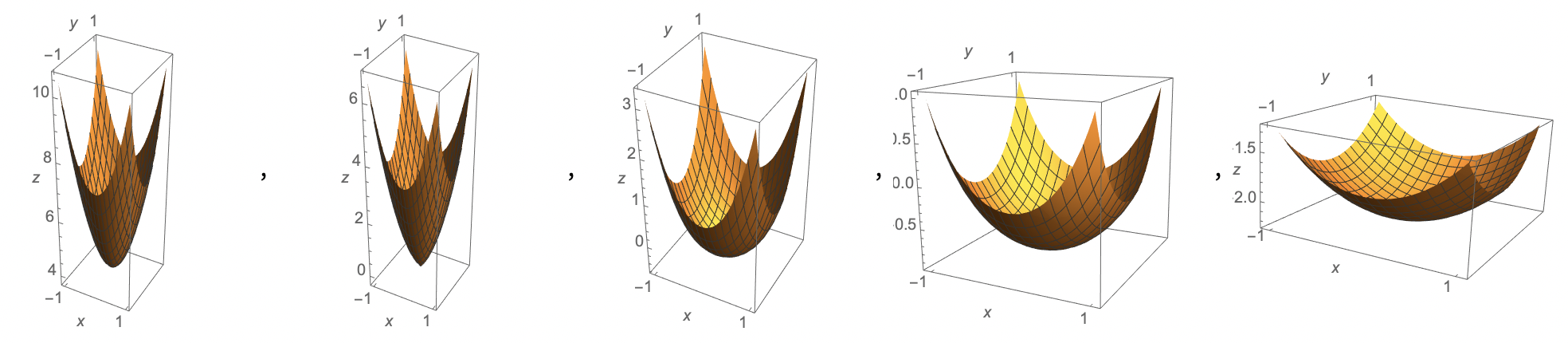}
    \caption{}
    \label{fig17}
\end{figure}
\noindent
To find the apparent shapes of fast-moving polyhedra, it is obvious that we must split the surface into $n$ points where $n$ is large. For a cube, we can find the apparent shapes of the 12 edges to form the outline of a cube. To define the {`}proper shape{'} of a cube, we can set the origin of \(S'\)as the center of the cube of side length \(l'\). To evaluate the apparent shape, we can compute $f$(\(\Vec{a} +\Vec{\beta}x_{\text{obs}}^0\), \((x',y',z')\)), \(\frac{-\text{\textit{$l$}}'}{2}\)
$<$ \(x',y',z'\) $<$ \(\frac{\text{\textit{$l$}}'}{2}\), where \((x',y',z')\) is a point on the cube. By evaluating the apparent shape of many such points and joining them together, we can retain the original mesh of the shape. Figure (\ref{fig18}) shows a series of snapshots of the cube with $\Vec{\beta}= (0.7, 0, 0)$, $\Vec{a} = (1, 0, 0)$, and $l' = 1$ at times \(x_{\text{obs}}^0 = -1\), \(x_{\text{obs}}^0\) = 0, \(x_{\text{obs}}^0\) = 0.5, \(x_{\text{obs}}^0\) = 1, and  \(x_{\text{obs}}^0\) = 3.
\begin{figure}
    \centering
    \includegraphics[scale = 0.41]{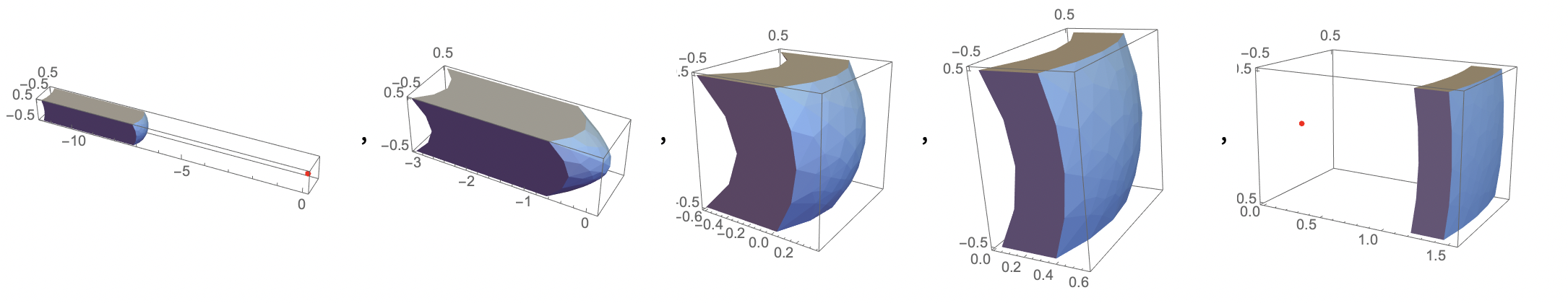}
    \caption{The red dot represents the stationary observer.}
    \label{fig18}
\end{figure}
\noindent In the case of an icosahedron, we can set \(S'\)as the center of the icosahedron. Let \(l'\) be the edge length of a regular icosahedron. Figure (\ref{fig19}) shows a series of snapshots of an icosahedron with $\Vec{\beta}$ = (0.9, 0, 0), $\Vec{a}$ = (0, 0, 0), and
\(l'\) = 1 at times $x_{\text{obs}}^0 = -2$, $x_{\text{obs}}^0 = -1$, \(x_{\text{obs}}^0\) = 0.5, \(x_{\text{obs}}^0\) = 1, and \(x_{\text{obs}}^0\)
= 3.
\begin{figure}
    \centering
    \includegraphics[scale = 0.32]{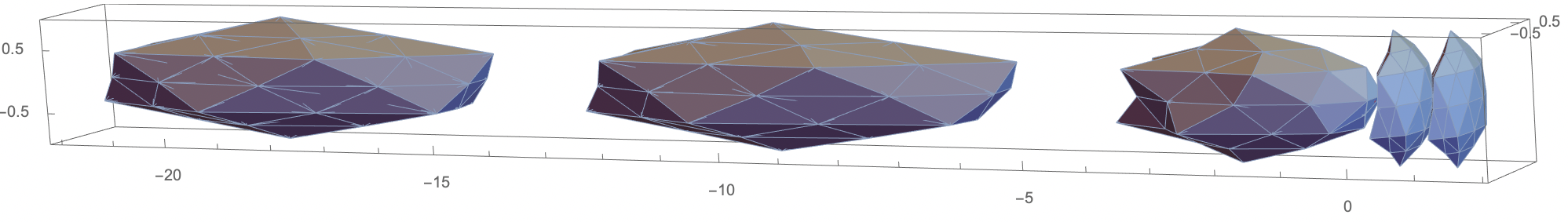}
    \caption{}
    \label{fig19}
\end{figure}
\noindent
Figure (\ref{figExotic}) shows three shapes$-$a cone, a torus, and the somewhat exotic shape of a cow$-$observed at time \(x_{\text{obs}}^0\) = 0, moving downwards with a speed of 0.95 times the speed of light.
\begin{figure}[H]
    \centering
    \includegraphics[scale = 0.25]{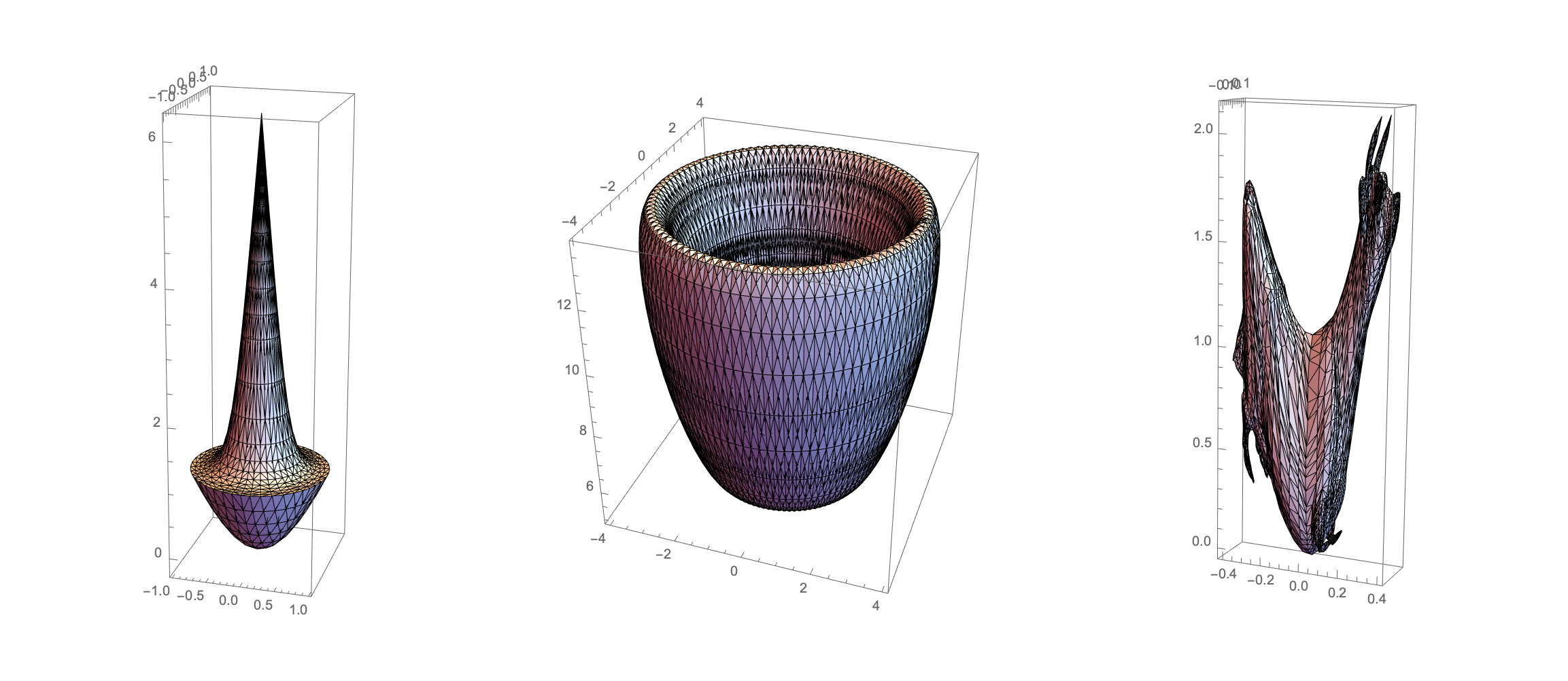}
    \caption{}
    \label{figExotic}
\end{figure}
\subsection{The Doppler Effect and Relativistic Aberration}\label{doppler_section}
As stated in \cite{Muller}, the Doppler factor emerges from equation (\ref{extra1}) when $\Vec{a} = 0$, $\Vec{\beta} = (\beta,0,0)$, and $\overset{\to }{x}'= 0$:
\begin{equation}\label{doppler1}
     x^{'0} = \sqrt{\frac{1 \pm \beta}{1 \mp \beta}}x_{\text{obs}}^0
\end{equation}
Exploring the regions of redshift and blueshift on the apparent shape of an object travelling near the speed of light may help us understand the distortion.  In \cite{illumination}, illumination in special relativity is explored and factors such as brightness and radiance are also considered. In this paper, however, we will only consider the apparent change in wavelength of received light due to the relative motion of the observer and source. These results may help us understand the effect of relativistic aberration. The frequency of light \textit{received} can be determined from the time period of the received wave, which is equal to the time between 2 consecutive wavefronts striking the observer. The longitudinal and transverse Doppler effect are explained in \cite{morin_2008}. Since we are dealing with arbitrary motion, we must be able to answer the following question: What is the frequency of light received by the stationary observer at $(0,0,0)$ in $S$ from a point located at $\overset{\to }{x}'$ with respect to the moving frame, which is travelling with a velocity of $\Vec{\beta}$ and is initially located at $\Vec{a}$ with respect to $S$. We must include the variable  $\overset{\to }{x}'$ since we are dealing with extended objects. Note that we will not denote the moving frame by $S'$\textemdash we use $S'$ to denote the frame of reference attached to $\overset{\to }{x}'$. Let $\Vec{d}$ denote the initial displacement of $S'$ from the origin of $S$ using equation (\ref{6}):
\begin{equation}\label{doppler2}
    d^{i} = x^{'i} + \frac{\gamma^2}{1 + \gamma} \beta^{i} (\beta_{j} x^{'j}) + a^{i}
\end{equation}
\par
\begin{figure}
    \centering
    \includegraphics[scale = 0.5]{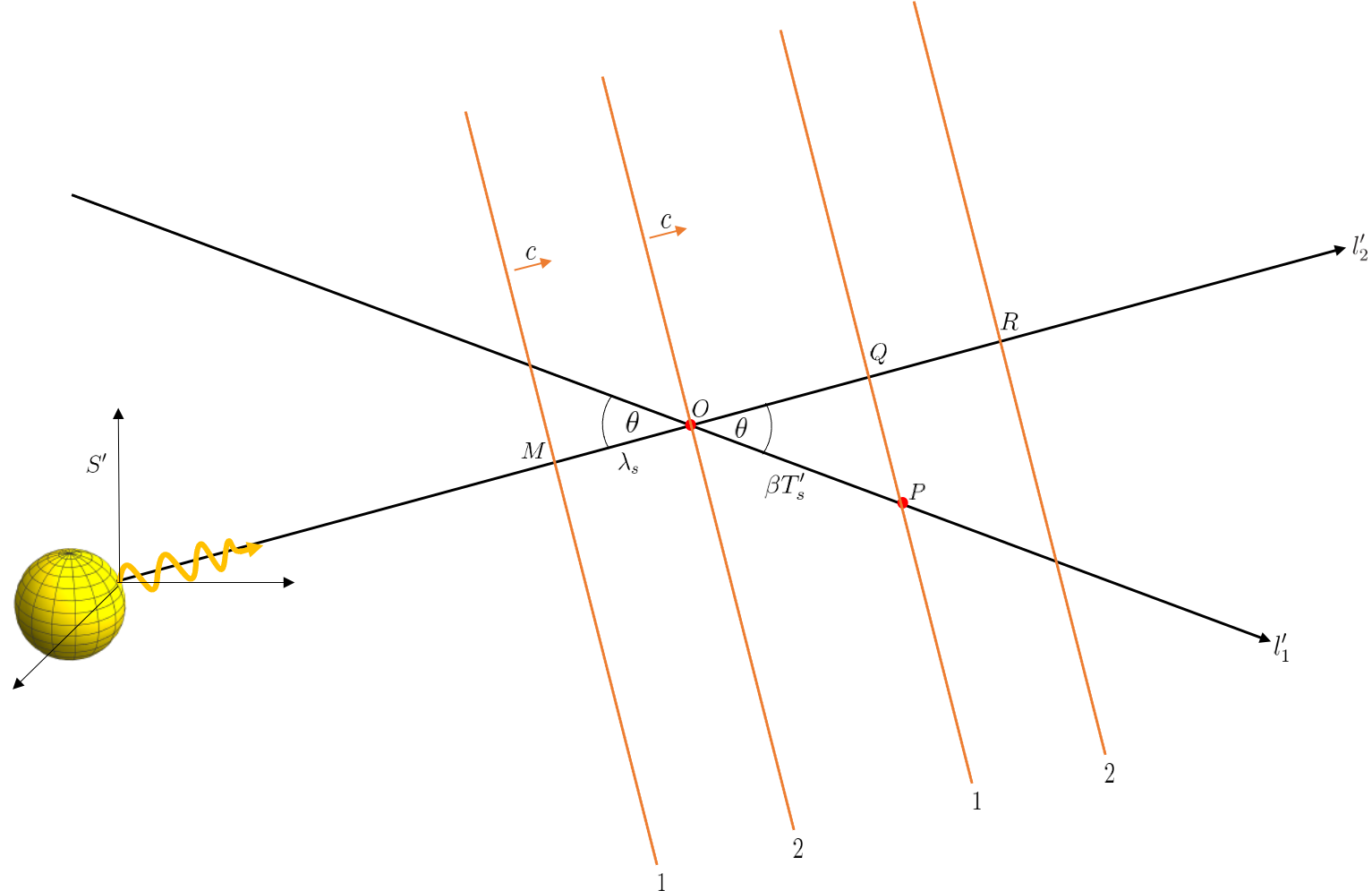}
    \caption{}
    \label{doppler_fig_1}
\end{figure}

\noindent
Imagine yourself attached to a random point on a sphere moving relative to a stationary observer. From your perspective, the observer is the one who is moving. Let $l_{1}'$ be the line representing the motion of the observer with respect to your frame $S'$. Let $x^{'0}_{\text{em}}$ be the time on your watch when you decide to switch on a torch (we will assume that the torch emits monochromatic light of frequency $f_{s}$). Figure (\ref{doppler_fig_1}) shows a light beam crossing paths with the moving observer at $O$ at a time $x^{'0}_{\text{obs}}$ on your watch. At time $x^{'0}_{\text{obs}}$ on your watch, a wavefront reaches the observer. The extra time it takes for wavefront behind (through $M$) to strike the moving observer again is the time period $T_{s}'$ of the wave perceived by the observer measured in \emph{your} reference frame at time $x^{'0}_{\text{obs}}$. Note that the time $T_{s}'$ will not be the same according to the moving observer's clock, which ticks slower with respect to your clock. In time $T_{s}'$, the observer would have move a distance $OP = \beta T_{s}'$ and the wavefront through $M$ would have travelled a distance $MQ = c T_{s}'$ to meet the observer at $P$. Therefore,
\begin{equation}\label{doppler3}
    \lambda_{s} + \mid \Vec{\beta} \mid T_{s}' \cos\theta = c T_{s}'
\end{equation}
\par\noindent
In units where $c = 1$, $T_{r}$, the time period measured by the moving observer, is given by $T_{s}'/\gamma$. Therefore, the received frequency can be found from equation (\ref{doppler2}):
\begin{equation}\label{doppler4}
    f_{r} = \gamma (1 - \mid \Vec{\beta} \mid \cos\theta) f_{s}
\end{equation}
\par
\begin{figure}
    \centering
    \includegraphics[scale = 0.38]{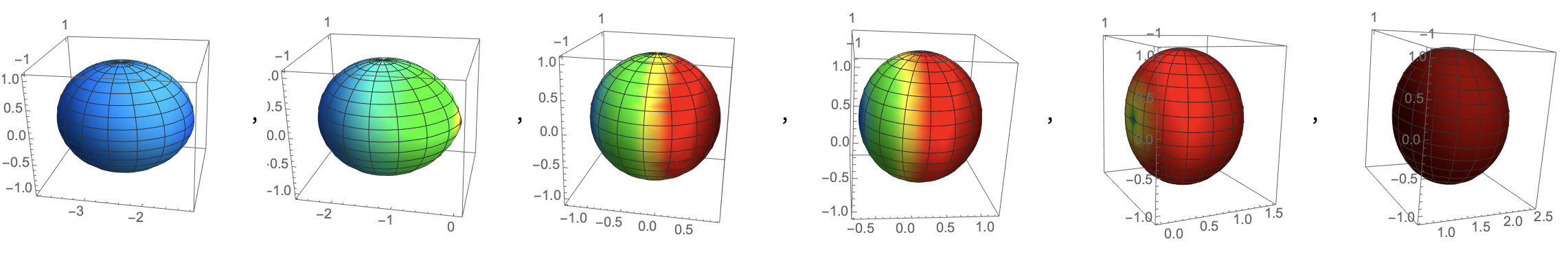}
    \caption{As shown above, at time $x_{\text{obs}}^0 = -10$, every point on the sphere is blueshifted. We also notice that the points approaching the observer are blueshifted, and the points receding away eventually become redshifted. At $x_{\text{obs}}^0 = 10$, every point on the sphere turns red.}
    \label{doppler_fig_2}
\end{figure}

\begin{figure}
    \centering
    \includegraphics[scale = 0.5]{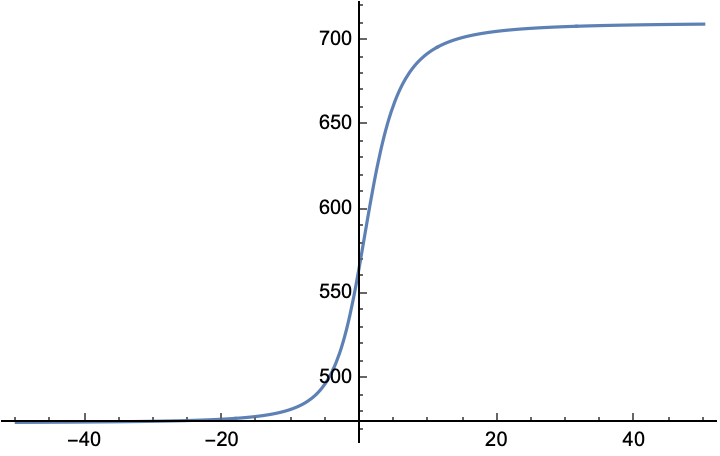}
    \caption{A graph of wavelength on the $y-$axis, measured in nanometres, against time on the $x-$axis.}
    \label{doppler_fig_3}
\end{figure}

\noindent
We can find the value of $\cos\theta$ using the direction unit vectors of $l_{1}'$ and $l_{2}'$. We can find both of these if we find the vector equation of $l_{1}'$ denoted by $\overset{\to }{r}'$ as a function of $x^{'0}$, the time on your watch. Then, the unit vector along $l_{1}'$  is merely the normalised direction vector of $\overset{\to }{r}'$ and the unit vector along $l_{2}'$ is $\hat{r}^{'}(x^{'0}_{\text{obs}})$. To find $\overset{\to }{r}'$, we use the inverse Poincar\'e transformation:
\begin{equation}\label{doppler5}
    x^{'\mu} = \overline{\Lambda}^{\mu }_{\nu }\left(x^\nu - d^\nu\right)
\end{equation}
where \(\overline{\Lambda} ^0_0\) = $\gamma$, \(\overline{\Lambda} ^0_i=- \gamma \beta _i\), \(\overline{\Lambda} ^i_0\)
= \(-\gamma \beta ^i\), \(\overline{\Lambda} ^i_j = \delta _j^i+ \frac{\gamma ^2}{1+\gamma }\beta ^i\beta _j\), and $d^\nu$ is given in equation (\ref{doppler2}). The spatial components of $x^{'\mu}$ are $r^{'i}$ when $x^{i} = 0$, as we are considering the trajectory of the stationary observer with respect to your frame $S'$. Substituting $x^{i} = 0$ in equation (\ref{doppler5}), we get:
\begin{equation}\label{doppler6}
    r^{'i}(x^{'0}) = -\beta^{i}x^{'0} + \beta^{i}\left(\beta_{j}d^{j}\right)\left(\frac{\gamma}{1 + \gamma}\right) - d^{i}
\end{equation}
where $x^{'0}$ is related to the time on the stationary observer's clock:
\begin{equation}\label{doppler7}
    x^{'0} = \gamma x^{0} + \gamma \beta_{i}d^{i}
\end{equation}
According to equation (\ref{doppler6}), the vector along $l_{1}'$ is $-\Vec{\beta}$, and the vector along $l_{2}'$ is $r^{'i}(x^{'0}_{\text{obs}})$. Using the dot product to find the value of $\cos\theta$ and substituting in equation (\ref{doppler4}), we can find $f_{r}$ as a function of $x^{'0}_{\text{obs}}$.
\begin{equation}\label{doppler8}
    f_{r} = \gamma (1 + \Vec{\beta}\cdot\hat{r}^{'}(x^{'0}_{\text{obs}}))f_{s}
\end{equation}
We can use equation (\ref{doppler7}) to compute the received frequency as a function of $x^{0}_{\text{obs}}$. When we set $\Vec{\beta} = (\beta \cos\theta_{r},0,\beta\sin\theta_{r})$,  $\overset{\to }{x}' = (0,0,0)$, $\Vec{a} = (a,0,0)$, and $x^{0}_{\text{obs}} = a$,  equation (\ref{doppler8}) becomes:
\begin{equation}\label{doppler9}
    f_{r} = \frac{\sqrt{1-\beta^2}}{1 + \beta\cos\theta_{r}} f_{s}
\end{equation}
Equation (\ref{doppler8}) reduces to the ordinary transverse Doppler effects in \cite{morin_2008}. To simulate the Doppler effect for extended objects, we can set $\overset{\to }{x}'$ as the required parametric form. Then, every point on the object glows monochromatically with a frequency $f_{s}$, but different points have a different perceived frequency $f_{s}$. Setting $\overset{\to }{x}' = (r \sin\theta\sin\phi, r\sin\theta\cos\phi, r\cos\theta)$, we can compute $f_{s}(\theta,\phi)$. In this paper, we will visualise the appearance of objects that emit monochromatic light of wavelength $\lambda_{s}$ = 580 nm (nanometres), corresponding to yellow light. Figure (\ref{doppler_fig_2}) shows a series of snapshots of a sphere (radius = 1) travelling to the right with a speed of $0.2$, initially positioned at $(0,0,0)$, at times \(x_{\text{obs}}^0\)= $-10$, \(x_{\text{obs}}^0\)= $-5$, \(x_{\text{obs}}^0\)= $0$, \(x_{\text{obs}}^0\)= $2$, \(x_{\text{obs}}^0\)= $5$, and \(x_{\text{obs}}^0\)= $10$.
\noindent
In figure (\ref{doppler_fig_2}), the wavelength of light of any point on the sphere does not necessarily increase linearly with time $x_{\text{obs}}^0$. It can be approximated to vary linearly over a certain time interval, which can be checked from the formula for $f_{s}(\theta,\phi)$ found from equation (\ref{doppler8}). For example, we can plot the variation of the wavelength of the north pole of the sphere above corresponding to $\theta = 0$ with respect to time as shown in figure (\ref{doppler_fig_3}).
\noindent
Intuitively, the frequency is asymptotic to a minimum and maximum frequency because the transverse distance $a$, which is equal to the radius of the sphere in figure (\ref{doppler_fig_3}), becomes negligible compared to the horizontal distance. Evaluating $\lambda_{r}(0,\phi)$,
\begin{equation}\label{doppler10}
    \lambda_{r}(0,\phi) = \lambda_{s} \frac{\sqrt{1-\beta^2}}{1 - \frac{t\beta^2}{\sqrt{1-\beta^2 + t^2\beta^2}}}
\end{equation}
Assuming $\beta > 0$, it can be seen from equation (\ref{doppler10}) that the ratio of the minimum received wavelength to $\lambda_{s}$ is the Doppler factor in equation (\ref{doppler1}) with minus and plus signs in the numerator and denominator respectively (the opposite signs for the maximum received wavelength). This Doppler factor is for motion in the $x-$direction without any initial displacement in an orthogonal direction \cite{morin_2008}. Below is a series of snapshots of a sphere (radius = 1) travelling to the right with a speed of $0.6$ and initially positioned at $(0,0,0)$ at times \(x_{\text{obs}}^0\)= $-2$, \(x_{\text{obs}}^0\)= $-1$, \(x_{\text{obs}}^0\)= $0$, \(x_{\text{obs}}^0\)= $1$, \(x_{\text{obs}}^0\)= $2$, and \(x_{\text{obs}}^0\)= $3$:
\begin{figure}[H]
    \centering
    \includegraphics[scale=0.38]{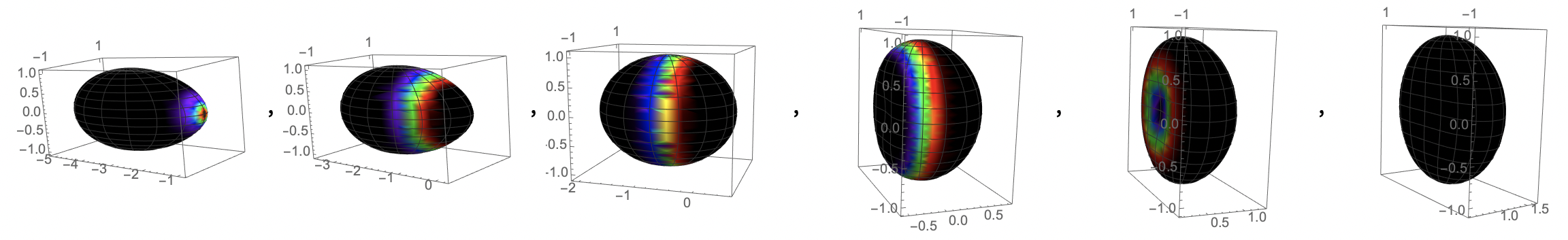}
    \caption{The same trend is observed in this figure. Furthermore, we can see that faster the velocity, faster the change in wavelength with respect to $\theta$ and $\phi$. The black areas represent wavelengths invisible to the human eye.}
    \label{doppler_fig_4}
\end{figure}
\noindent
Figure (\ref{doplol}) shows a series of snapshots of a sphere of radius 1 travelling with a speed of $(0.15, 0.15, 0.3/\sqrt{2})$ and initially positioned at $(1,0,0)$ at times \(x_{\text{obs}}^0\)= $-10$, \(x_{\text{obs}}^0\)= $-5$, \(x_{\text{obs}}^0\)= $0$, \(x_{\text{obs}}^0\)= $1$, \(x_{\text{obs}}^0\)= $2$, and \(x_{\text{obs}}^0\)= $5$.
\begin{figure}[H]
    \centering
    \includegraphics[scale=0.38]{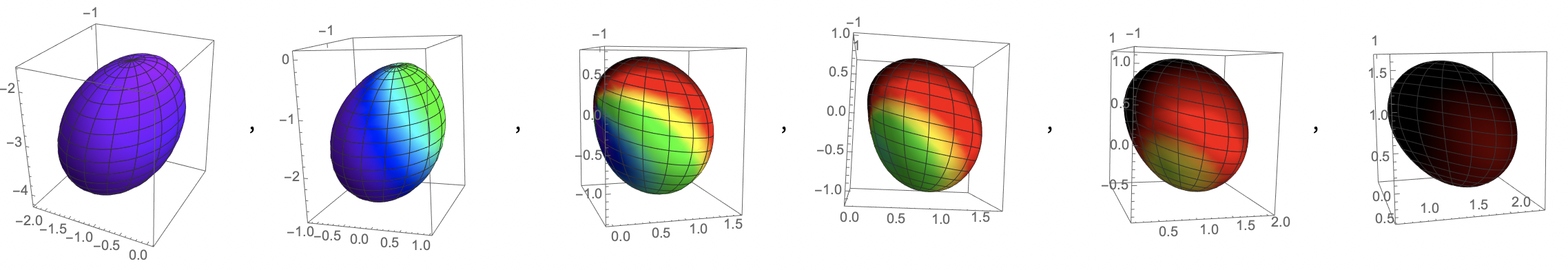}
    \caption{A sphere travelling with a speed of 0.3 in the direction described by $
\theta = \pi/4$ and $\phi = \pi/4$. The same trend is observed, and a more interesting distribution colours is obtained.}
    \label{doplol}
\end{figure}
\noindent
Aberration is the difference in the direction of propagating light in two inertial frames of reference. It can be thought of as the analogous `tilt' seen by an observer moving through falling rain, where the direction of rain in the stationary frame of reference is vertically downwards. In the context of this paper, aberration arises out of the fact that the stationary observer doesn't agree with the source on the direction of emitted light. In $S'$, the direction of emitted light is the vector along $l_{2}'$ (see figure (\ref{doppler1})), which is explicitly $\hat{r}^{'}(x^{'0}_{\text{obs}})$. In $S$, the direction of emitted (incoming) light is opposite to the unit vector pointing to the apparent position of the point (given in equation (\ref{10})). In fact, we can use the tools of relativistic aberration to re-derive the equation for $f_{r}$ in equation (\ref{doppler8}) and the apparent position of a point in $S$ quickly.
\par\noindent
As explained in \cite{CFT}, we can derive the formula describing relativistic aberration by considering a null four-vector $k^{\mu} = (\omega, \omega \hat{n})$  that represents a photon propagating in the direction $\hat{n}$ with a frequency $\omega$. Then, the components of the four vector in another frame of reference will be described by the Lorentz transformation, describing the same photon (null directions are preserved). Since the components of $k$ determine both the frequency of light and the direction of propagation, we expect the transformed four-vector to encode the equation for received frequency of light $f_{s}$ and some information about the apparent position of a moving point described in equation (\ref{10}). Specifically, we declare the four vector in $S'$ as representing the emitted photon (which ultimately reaches $S$ at time $x^{'0}_{\text{obs}}$ according to the source's clock:
\begin{equation}\label{doppler11}
    k^{'\mu}= (\omega', \omega' \hat{r}^{'}(x^{'0}_{\text{obs}}) )
\end{equation}
The vector in $S$, $k^{\mu}$, will be given by the Lorentz transformation (the initial separation $a^{\mu} = (0, \Vec{a})$ will not included as we are transforming a vector, not coordinates). Then, $k^{0} = \omega$ is given by:
\begin{equation} \label{doppler12}
    \omega = \gamma\omega' + \gamma\omega'\beta_{i}\hat{r}^{'i}(x^{'0}_{\text{obs}})
\end{equation}
which is the same as equation (\ref{doppler8}). Similarly, we can find the spatial components $k^{i}$. Since $k^{\mu}$ is a null four vector, its spatial components $k^{i}$ must be equal to $\omega \hat{n}^{i}$, where $\hat{n}^{i}$ is the unit vector describing the direction of incoming light in $S$ and $\omega$ is given in equation (\ref{doppler12}). Explicitly, 
\begin{equation} \label{doppler13}
    \hat{n}^{i} = \frac{\gamma\beta + \hat{r}^{'i} + \frac{\gamma^2}{1 + \gamma}\beta^{i}(\beta_{j}\hat{r}^{'j})}{\gamma + \gamma\beta_{i}\hat{r}^{'i}}
\end{equation}
where $\hat{r}^{'i}$ is evaluated at $x^{'0}_{\text{obs}}$ (note that the frequency in $S'$, $\omega'$, cancels out in the above equation). It can shown by explicit calculation that $\hat{n}^{i}$ in equation (\ref{doppler13}) is, as anticipated, equal to $-\frac{\Vec{x}_{\text{app}}}{|\Vec{x}_{\text{app}}|}$, where $\Vec{x}_{\text{app}}$ is $f\left(\Vec{a} + \Vec{\beta}x^{0}_{\text{obs}},\overset{\to }{x}'\right)$ given in equation (\ref{10}) (for example, in equation (\ref{doppler9}), the source was moving at an angle of $\theta_{r}$ with respect to the positive $x$-axis, emitting light at time $x^{0}_{\text{em}} = 0$). If we substitute these values in equation (\ref{doppler13}), we obtain that $\hat{n}^{i} = (-1,0,0)$. An alternative equation describing aberration is to find the angle $\theta_{s}$ the light beam makes with observer $S$ in $S'$. This is analogous to the angle $\theta$ shown in figure (\ref{doppler_fig_1}). As explained before, $\cos\theta_{s}$ is merely $-\hat{\beta}\cdot\hat{r}^{'i}(x^{'0}_{\text{obs}})$. When we set $\Vec{\beta} = (\beta \cos\theta_{r},0,\beta\sin\theta_{r})$,  $\overset{\to }{x}' = (0,0,0)$, $\Vec{a} = (a,0,0)$, and $x^{0}_{\text{obs}} = a$ as before, we can express $\cos\theta_{s}$ in terms of $\cos\theta_{r}$ and hence, $\cos\theta_{r}$ in terms of $\cos\theta_{s}$. After some algebra, we obtain:
\begin{equation}\label{doppler14}
    \cos\theta_{r} = \frac{\beta - \cos\theta_{s}}{\beta\cos\theta_{s} - 1}
\end{equation}
which is the equation describing relativistic aberration in Einstein's 1905 paper on special relativity. 
\subsection{Other Observations}\label{observe}
Let{'}s find the apparent position \(\overset{\to }{x}\) of a point moving with velocity \(\Vec{\beta} =(\beta, 0 ,0) \) crossing the origin of \(S\) at time \(t\) = 0 i.e. \(\text{\textit{$\Vec{a}$}}\) = 0 = \(\overset{\to }{x}'\): 
\begin{equation}\label{24}
\overset{\to }{x}(x_{\text{obs}}^0)=\left(\frac{x_{\text{obs}}^0 \beta }{1 \pm  \beta },0,0\right)
\end{equation}
\noindent
Plotting \(\frac{d\text{\textit{$x$}}}{d\text{\textit{$t$}}}\) on the $y$-axis and \(x_{\text{obs}}^0\) on the $x$-axis with $-5$ < \(x_{\text{obs}}^0\) < $5$ for $\beta $ = 0.1, 0.3, 0.5, 0.7, 0.9:
\begin{figure}[H]
    \centering
    \includegraphics[scale = 0.4]{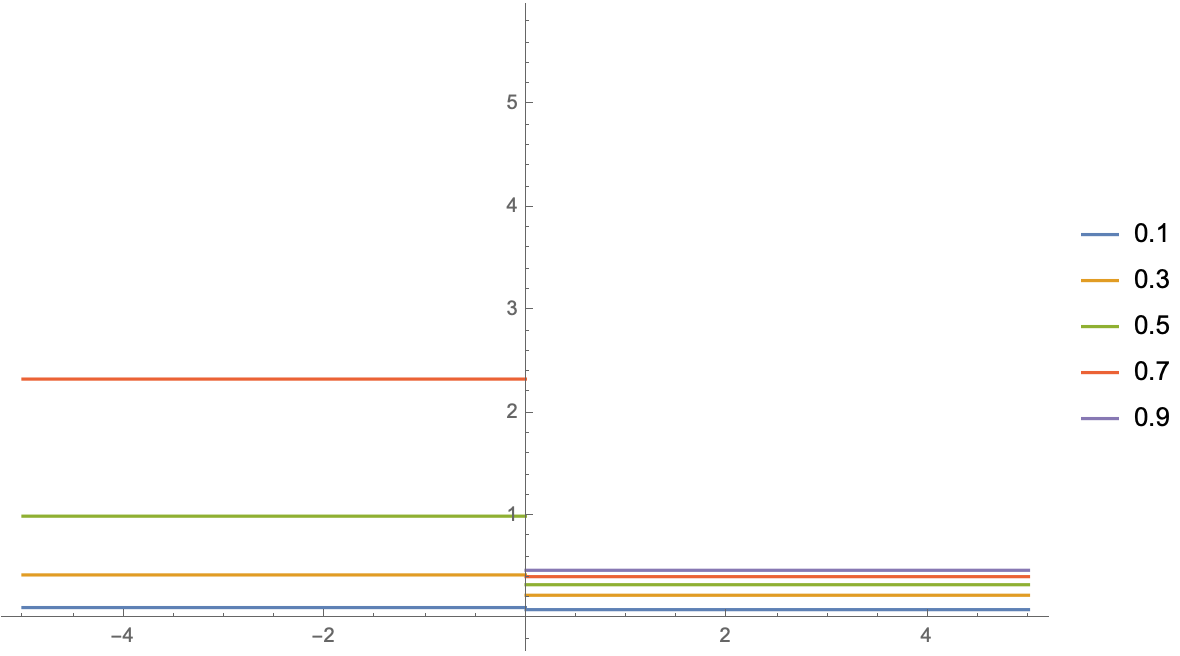}
    \caption{}
    \label{fig20}
\end{figure}
\noindent
Note that for negative times i.e. when the object approaches, actual velocities of 0.7 and 0.9 can appear to travel at speeds much greater than the
speed of light. Taking the time derivative of equation (\ref{24}), we obtain:
\begin{equation}\label{25}
\overset{\to }{v}(x_{\text{obs}}^0\pmb{)} = \left(\frac{t \beta }{t+\sqrt{t^2} \beta },0,0\right)
\end{equation}
For positive times, the speed is \(\frac{ \beta }{1+ \beta }\) and for negative times, the speed is \(\frac{ \beta }{1- \beta }\). This is also shown
in figure (\ref{fig20}) above. We can also analyse apparent shapes of parametrizable objects, such as spheres, geometrically.
\par\noindent
Recall the parametric equation of a sphere $\overset{\to }{x}_s(\theta ,\phi )$ at \(x_{\text{obs}}^0\)= 0 and \(\Vec{a}=0\) with \(\overset{\to}{\beta}=(\beta , 0,0)\). The line element $ds^2 = dx^2 + dy^2 + dz^2$ when the euclidean coordinates are parametrized by the curvilinear coordinates (\(u^{ 1},u^2\)) = ($\theta $, $\phi $) above can be computed using the metric tensor $g'_{ij} = e'_{i} \cdot e'_{j}$, where $e'_{i}$ are the new coordinate basis vectors. The induced metric tensor $g'_{ij}$ on the parametric surface $\Vec{x}_{s}$ in equation (\ref{16}) is given by:
\begin{equation}\label{newMetric}
    g'_{ij} = \sum_{k=1}^{3} \frac{\partial x'_{k}}{\partial u^{i}} \frac{\partial x'_{k}}{\partial u^{j}}
\end{equation}
where $\Vec{x}_{s} = (x'_{1},x'_{2},x'_{3})$. Note that are not changing the curvilinear coordinate parametrization; we are changing the surface itself. Let's express the transformation in equation (\ref{16}) as \(f\):\text{ }\(\mathbb{R}^3\to \mathbb{R}^3\). From equation (\ref{10}), we can express equation \ref{16} (the apparent position of a point $(x_{1},x_{2},x_{3})$ travelling with velocity $(\beta,0,0)$) in the following way:
\begin{equation}
    \left(x'_{1},x'_{2},x'_{3}\right) = \left(\frac{x_{1} - \beta\sqrt{x_{1}^2 + x_{2}^2 + x_{3}^2}}{\sqrt{1-\beta^2}}, x_{2},x_{3}\right)
\end{equation}
Denoting $x'_{i} = f_{i}(\textbf{x})$ and using the chain rule, it's easy to rewrite equation (\ref{newMetric}) in terms of the Jacobian matrix $\textbf{J}$ with components $J_{ij} = \frac{\partial f_{i}}{\partial x_{j}}$:
\begin{equation}
      g'_{ij} = \sum_{k=1}^{3} J_{kl} J_{km} \left(\textbf{e}_{i}\right)^{l} \left(\textbf{e}_{j}\right)^{m}
\end{equation}
where we use the summation convention over indices $l$ and $m$.
\begin{equation}\label{26}
\left[g_{\text{\textit{ij}}}\right] =\left(
\begin{array}{ccc}
\frac{\text{\textit{$r$}}^2-\text{\textit{$r$}}^2
\beta ^2 \cos ^2\theta \cos ^2\phi - \text{\textit{$r$}}^2 \beta ^2 \sin ^2\theta }{1-\beta ^2} & \frac{\text{\textit{$r$}}^2 \beta ^2 \sin
2\theta  \sin 2\phi }{4\left(1-\beta ^2\right)} \\ \frac{\text{\textit{$r$}}^2
\beta ^2 \sin 2\theta  \sin 2\phi }{4\left(1-\beta ^2\right)} & \frac{\text{\textit{$r$}}^2 \sin ^2\theta  \left(1-\beta ^2 \sin ^2\phi \right)}{1-\beta
^2} \\
\end{array}
\right)
\end{equation}
\noindent
We can also compute the metric tensor through equation (\ref{16}).
To quantify the curvature of the surface of the sphere, we compute the scalar curvature \(R\) for the resulting 2-dimensional metric above:
\begin{equation}\label{27}
R\text{($\theta $, $\phi $)}=\frac{2 \left(1-\beta ^2\right)}{\text{\textit{$r$}}^2 \left(1-\beta ^2 \sin ^2\theta  \sin ^2\phi \right)^2}
\end{equation}
\noindent
We can calculate the Gaussian curvature $K(\theta,\phi)$ as well, which satisfies the relation \(K\) = \(\frac{R}{2}\). We can verify the equation above by substituting $\beta = 0$\textemdash the Gaussian curvature becomes
\(\frac{1}{r^2}\), which is the curvature for an ordinary sphere.
\begin{figure}
    \centering
    \includegraphics[scale = 0.35]{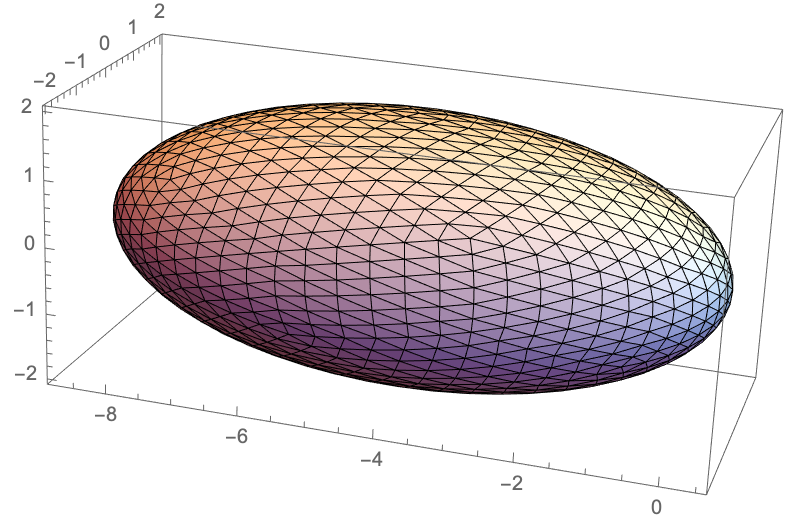}
    \caption{The apparent mesh of a sphere described in equation (\ref{27}) with $\beta $= 0.9 and \(r\) = 2.}
    \label{fig21}
\end{figure}
\noindent
We obtain \(K\left(\frac{\pi }{2},\frac{\pi }{2}\right)\) = \(\frac{1}{\text{\textit{$r$}}^2\left( 1-\beta ^2\right)}\) from equation (\ref{27}). \(K\left(\frac{\pi }{2},\frac{\pi }{2}\right)\) increases in value as $\beta $ increases, which is what we expect as the sphere is more
stretched at time \textrm{\(x_{\text{obs}}^0\)} = 0. Moreover, \(K\) at the poles approaches $\infty $ as \(\beta\)\(\to\)1. Note that $\theta $
is the polar angle and $\phi $ is the azimuthal angle. The gaussian curvature on the north and south poles, which is equal to \(K(0,\phi )\) and
\(K(\pi ,\phi ) ,\) is equal to \(\frac{1-\beta ^2}{\text{\textit{$r$}}^2}\). The curvature at the poles decrease with increasing $\beta $, which
can also be seen from the figure as the sphere gets flattened out at the top and bottom. Note that the metric tensor in equation (\ref{27}) is a symmetric
matrix i.e. \(g_{\text{\textit{ij}}}\) = \(g_{\text{\textit{ji}}}\). This shows the symmetrical structure of the sphere in figure (\ref{fig21}), specifically
that the apparent shape of the sphere is symmetrical about the horizontal and vertical planes passing through its center. Equation (\ref{27}) implies that the curvature varies across the surface. In fact, we expect the curvature to change under the transformation $\phi$ \(\to
 \frac{\pi }{2} - \phi\) and $\theta \to  \frac{\pi }{2} - \theta $, which can seen intuitively from figure (\ref{fig21}) as well.
\par \noindent
Let{'}s compute the curvature and torsion of the curve represented in equation (\ref{23}). Let{'}s parametrize equation (\ref{23}) by arc length \(s\). Since
\(s\) as a function of $\lambda $ satisfies \(\frac{\text{ds}}{\text{d$\lambda $}}\)= $|$\(\overset{\to }{r}'(\lambda )\)$|$, we can find $\lambda
$(\(s\)). After some algebra, we obtain the curvature and torsion as a function of $\lambda $: 
\begin{equation}\label{28}
\kappa (\lambda ) = \frac{\left(x_{\text{obs}}^0\right)^2 \beta ^3}{\left(\left(x_{\text{obs}}^0\right)^2 \beta ^2+(\text{\textit{$x'$}}\text{\textit{$
$}})^2 \lambda ^2\right)^{3/2}}, \tau  = 0
\end{equation}
Since the curve described by equation (\ref{23}) is a plane curve, we expect the torsion $\tau $ to be 0, which is indeed the case. $\kappa $ at $\lambda
$ = 0 is given by \(\frac{1}{x_{\text{obs}}^0}\), which is an intuitive result$-$as seen in figure (\ref{fig13}), the curvature of the curve at $\lambda
$ = 0 decreases as $|$\(x_{\text{obs}}^0\)$|$ increases. At \(x_{\text{obs}}^0\)= 0, we get infinite curvature as the curve is not differentiable
at $\lambda $ = 0. Moreover, $\kappa $(0) is independent of $\beta $. Computing the torsion of a line with \(\Vec{a} = 0\), velocity \(\overset{\to
}{\beta }\) = ($\beta $, 0, 0), and position vector \(\overset{\to }{x}'\) = $\lambda $(\(a\), \(b\), \(c\))with respect to the origin of \(S'\),
we obtain that $\tau $ = 0. This indicates the apparent shape of any line travelling along the axes will be a plane curve. Furthermore, computing
the torsion of a line with \(\Vec{a} = 0\), velocity $\Vec{\beta}$ = \(\left(\beta _1\right.\),\(\beta _2\), \(\beta _3\)),
and position vector \(\overset{\to }{x}'\) = $\lambda $(\(a\), \(b\), \(c\)) with respect to the origin of \(S'\), we again obtain that $\tau $ =
0. So the torsion of the curves in figure (\ref{fig14}) is 0 for every $\lambda $. In fact, for every curve whose proper shape lies in the plane, its visual
appearance when moving in a straight line will also be in a plane i.e. $\tau $ = 0.
\section{Acceleration}\label{section3}
\subsection{Bell's Spaceship Paradox and Born Rigidity}
Much of the controversy surrounding the Bell{'}s Spaceship Paradox was clarified in \cite{paradox}. The authors stated that the string connecting the spaceships
will break even if the stationary observer measures its length as constant throughout. In fact, when the rightmost and leftmost ends of the strings
have the same proper acceleration, the actual length of the string increases with the time because the spaceship on the farther end accelerates faster.
Therefore, in reality, there \textit{is} indeed a tension in the string which does cause it to break. To make sure there are no stresses within the string, we need to make sure that the proper acceleration of each point falls with distance in such a way as to constitute a uniformly accelerating reference system. Generally, a rigid object exhibits motion in special relativity if the infinitesimal proper distance (the distance measured by a co-moving inertial observer) between 2 near points on an object is constant throughout for every point on the object. Thus, in Bell's spaceship paradox, the proper acceleration must fall with distance. If the endpoints of a rod have the same proper acceleration, then the proper length of the rod will increase in a co-moving frame due to relativity of simultaneity.
\par\noindent These conditions that must be satisfied by a rigid object are known as the \textit{Born rigidity} conditions \cite{born}. We may classify Born rigid motions using the Herglotz–Noether theorem, as mentioned in \cite{herglotz} and \cite{noether}, which states that a body can be brought into translational motion from rest by applying a force without violation of the Born Rigidity conditions, whereas we cannot bring a rigid object from rest into rotational motion. In the next section, we derive the necessary equations of motion that satisfy the Born Rigidity conditions (a basic derivation is given in 
\cite{franklin}). As mentioned in \cite{CFT}, \emph{rigid bodies} can't actually exist in relativity$-$we assume that force propagates instantaneously across the object, which obviously can't be the case. But we can define rigid-body \emph{motion} instead.
\subsection{Defining the Problem}
In section (\ref{prob_object}), we utilised \(\overset{\to }{x}'\)to define the proper shape of the object. In the case of acceleration, the proper shape of the
object can be defined in a uniformly accelerating reference frame \(S'\), in which the proper acceleration of each point on the extended body falls
with distance away from the origin. Note that the observer located in \(S'\) will perceive the same shape of the object at all times whereas the
observer in \(S\) will measure the object getting shorter and shorter each time. 
\par\noindent
In other words, a uniformly accelerating reference frame implies that no stresses act within the object i.e. every point on the object experiences
the same force that causes it to accelerate. On the other hand, if each point on the extended object has the same proper acceleration as measured
in \(S'\), then the object will stretch i.e. increase in length from the perspective of a person sitting on the object located in \(S'\). Even if
the object will have a constant length and shape from the perspective of \(S\), as shown in \cite{paradox}, the shape of the object as viewed from the observer
sitting on the extended object will change (stretch), meaning that every point on the object experiences different forces. Specifically, the points
on the object further from \(S\) will experience greater forces. This is obviously not feasible. Hence, the only logical conclusion is to find the visual appearance of objects accelerating in a \textit{ uniformly accelerated reference frame}.
\par\noindent
In Section (\ref{section2}), the independent variables for determining the apparent visual shape of the object were the time of observation \(x_{\text{obs}}^0\),
the velocity $\Vec{\beta}$, the initial position of \(S'\)($\Vec{a}$). In the case of acceleration, the independent variables
should be the time of observation \(x_{\text{obs}}^0\), the proper acceleration of an \textit{arbitrary point} of the object $\alpha $, the initial
three velocity \(v_i\) of that arbitrary point, and the initial position $\Vec{a}$ of that arbitrary point . The reason for choosing
an arbitrary point is straightforward; if we want to define the proper shape of a horizontal line of length \(l\), we can declare the leftmost end
as the arbitrary point with proper acceleration $\alpha$. Then, to constitute a uniformly accelerated reference frame, the coordinates of the left
most point on the spatial axis of \(S'\) must be \(\left(\frac{1}{\alpha },0,0\right)\). By defining the initial velocity \(v_i\) the proper acceleration $\alpha $, and the initial position $\Vec{a}$ of the leftmost point, we can figure out the trajectories for all the other points on the object (in section (\ref{app_accel}), we choose the midpoint of a horizontal line instead of the leftmost one). These conclusions are proven formally in section (\ref{feqm}). For a sphere, the arbitrary point can be defined as the centre.
\subsection{Finding the Equations of Motion}\label{feqm}
Let{'}s find the equation of motion of every point on a horizontal rod of proper length \(l'\) moving with an initial velocity of 0. Let{'}s say
that the initial position of the midpoint of the rod i.e. at time \(x^0 = 0\) is the origin of \(S\), and that the constant proper acceleration of
the midpoint is $\alpha $. From this information, we can easily find the trajectory of the midpoint with respect to the stationary observer at the
origin of \(S\):
\begin{equation}\label{29}
(t, x) = \left( \frac{1}{\alpha }\sinh \alpha \tau , \frac{1}{\alpha }\cosh \alpha \tau -\frac{1}{\alpha }\right),
\end{equation}
\noindent
To find the equation of motion of the rightmost end of the rod, which is at a distance \(\frac{l'}{2}\) light seconds from the midpoint, we must
remember that the extended rod constitutes a uniformly accelerated reference frame. That is, the distance of the midpoint from the origin of \(S'\)
is \(\frac{1}{\alpha }\) light seconds. Therefore, the distance of the right most point should be \(\frac{1}{\alpha }\) + \(\frac{l'}{2}\) light
seconds, implying that equation of motion of the rightmost end is:
\begin{equation}\label{30}
(t, x) = \left( \frac{2 + \alpha  l'}{2 \alpha  }\sinh  \left(\frac{2 \alpha }{2 + \alpha  l'}\tau \right) , \frac{2 + \alpha  l'}{2 \alpha  }\cosh \left(
 \frac{2 \alpha }{2 + \alpha  l'}\tau \right)  - \frac{1}{\alpha }\right),
\end{equation}
\noindent
The proper acceleration of the right most point is clearly \(\frac{2 \alpha }{2 + \alpha  l'}\), which is shown in the equation above. The reason
why we subtract \(\frac{1}{\alpha }\) instead of \(\frac{2 + \alpha  l'}{2 \alpha  }\) from \(x(\tau\)) is because we want the initial position (at
time \(x^0 = 0\)) of the rightmost point to be \(\frac{l'}{2}\) light seconds away from the midpoint in \(S\) and not at the origin of \(S\). One
may ask that \(\frac{l'}{2}\) metres is defined in \(S'\), not in \(S\). However, note that since the initial velocity \(v_i = 0\), there is no Lorentz
contraction at time \(x^0 = 0\). Hence, the length of the horizontal rod at time \(x^0 = 0\) is equivalent to the proper length \(l'\). This can be seen in the figure below which consists of the world-lines represented by equation (\ref{30}) when \(l'\) = 10 and $\alpha $ = 0.15 in \(S\):
\begin{figure}[H]
    \centering
    \includegraphics[scale = 0.4]{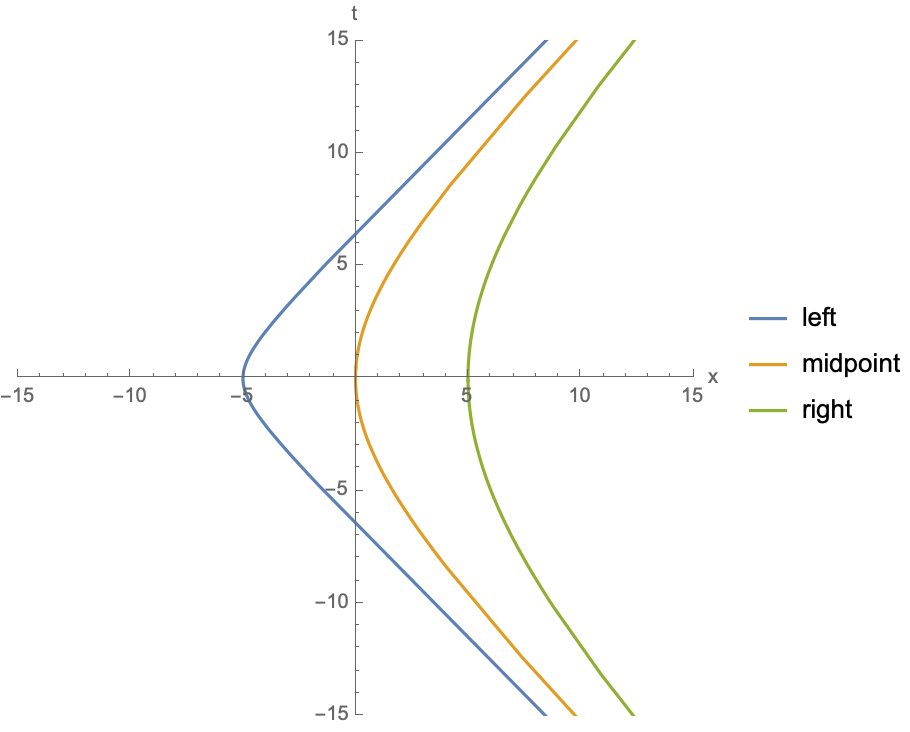}
    \caption{}
    \label{fig22}
\end{figure}
\noindent
The orange world line is the world-line of the rightmost point, and the blue world-line is that of the midpoint. As seen in the figure above, the
length of rod decreases with time from the perspective of the stationary observer \(S\) (the horizontal distance between the word-line falls). In
a similar way, we can find the world line of the leftmost point, which is -\(\frac{l'}{2}\) light seconds away from the midpoint on the spatial axis
of \(S'\).  Now, let{'}s find the word-line of each point on a rod of proper length \(l'\) whose midpoint has an initial position of \(x_0\), proper acceleration
$\alpha $, and initial velocity \(v_i\). It is obvious that the new trajectory of the midpoint is merely a translation of the curve such that \(\frac{d\text{\textit{$x$}}}{d\text{\textit{$t$}}}(t
= 0)\) = \(v_i\) and \(x\)(t = 0) = 0. Let the new curve be \((t,x)\) = (\(\frac{1}{\alpha }\)\(\sinh\)\(\alpha \tau\) + \(a\), \(\frac{1}{\alpha
}\)\(\cosh\)\(\alpha \tau\) - \(\frac{1}{\alpha }\) + \(b\)). In this case, \(\frac{dx}{d\text{\textit{$t$}}}\) is given by \(\tanh (\alpha \tau
)\). \(\frac{d\text{\textit{$x$}}}{d\text{\textit{$t$}}}(\tau\)) = \(\tanh (\alpha \tau )\) at \(t = 0\) is given by $\frac{dx}{dt}$ at $\tau = \frac{\text{arcsinh}(-\text{a$\alpha $})}{\alpha }$:
\begin{equation}\label{31}
\frac{d\text{\textit{$x$}}}{dt} = -\frac{\text{\textit{$a$}} \alpha }{\sqrt{1+\text{\textit{$a$}}^2 \alpha ^2}} = v_i,\text{at } \tau  =\frac{\text{arcsinh}(-a
\alpha )}{\alpha }
\end{equation}
\noindent
Solving for \(a\), we get 2 solutions. If \(v_i>0\), then we know that the curve defined by equation (\ref{29}) has to shift downwards. Therefore, \(a\)
has to be the opposite sign of \(v_i\):
\begin{equation}\label{32}
    a =-\frac{v_i}{\alpha \sqrt{1-v_i^2}}
\end{equation}
Since \(x(t = 0) = x(\tau  =\)\(\frac{\text{arcsinh}(-\text{a$\alpha $})}{\alpha }\)) = 0, we can solve for \(b\) and obtain that $b = \frac{1-\frac{1}{\sqrt{1-v_i^2}}}{\alpha }$.
Now, we can easily evaluate the equation of motion of the central point. Since every word-line should be translated by the same point, we can apply
the same translation \((a,b)\) to equation (\ref{30}): 
\begin{equation}\label{33}
(t, x) = \left(\frac{2 + \alpha  l'}{2 \alpha  }\sinh  \left(\frac{2 \alpha }{2 + \alpha  l'}\tau \right) -\frac{v_i}{\alpha \sqrt{1-v_i^2}}, \frac{2
+ \alpha  l'}{2 \alpha  }\cosh  \left(\frac{2 \alpha }{2 + \alpha  l'}\tau \right) - \frac{1}{\alpha \sqrt{1-v_i^2}}+ x_0\right)
\end{equation}
\noindent
In the above equation, we have introduce the final variable: the initial position of the midpoint \(x_0\). This is equivalent to translating every
world-line by \(x_0\) units in the \(x\)-direction. Therefore, we can just add \(x_0\) to the parametric form in equation (\ref{33}). 
\noindent
We can generalise equation (\ref{33}): to find the equation of motion of a point \(l'\) light seconds away from another point which moves with proper acceleration
$\alpha $, has an initial position \(x_0\), and has an initial 3-velocity \(v_i\). To do so, we must replace \(l'\) by \(2l'\) in equation (\ref{33}). We find $x(\tau)$ and $t(\tau)$:
\begin{equation}\label{34}
x(\tau ) = \frac{x_0\alpha -\frac{1}{\sqrt{1-v_i^2}}+\cosh \left(\frac{\alpha  \tau }{1+\alpha  l'}\right)\left(1+\alpha  l'\right)}{\alpha }
\end{equation}
\begin{equation}\label{35}
t(\tau ) = -\frac{v_{i}}{\alpha\sqrt{1 - v_{i}^2}} + \left(\frac{1}{\alpha} + l'\right)\sinh\left(\frac{\alpha\tau}{1 + \alpha l'}\right)
\end{equation}
\noindent
Figure (\ref{fig23}) shows the word-lines of the leftmost, central, and the rightmost points of a horizontal line (\(l' = 6)\) when the central point has a proper acceleration of 0.9, \(v_i\) = 0.5, and \(x_0 = 1\).
\begin{figure}
    \centering
    \includegraphics[scale = 0.35]{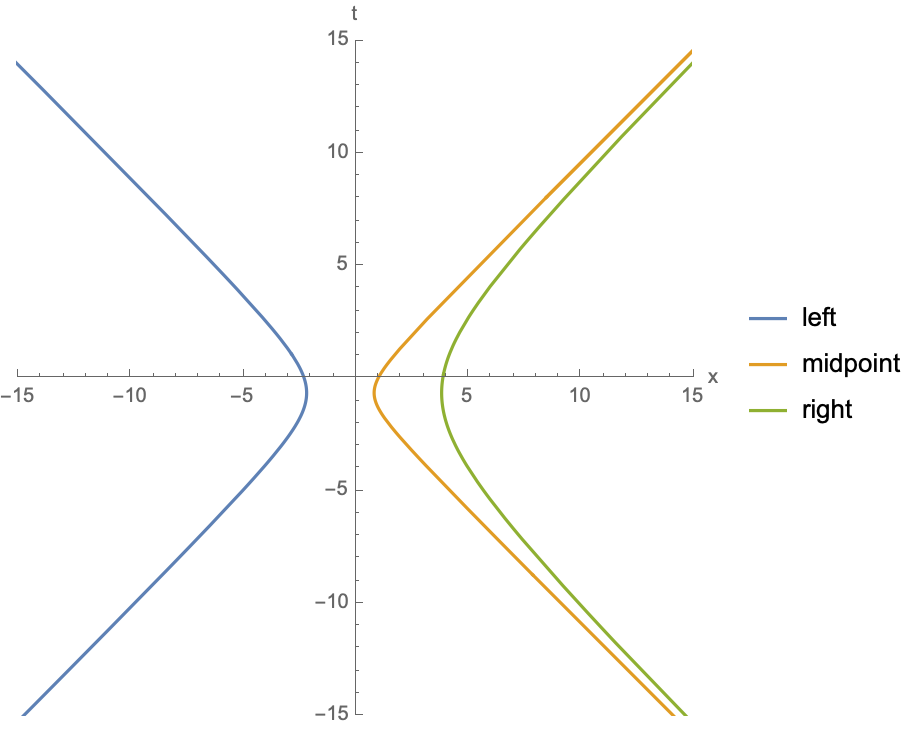}
    \caption{}
    \label{fig23}
\end{figure}
\noindent
Note that the world line of the leftmost point is very different from the other two points. This implies that the origin of \(S'\) lies somewhere
in between $x = -2$ and $x=1$. From equation (\ref{34}) and (35), we can easily find $\tau $(t) and \(x\)(\(t\)). Let \(\beta _i\) =\(\frac{1}{\sqrt{1-v_i^2}}\). After some algebra, we obtain:
\begin{equation}\label{36}
    x(t,l') = \frac{\pm\sqrt{ (1+l'\alpha)^2 + (v_{i} + t\alpha\sqrt{1 - v_{i}})^2} - 1}{\sqrt{1 - v_{i}^2}\alpha} + x_{0}
\end{equation}
\noindent
We have explicitly made \(l'\) a variable because we will consider different points along an extended object. Note that the sign above is positive when $1 + l'\alpha > 0$, and negative when  $1 + l'\alpha < 0$. In other words, $x(t, 1 - \frac{l}{2}\alpha)$ in represents the trajectory for the leftmost point in figure (\ref{fig23}), in which $1 - \frac{l'}{2}\alpha < 0$. When  $1 + l'\alpha = 0$, the point, which is a distance  $l'$ light-seconds away from the reference point, has a proper acceleration of $\infty$, degenerating to the light-like asymptote at the origin (which can also be called the `pivot point' for obvious reasons). Therefore, if we declare that the right end of a rod has a proper acceleration of $\alpha$, then its maximum length is $\frac{1}{\alpha}$ towards the left, i.e. if the reference point is the right end of the rod, then we should $not$ consider $l' < -\frac{1}{\alpha}$. This is simply because any point on a rigid body cannot travel at the speed of light as it has a certain mass. So, we will consider the positive sign in equation (\ref{36}) unless explicitly stated otherwise. As mentioned in \cite{born}, an accelerating rigid body has a maximal spatial extension depending on its acceleration.
\subsection{Simulating the Apparent Shapes of Accelerating Objects}\label{app_accel}
\begin{figure}
    \centering
    \includegraphics[scale = 0.28]{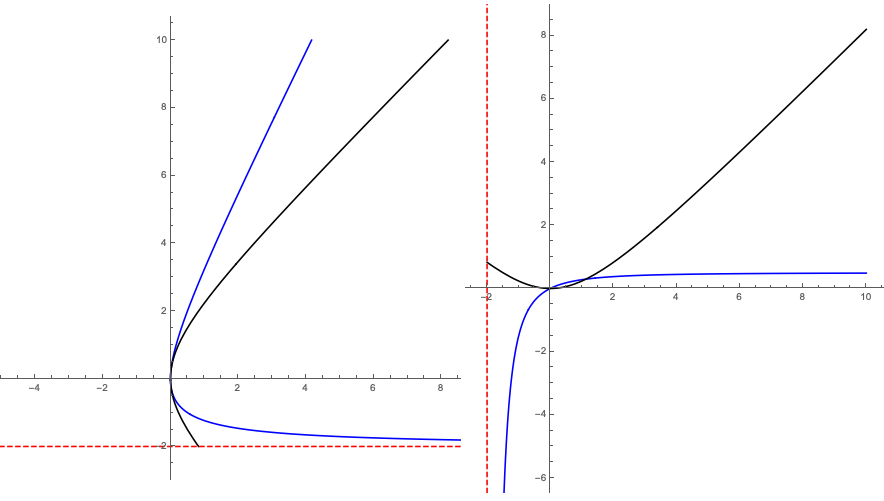}
    \caption{Time is on the $y-axis$ and the position is on the $x-axis$. The blue curve represents the apparent position whereas the black hyperbola represents the point's actual position on the left. It is clear from the figure that $x_{\text{obs}}^0 = -2$ is the horizontal asymptote. The figure on the right shows the apparent speed as a function of time. The black curve represents the actual velocity $\dot{x}$, where $x$ is given in equation (\ref{36}), and the blue curve represents the apparent velocity.}
    \label{fig24}
\end{figure}

\begin{figure}
    \centering
    \includegraphics[scale = 0.3]{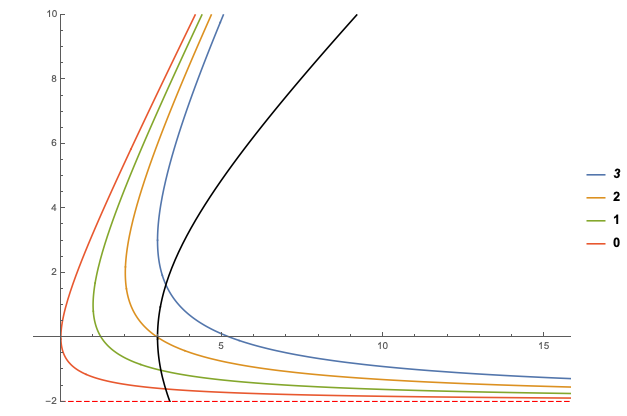}
    \caption{Time on the \textit{y}-axis and position on the \textit{x}-axis. The black curve is the \textit{actual} trajectory of a point $3$ light metres away from a reference point with proper acceleration $\alpha = 0.5$.}
    \label{fig25}
\end{figure}

We begin by exploring time delay and finding the equivalent $`f`$ for accelerated objects. Using (\ref{1}), we can find an expression for
\(x_{\text{em}}^0\) in terms of \(x_{\text{obs}}^0\). Let \(l' = x_0 = v_i = 0\). From now on, we will denote \(x_{\text{em}}^0\) by \(t\). Solving \(x_{\text{obs}}^0\)- \(t\) = $|$\(x(t,0)\)$|$ for \(t\), we obtain:
\begin{equation}\label{37}
t= x_{\text{obs}}^0 \left(\frac{1}{2}\, +\frac{1}{2\, + 2x_{\text{obs}}^0 \alpha }\right) \forall \text{ } x_{\text{obs}}^0 > -\frac{1}{\alpha}
\end{equation}
\noindent
Since $\alpha > 0$, there exists a unique solution $t$ for all $x_{\text{obs}}^0 > 0$. For negative times, however, the situation is different. Consider $\alpha = 0.5$. This means that the point coming in from the left will be first  visible at time $x_{\text{obs}}^0 = \frac{-1}{0.5} = -2$ seconds. The apparent position of the point, however, will be at $\infty$ at $-2$ seconds. At time $x_{\text{obs}}^0 = 0$, the apparent position of the point, given that the actual position of the point at $x_{\text{obs}}^0$ is $0$, is also $0$. So, we expect the point's apparent speed to be greater the speed of light for times until it decelerates quickly. To do so, let's first substitute equation (\ref{37}) into equation (\ref{36}) to evaluate the apparent position of the point as a function of time $x_{\text{obs}}$. Note that we use the positive sign as $l' = 0$. We obtain:
\begin{equation}\label{38}
    x_{\text{app}}(x_{\text{obs}}^0) = \frac{(x_{\text{obs}}^0)^2\alpha}{2 + 2x_{\text{obs}}^0\alpha} \text{ } \forall \text{ } x_{\text{obs}} > -\frac{1}{\alpha}
\end{equation}
\noindent
Note that, in the above derivation, we used the fact that $2 + 2x_{\text{obs}}^0\alpha + (x_{\text{obs}}^0)^2\alpha^2 > 0$ for all pairs of $x_{\text{obs}}^0$ and $\alpha$ that satisfy the condition given in equation (\ref{37}). We can consequently find the first time derivative to find the apparent velocity:
\begin{equation}\label{39}
    \dot{x}_{\text{app}}(x_{\text{obs}}^0) = \frac{1}{2} - \frac{1}{2(1 + x_{\text{obs}}\alpha)^2} \text{ } \forall \text{ } x_{\text{obs}} > -\frac{1}{\alpha}
\end{equation}
\noindent
We plot the apparent trajectory and apparent velocity, given by equations (38) and (39), against time for $\alpha = 0.5$, as show  in figure (\ref{fig24}).
\noindent
Note that in figure (\ref{fig24}), the apparent velocity approaches $-\infty$ as $x_{\text{obs}}^0$ approaches $0$.  Another surprising result is that the point is always accelerating, even if its \textit{actual} motion for negative times is deceleration, as shown by the black curve. Now let{'}s compute \(t\) (the time of emission) for a non-zero \(l'\) in the same way by solving \(x_{\text{obs}}^0-\)
\(t\) = $|$\(x(t,l')\)$|$. For \(1 + l'\alpha> 0\), we obtain after some algebra:
\begin{equation}\label{40}
t =\frac{-2l' - (l')^2 \alpha  + 2x_{\text{obs}}^0 + (x_{\text{obs}}^0)^2\alpha}{2+2 x_{\text{obs}}^0 \alpha } \text{ } \forall \text{ }  x_{\text{obs}}^0 \text{ iff } \alpha > -\frac{2l'}{(l')^2 + (x_{\text{obs}}^0)^2} \text{ and }   x_{\text{obs}}^0 > -\frac{1}{\alpha} 
\end{equation}
\begin{equation}\label{40lol}
t =\frac{-2l' - (l')^2 \alpha  - 2x_{\text{obs}}^0 + (x_{\text{obs}}^0)^2\alpha}{-2 +2 x_{\text{obs}}^0 \alpha } \text{ } \forall \text{ }  x_{\text{obs}}^0 \text{ iff } \alpha < -\frac{2l'}{(l')^2 + (x_{\text{obs}}^0)^2} \text{ and }   x_{\text{obs}}^0 > -\frac{1}{\alpha} 
\end{equation}
\noindent
Note that we get the same restrictions on time of observation $x_{\text{obs}}^0$. Namely, if $\alpha = 0.5$, then an entire horizontal rod will be visible \textit{at once} at time $x_{\text{obs}}^0 = -2$ seconds. This is because both equations (\ref{40}) and (\ref{40lol}) do not yield a solution when $x_{\text{obs}}^0 < -\frac{1}{\alpha}.$ This is also a somewhat non-trivial result, but it paves the way to making some intuitive predictions. When we introduce non-zero values of \(x_0\) and \(v_i\), more restrictions on the solutions will be needed. Equations (\ref{40}) and (\ref{40lol}) find the time of emission. Let's find the apparent position by computing \(x(t,l')\) for equation (\ref{40lol}):
\begin{equation}\label{xapp2}
 x_{\text{app}}(t,l') = \frac{2l' + (l')^2\alpha + (x_{\text{obs}}^0)^2 \alpha}{2 - 2x_{\text{obs}}^0\alpha} \text{ } \forall -\sqrt{\frac{-2l' - (l')^2\alpha}{\alpha}} < x_{\text{obs}}^0< \sqrt{\frac{-2l' - (l')^2\alpha}{\alpha}}
 \end{equation}
 We can find the apparent position for equation (\ref{40}): 
\begin{equation}\label{xapp1}
 x_{\text{app}}(t,l') = \frac{2l' + (l')^2\alpha + (x_{\text{obs}}^0)^2 \alpha}{2 + 2x_{\text{obs}}^0\alpha} \textit{ otherwise}
\end{equation}
\noindent
Note that in deriving the above equations, we used the fact the numerator of both equations is always greater than 0 $\forall$ $x_{\text{obs}}^0, \alpha, l'$ and that $1 + t\alpha > 0$, according to the condition in equation (\ref{40}).
Note that an obvious case of equation (\ref{xapp1}) and equation (\ref{40}) is when $l' > 0$ since we are only considering $\alpha > 0$. We can plot a graph showing the apparent position of a point as a function of time a distance of $l' = 0, 1, 2, 3$ light seconds away from a reference point with proper acceleration $\alpha = 0.5$. This means that we will use equation (\ref{xapp1}). We compare the apparent trajectory with the actual trajectory as well, as shown by the black curve (figure (\ref{fig25})). Figure (\ref{fig25}) yields the expected result, namely that the apparent position of \textit{every} point on the rod at time $x_{\text{obs}}^0 = -2$ seconds is $\infty$, i.e. all curves have the same horizontal asymptote. This also gives some intuitive observations. According to figure (\ref{fig25}), the apparent length of a horizontal rod must be $\infty$ at time $x_{\text{obs}}^0 = -2$ seconds, and should decrease consistently afterwards. 
We can find the apparent length \(l_{\text{app}}\) of a horizontal line whose midpoint has proper acceleration $\alpha $, with \(x_0 = v_i = 0\)
as a function of \(x_{\text{obs}}^0\); \(\left.l_{\text{app}}\right(\)\(x_{\text{obs}}^0\)) = \(\left.x_{\text{app}}\right(\)\(x_{\text{obs}}^0\),
\(\frac{l'}{2}\)) - \(\left.x_{\text{app}}\right(\)\(x_{\text{obs}}^0\),-\(\frac{l'}{2}\)). After some algebra, we obtain:
\begin{equation}\label{42}
    l_{app}(x_{\text{obs}}^0) = \frac{l'}{1 + x_{\text{obs}}^0 \alpha}
\end{equation}
Note that the apparent length of a horizontal rod is indeed $\infty$ at time $x_{\text{obs}}^0 = -\frac{1}{\alpha}$. We can plot a graph showing the relationship of the apparent
length a rod against \(x_{\text{obs}}^0\), where $\alpha $ = 0.5 and \(l'\) (the proper length) is 6:
\begin{figure}[H]
    \centering
    \includegraphics[scale = 0.4]{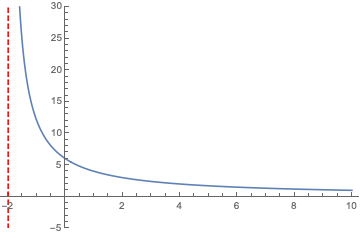}
    \caption{}
    \label{fig26}
\end{figure}
\noindent
Note that the reference point is the midpoint of the rod. To include both equations (\ref{xapp2}) and (\ref{xapp1}), we can plot the apparent length of a rod whose midpoint has a proper acceleration of $\alpha = 0.5$, with $x_{0} = v_{i} = 0$, with its left and right ends defined by $l' = -1$ and $l' = 1$ respectively. The apparent position of the left end would require use of equation (\ref{xapp2}) for $-\sqrt{3} < x_{\text{obs}}^0 < \sqrt{3}$. As anticipated, we get a discontinuity at $x_{\text{obs}}^0 = \sqrt{3}$:
\begin{figure}[H]
    \centering
    \includegraphics[scale = 0.22]{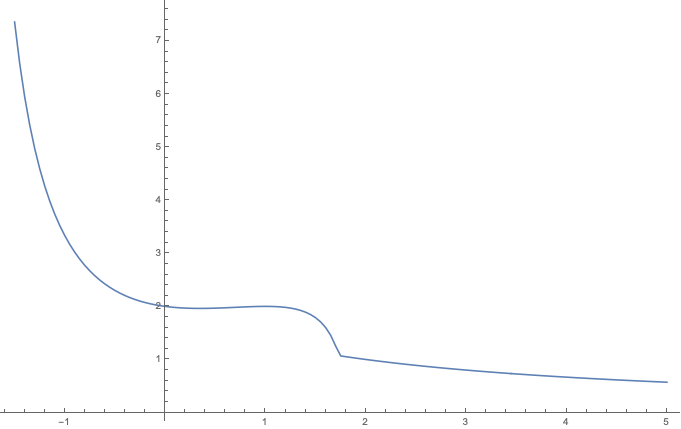}
    \caption{}
    \label{funky}
\end{figure}
\par
\begin{figure}
    \centering
    \includegraphics[scale = 0.22]{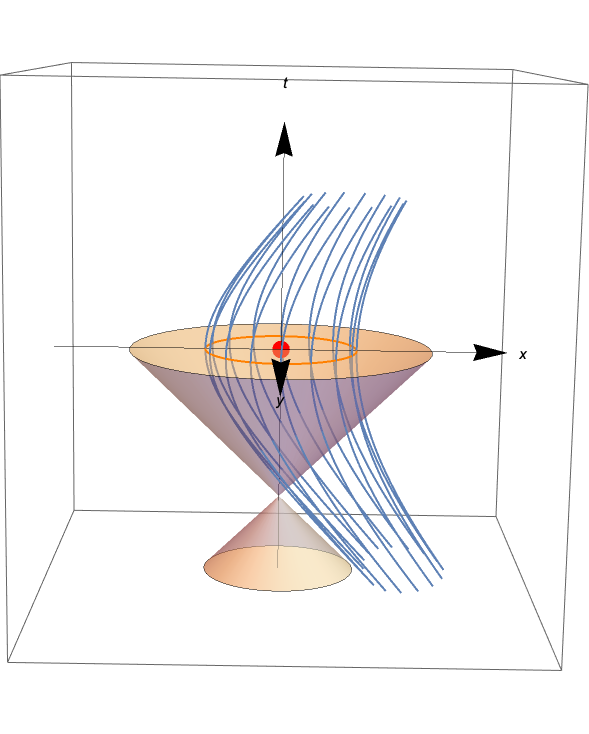}
    \caption{The light cone drawn at $x_{\text{obs}} = -2$ seconds along the \textit{t}-axis. The world lines of points equally spaced along the disk are asymptotic to the surface of past light cone of the observer. This indeed indicates the object appears at $\infty$ at time $x_{\text{obs}} = -\frac{1}{\alpha}$ seconds.}
    \label{fig27}
\end{figure}

\noindent
For a vertical line moving along the $x$-axis, the trajectory of each point on the vertical line will be the same i.e. each point will have the same proper acceleration \(\alpha\). To find the parametric form of the vertical line with initial $y$-coordinate of $y_{i}$ and final $y$-coordinate $y_{f}$, we must find the time of emission of a point with proper acceleration $\alpha $ with a constant $y$-coordinate \(y\). Note that \(l'\) will be 0. For \(x_0 = v_i = 0\), solving \(x_{\text{obs}}^0-\) \(t\) =\textrm{ \(\sqrt{x (t, 0)^2+y^2}\)}, we obtain,
for some arbitrary \(y\):
\begin{equation}\label{43}
t = \frac{2 x_{\text{obs}}^0-\left(x_{\text{obs}}^0\right)^3 \alpha ^2+x_{\text{obs}}^0 y^2 \alpha ^2}{2 \left(1-t^2 \alpha ^2\right)}
+ \frac{\sqrt{4 y^2+\left(\left(x_{\text{obs}}^0\right)^2-y^2\right)^2 \alpha ^2}}{2 \left(1-t^2 \alpha ^2\right)} \text{ } \forall \text{ } x_{\text{obs}}^0 > \frac{1}{\alpha}
\end{equation}
\begin{equation}\label{43second}
t = \frac{2 x_{\text{obs}}^0-\left(x_{\text{obs}}^0\right)^3 \alpha ^2+x_{\text{obs}}^0 y^2 \alpha ^2}{2 \left(1-t^2 \alpha ^2\right)}
- \frac{\sqrt{4 y^2+\left(\left(x_{\text{obs}}^0\right)^2-y^2\right)^2 \alpha ^2}}{2 \left(1-t^2 \alpha ^2\right)} \text{ } \forall \text{ } -\frac{1}{\alpha} < x_{\text{obs}}^0 < \frac{1}{\alpha}
\end{equation}
This implies, again, that the entire vertical rod will be visible first at time $x_{\text{obs}} = -\frac{1}{\alpha}$ at $\infty$ \textit{at once}. This may seem a bit non-intuitive however, as this `invisibility criterion' is independent of the $y-$coordinate. Since the invisibility is only based on the proper acceleration and not on the horizontal displacement $l$ away from the reference point nor on the vertical displacement $y$, we can geometrically analyse the fact that an object with a reference point traveling with a proper acceleration of $\alpha$ is not visible before $x_{\text{obs}} = -\frac{1}{\alpha}$ by plotting the trajectories of a circular disk on the $x-y$ plane. For more details, refer to figure (\ref{fig34}). Let's try to the intersect the trajectories of a circular disk, if the center of the disk travels with a proper acceleration of $\alpha = 0.5$, with backward light cone of the observer at $x_{\text{obs}} = -2$ seconds (figure (\ref{fig27})).
\par
\begin{figure}[H]
    \centering
    \includegraphics[scale = 0.4]{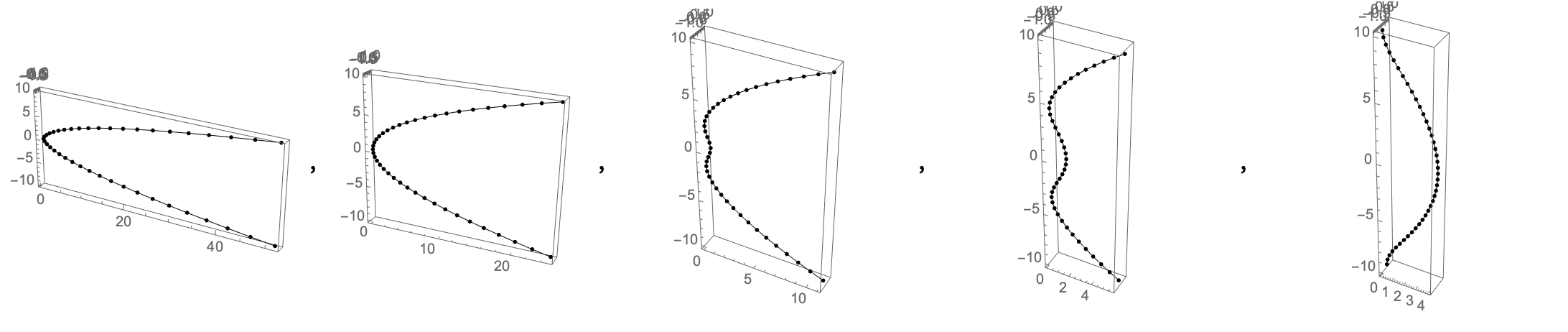}
    \caption{}
    \label{fig28}
\end{figure}

\begin{figure}
    \centering
    \includegraphics[scale = 0.35]{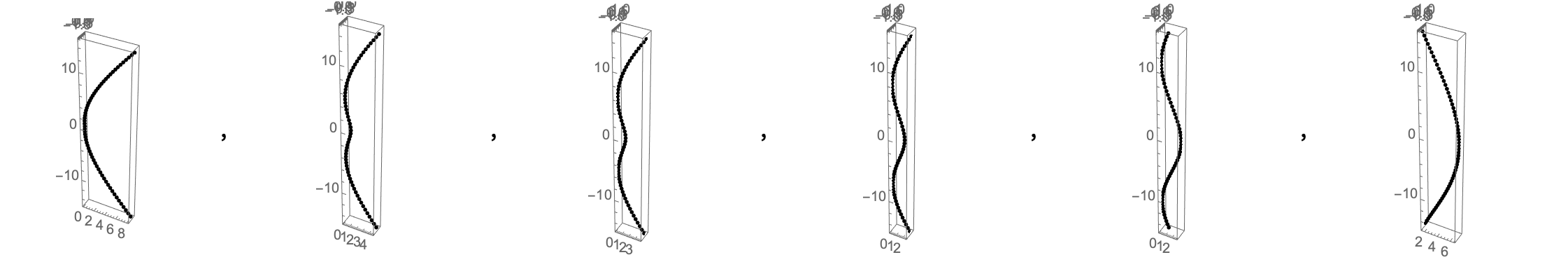}
    \caption{}
    \label{fig29}
\end{figure}

\noindent We can make some preliminary observations. Let{'}s find the parametric
equation of the apparent view of the vertical line at time \(x_{\text{obs}}^0\)= 0, parametrized by \(y\). Equation (\ref{43second}), when \(x_{\text{obs}}^0\)
is substituted as 0, yields \textrm{ \(\text{\textit{$t$}}=\pm  \sqrt{\text{\textit{$y$}}^2+0.25 \alpha ^2 \text{\textit{$y$}}^4}\)} respectively.
In order to find the apparent position \(x_{\text{app}}\) of the point, we must calculate \(x(t,0)\), in equation (\ref{36}), for both solutions. It can
be computed that for both \(t\), \(x_{\text{app}} = \alpha \frac{y^2 }{2}\). Therefore, the parametric form of the vertical line travelling with
proper acceleration $\alpha $ at \(x_{\text{obs}}^0\) = 0 is given by (\(\alpha \frac{y^2 }{2}\), \(y\)), \(y\) $\in $ [\(y_i\), \(y_f\)]. To find
the apparent path of a point on a line, we can explicitly evaluate \(x\left(t_1,0\right)\) and \(x\left(t_2,0\right)\). A series of snapshots
of a vertical line of proper length 20, with its midpoint having a proper acceleration of 0.5, initial velocity $v_{i} = 0$, observed at times \(x_{\text{obs}}^0\)= -1, \(x_{\text{obs}}^0\)= 0, \(x_{\text{obs}}^0\)= 2, \(x_{\text{obs}}^0\)=
4, and \(x_{\text{obs}}^0\)= 10 is shown in figure (\ref{fig28}).
\noindent
We notice that for \(x_{\text{obs}}^0\) $>$ 0, the \(y -\text{intercepts}\) are $\pm $\(x_{\text{obs}}^0\), which is an expected result. This is
because the time it takes for light to reach from a point \(x_{\text{obs}}^0\) light seconds away from the origin is \(x_{\text{obs}}^0\) itself,
meaning that the time at which the light must have been emitted is 0. Let{'}s compute the curvature and torsion for a vertical line for \(x_{\text{obs}}^0\)\(=
0\). Note that the torsion will obviously equal 0 because the curve lies in a plane. The parametric equation is (\textrm{\(\frac{\text{\textit{$y$}}^2
\alpha }{2}\)}, \(y\)) for \(y\) $\in $ [\(y_i\), \(y_f\)]. We can express \(y\) as a function of the arc length \(s\), which enables analysis of
local properties of the apparent shape of a vertical line. We find that the curvature $\kappa $(\(y\)) is given by:
\begin{equation}\label{44}
     \kappa(y) = \frac{\alpha}{\left(1+y^2\alpha^2\right)^{3/2}}
\end{equation}
It's easy to see that $\kappa $(\(y\)) \(\to\) 0 as $\alpha$  \(\to\) 0, which is an expected result. Figure (\ref{fig29}) shows a series of snapshots of a vertical line moving with initial velocity \(v_i\) = 0.9, proper acceleration 0.5, initial position \(x_0\) = 0 at times \(x_{\text{obs}}^0\)= $-5$, \(x_{\text{obs}}^0\)= 1, \(x_{\text{obs}}^0\)= 0, \(x_{\text{obs}}^0\)= 2, \(x_{\text{obs}}^0\)= 5, and \(x_{\text{obs}}^0\)= 15. 
\noindent
As we can see, changing the initial velocity affects the apparent shape significantly. Moreover, we notice that the curve at \(x_{\text{obs}}^0\)=
0 above seems non-differentiable at \(y = 0\). Solving \(x_{\text{obs}}^0-\) \(t\) = \textrm{ \(\sqrt{\text{\textit{$x$}}^2+\text{\textit{$y$}}^2}\)} for
an arbitrary \(v_i\) and \(x_{\text{obs}}^0\) = 0, where \(x\) is given in equation (\ref{36}), we obtain:
\begin{equation}\label{45}
t_{\text{em}} = \frac{\text{\textit{$y$}} \sqrt{4+\text{\textit{$y$}}^2 \alpha ^2} - \text{\textit{$v_i$}} y^2 \alpha }{2 \sqrt{1-v_i} \sqrt{1+\text{\textit{$v_i$}}}}
\end{equation}
\noindent
Evaluating \(x\left(t_{\text{em}},0\right)\), we can compute the unsimplified form of \(x(y)\). The apparent shape of the vertical line at \(x_{\text{obs}}^0\)=
0 is given by \((x(y),y)\). After some algebra, we find that \(\frac{d x}{d y}(y = 0)\) is not defined, indicating that the curve is indeed not differentiable.
\noindent
Below is a series of snapshots of a vertical line of proper length 30 moving with initial velocity \(v_i\) = 0.2, proper acceleration $\alpha $ = 10, initial position of the midpoint $(x_{0},y_{0},z_{0}) = (1,3,3)$ at times \(x_{\text{obs}}^0\)= -10, \(x_{\text{obs}}^0\)= -5, \(x_{\text{obs}}^0\)= 0, \(x_{\text{obs}}^0\)= 5, \(x_{\text{obs}}^0\)= 10:
\begin{figure}[H]
    \centering
    \includegraphics[scale = 0.45]{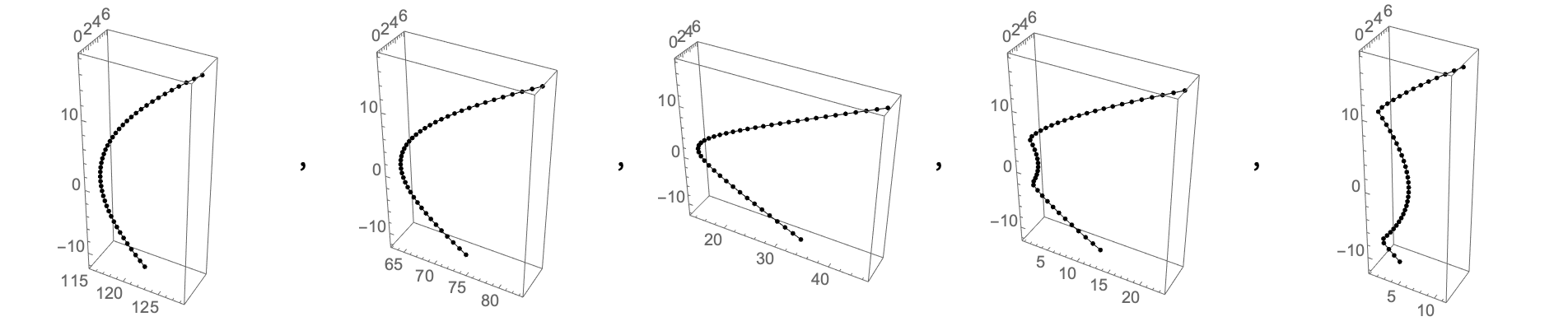}
    \caption{}
    \label{fig30}
\end{figure}
\noindent
Note that in both figures (\ref{fig29}) and (\ref{fig30}), the vertical line \textit{will} be visible at times $x_{\text{obs}}^0 > -\frac{1}{\alpha}$ as we are considering non-zero values for the initial velocity $v_{i}$ and initial position $(x_{0},y_{0},z_{0})$. Also note that the proper acceleration can indeed be greater than 1. Increasing the proper acceleration implies that the trajectory of the reference point gets closer to the origin of our uniformly accelerated reference frame, but is still asymptotic to the 45-degree lines.
\par\noindent
To find the apparent shape of a sphere accelerating in the \(x\)-direction, we first need to find the time of emission for an arbitrary point on
the sphere. Let{'}s parametrize the sphere using Spherical Polar coordinates \((r\text{, $\theta $, $\phi $)}\). Since the sphere must constitute
a uniformly accelerated reference, the proper acceleration will fall with distance away from the stationary observer. To find the time of emission
\(t_{\text{em}}\)for a point with coordinates $(r\sin\theta\sin\phi, r\sin\theta\cos\phi, r\cos\theta)$, we must solve the equation:
\begin{equation}\label{46}
    x_{\text{obs}}^0 - t_{\text{em}} =\sqrt{x(t, r\sin\theta\sin\phi)^2 + r^2 \cos ^2\theta +r^2 \cos ^2\phi \sin^2\theta }
\end{equation}
where $\alpha $ is the proper acceleration of the center of the sphere, and, consequently, \(l'\) is $r\sin\theta\sin\phi$.
When \(v_i = x_0 = 0\), we obtain, after some algebra, a formula for \(t\) when \(x_{\text{obs}}^0\) = 0:
\begin{equation}\label{47}
t_{\text{em}} = \frac{1}{2} r \sqrt{4\, + r^2 \alpha ^2+4r \alpha \sin\theta\sin\phi}
\end{equation}
Note that $r\sin\theta\sin\phi < \frac{1}{\alpha}$ for all $\theta$ and $\phi$. This implies we need to impose the length limit bounding the radius of the sphere in order to ensure rigid motion:
\begin{equation}\label{48}
    r < \frac{1}{\alpha}
\end{equation}
Note that $\alpha$ is the proper acceleration of the \textit{center} of the sphere. Calculating the apparent position by substituting both solutions for $t$ in $x(t_{\text{em}},r\sin\theta\sin\phi)$
\begin{equation}\label{49}
x_{\text{app}}\left(x_{\text{obs}}^0= 0\right) = r\sin\theta\sin\phi + \frac{r^2\alpha}{2}
\end{equation}
Note that, in this case, the sphere will appear \textit{shifted} towards the right by an amount $\frac{r^2\alpha}{2}$. Therefore, we expect the gaussian curvature to be $\frac{1}{r^2}$, which is indeed the case. Below is a series of snapshots of a sphere of radius \(r = 1\) moving
with proper acceleration $\alpha $ = 0.5 with \(x_0 = v_i = 0\) at times \(x_{\text{obs}}^0\)= -1, \(x_{\text{obs}}^0\)= 0, \(x_{\text{obs}}^0\)=
2, \(x_{\text{obs}}^0\)= 5, \(x_{\text{obs}}^0\)= 10:
\begin{figure}[H]
    \centering
    \includegraphics[scale = 0.32]{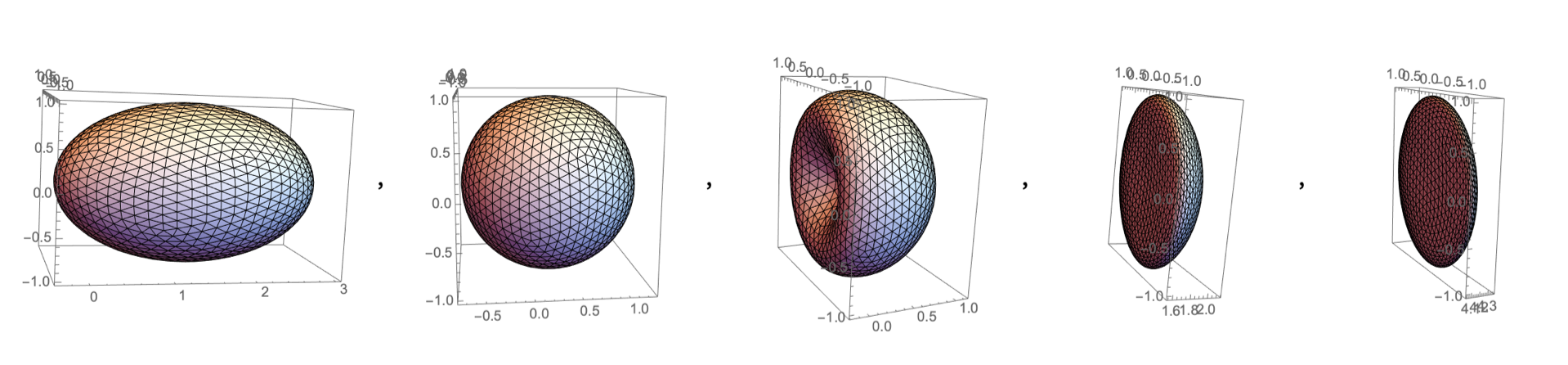}
    \caption{}
    \label{fig31}
\end{figure}
\noindent
Note that the sphere will not be visible before $x_{\text{obs}}^0 = -2$ seconds. When the radius becomes too large, we violate the length limit of an accelerating rigid body, as mentioned in section (\ref{feqm}). If the \textit{center} of the sphere has a proper acceleration of $\alpha$, then the radius of the sphere should be at most $\frac{1}{\alpha}$. When $
\alpha = 0.5$, the maximum radius is 2. In the case of a sphere with radius $5$, we expect to get an incorrect apparent shape, which is indeed the case (shown in figure (\ref{fig32})).
\begin{figure}
    \centering
    \includegraphics[scale = 0.36]{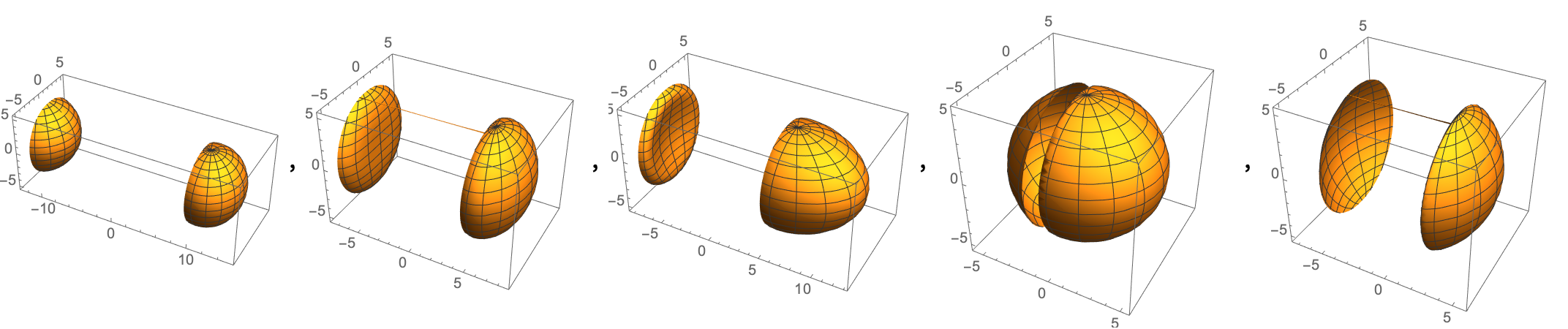}
    \caption{}
    \label{fig32}
\end{figure}\noindent
The sphere appears `cut-off', simply because the origin of our accelerated reference frame, or the pivot point, lies between the center and leftmost end. Figure (\ref{fig33}) shows the trajectories of the leftmost, center, and rightmost points on the sphere of radius 5 when the proper acceleration of the central
point is $\alpha = 0.5$.
\begin{figure}[H]
    \centering
    \includegraphics[scale = 0.3]{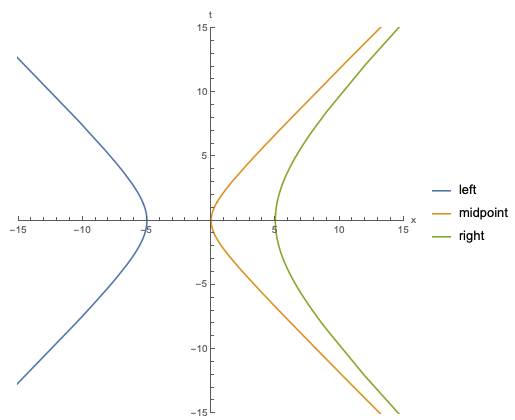}
    \caption{}
    \label{fig33}
\end{figure}
\noindent
To see that the visual appearance of an accelerating sphere at time $x_{\text{obs}}^0 = 0$ is an ordinary sphere, we can use the light cone formalism
in section (\ref{ff}) to find the apparent shape of 2 dimensional circle initially positioned at \(x_0=0\) with \(v_i=\) 0. We will plot the hyperbolic
trajectories of equally spaced points along the circular disk by using polar coordinates with constant \(r\) and $\theta $ $\in $ [0, 2$\pi $]. The world-line of an arbitrary point at an angle of $\theta$ is given by $(t, x(t, r\cos\theta,r\sin\theta),r\sin\theta)$ where \(x\)(\(t\)) is given in equation (\ref{36}). Note that the \(y\)-coordinate will remain constant as we are only dealing
with one-dimensional motion. Since we are determining the apparent shape of the circle at time \(x_{\text{obs}}^0\)= 0, we find the intersection
of the world-lines with the past light cone of the observer located at the origin of \(S\):
\begin{figure}[H]
    \centering
    \includegraphics[scale = 0.3]{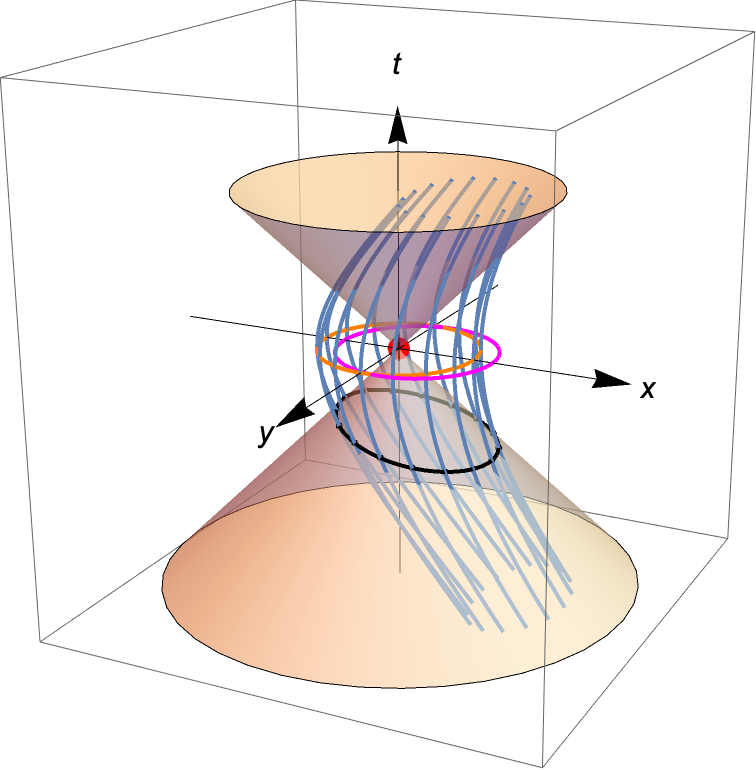}
    \caption{}
    \label{fig34}
\end{figure}
\noindent
The orange circle represents the actual shape of the sphere of proper radius \(r = 1\). Note that we do not need to start with the Lorentz contracted
sphere at time \(x_{\text{obs}}^0\) = 0 because \(v_i = 0\) for all points on the circle. Moreover, equation (\ref{36}) encodes
Lorentz contraction for non-zero \(v_i\). The blue lines represent the trajectories of equally spaced points on the circle for $\alpha $ = 0.5.
Note that the magenta circle represents the apparent shape of the circle, as we have projected the sphere back from the light cone onto the \(\text{\textit{$x-y$}}\)
plane. The circle appears to be shifted ahead slightly while maintaining its shape, indicating that the apparent shape is similar to the original
one. By looking at the light cone above, we can prove this mathematically. For a point on the circle $(r\cos\theta, r\sin\theta)$, we can find the time of emission \(t_{\text{em}}\) by solving -\(t_{\text{em}}\) = \(\sqrt{x(t, r \text{cos$\theta $})^2+ r^2\sin ^2\theta }\). For \(r =1\), we obtain that:
\begin{equation}\label{50}
    x_{\text{em}} = \frac{1}{4} + \cos\theta
\end{equation}
where $\frac{1}{4}$ is equal to $\frac{r^2\alpha}{2}$, agreeing with the result obtained in equation (\ref{49}).
Below is a series of snapshots of a sphere of radius \(r = 1\) moving with proper acceleration $\alpha $ = 0.5 with \(x_0 = 0,v_i = 0.9\) at times
\(x_{\text{obs}}^0\)= $-5$, \(x_{\text{obs}}^0\)= $-3$, \(x_{\text{obs}}^0\)= 0, \(x_{\text{obs}}^0\)= 3, and \(x_{\text{obs}}^0\)= 5: 
\begin{figure}[H]
    \centering
    \includegraphics[scale = 0.29]{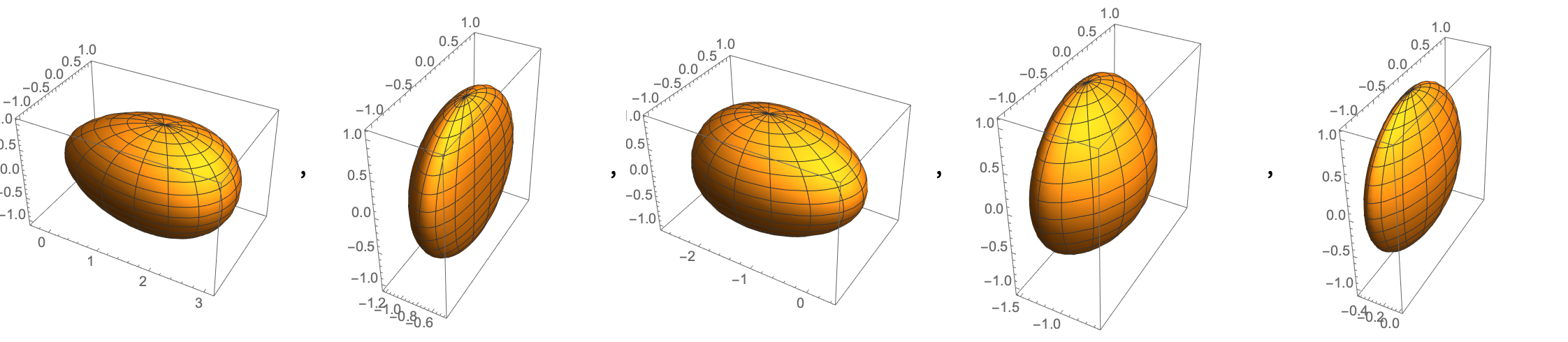}
    \caption{}
    \label{fig35}
\end{figure}
\noindent
In this case, we can see that the scalar curvature is not \(\frac{2}{r^2}\) at time \(x_{\text{obs}}^0\) = 0 as \(\text{\textit{$v_i$}}\) is non-zero. Below is a series of snapshots of a  sphere with $\alpha = 0.5$,$v_{i} = 0$,and initial position vector $(x_{0},y_{0},z_{0}) = (0,2,1)$ observed at times
\(x_{\text{obs}}^0\)= -1, \(x_{\text{obs}}^0\)= 0, \(x_{\text{obs}}^0\)= 2, \(x_{\text{obs}}^0\)= 5, and \(x_{\text{obs}}^0\)= 10: 
\begin{figure}[H]
    \centering
    \includegraphics[scale = 0.29]{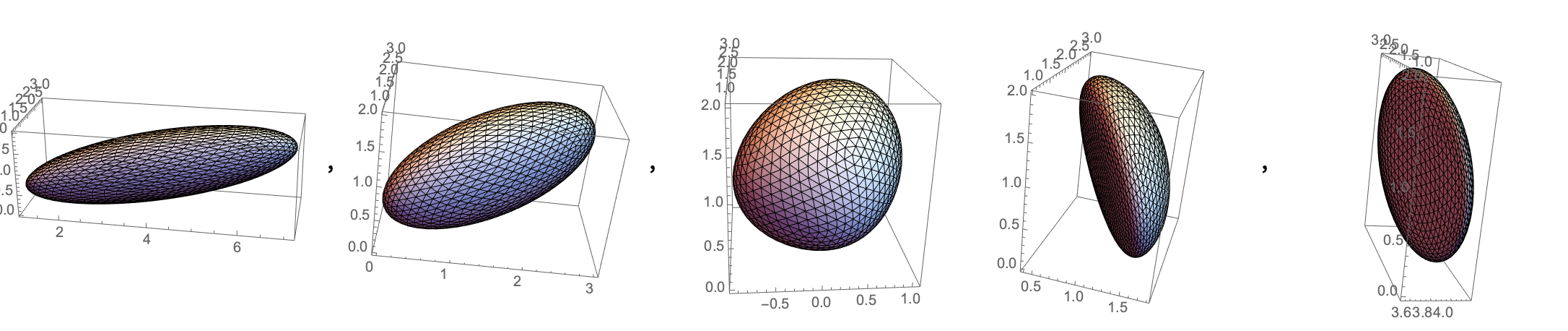}
    \caption{}
    \label{fig35_new}
\end{figure}
\noindent 
Here is an interesting result of the apparent position of a sphere with $v_{i}= 0$, $\alpha = 0.9$, and initial position = $(1,2,1)$, observed at  time \(x_{\text{obs}}^0\)= 2:
\begin{figure}[H]
    \centering
    \includegraphics[scale = 0.22]{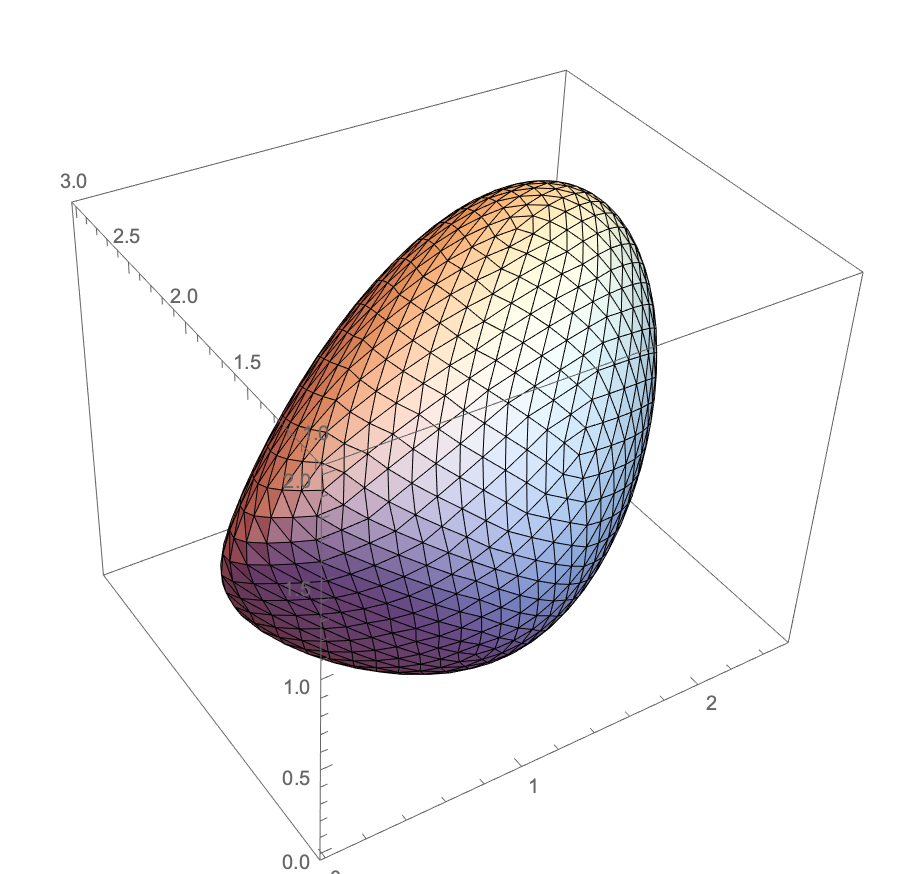}
    \caption{}
    \label{fig35_new_new}
\end{figure}
\noindent
This apparent shape is highly unsymmetrical, and it would be worthwhile to compute its geometric properties such as the metric on its surface, which could allow computation of the affine connection. Figure (\ref{cones}) shows the shape of a cone travelling in different scenarios, which can teach us more about the apparent shape of accelerating objects.
\begin{figure}
    \centering
    \includegraphics[scale = 0.3]{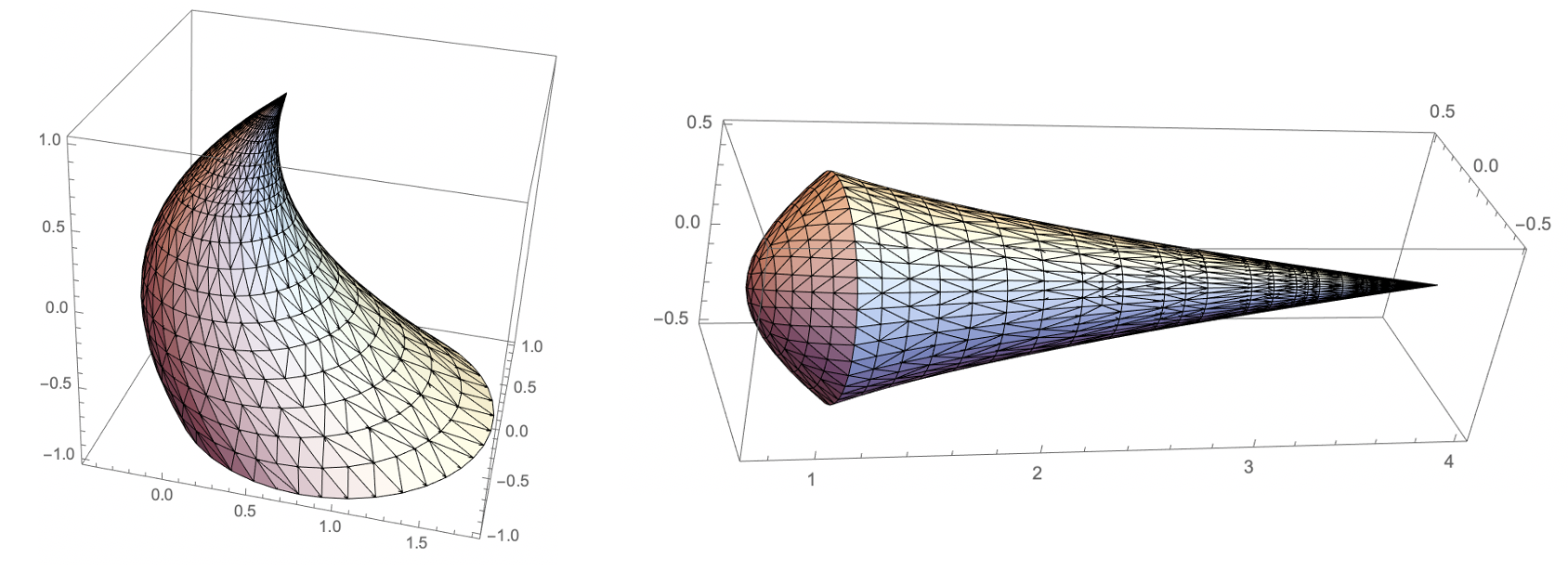}
    \caption{To the left is the apparent shape of a cone with unit radius and a height of 2, with $v_{i} = 0$, proper acceleration $\alpha = 0.8$, and initial position at $(0,0,0)$, observed at time \(x_{\text{obs}}^0\)= 0, whose reference point is located at (0,1,0). To the right is the apparent shape of a cone with a radius of $0.5$ and height of 1, lying along the \textit{x}-axis, at time \(x_{\text{obs}}^0\)= 0, with $v_{i} = 0$, proper acceleration $\alpha = 0.8$, and initial position at $(0,0,0)$, whose reference point is the center of the circular base.}
    \label{cones}
\end{figure}
\subsection{Verifying Results}
The purpose of this section is to highlight tests to verify our results pertaining to the visual appearance of accelerating objects. Since we have
allowed freedom to choose arbitrary values of the initial velocity \(v_i\) and proper acceleration $\alpha $, we expect to retrieve the results described
in section (\ref{section2}) as $\alpha $ \(\to\) 0 for a non-zero \(v_i\). Let{'}s analyse equation (\ref{45}), which gives \(t_{\text{em}}\) that yields the parametric
form \(x\)(\(y\)) of the apparent shape of a vertical line at \(x_{\text{obs}}^0\) = 0. If we take the limit of \(x(y)\) as $\alpha $ \(\to  0\), we obtain:
\begin{equation}\label{51}
    x(y) \to \pm \frac{v_i}{\sqrt{1-v_i^2}}y
\end{equation}
These are the equations of the
asymptotes described in section (\ref{other_constant}), which indicates our equations for acceleration with initial velocity reduce to the constant velocity case as the proper acceleration of each point goes to 0. From section (\ref{observe}), we can see that Gaussian curvature of a sphere of radius 1 for the point with spherical polar coordinates ($\pi $, $\pi $) is
given by $1 - \beta^2$. We expect that the Gaussian curvature at time \(x_{\text{obs}}^0\) = 0 of a sphere of radius 1 at the point ($\pi $, $\pi
$), moving with an initial velocity \(v_i\) and proper acceleration $\alpha $, approaches the value $1 - v_i^2$ as $\alpha $ \(\to\) 0. Although
we cannot perform the exact computation with arbitrary $\alpha $ and \(v_i\), we can indeed observe this result. Below is a
series of images showing the apparent shapes of spheres each of radius 1 moving at \(v_i\) = 0.7 at times \(x_{\text{obs}}^0\)\(=\) 0 for $\alpha
$ = 0.9, 0.3, 0.1, and 0.01:
\begin{figure}[H]
    \centering
    \includegraphics[scale = 0.35]{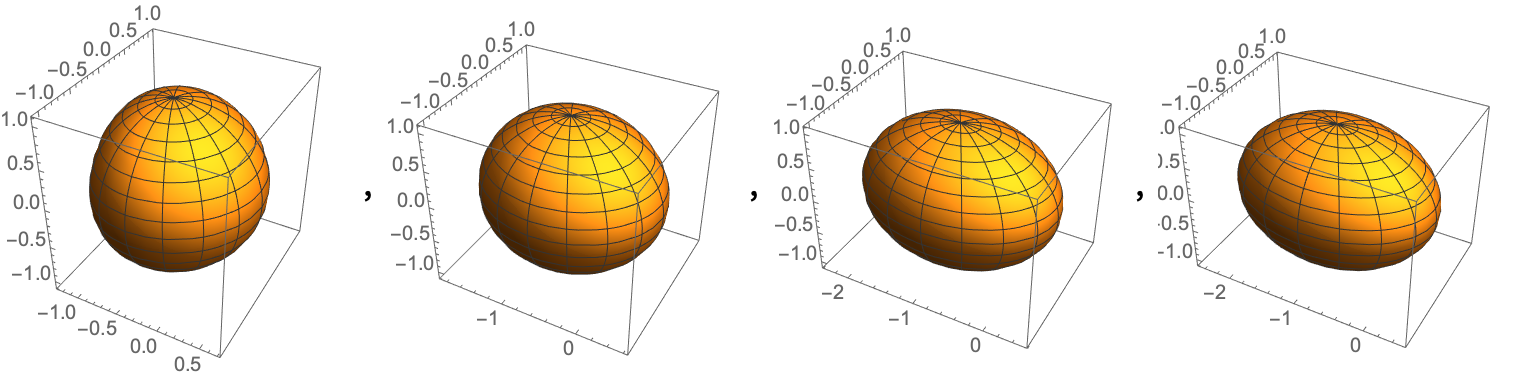}
    \caption{}
    \label{fig36}
\end{figure}
\noindent
We can see from above that the spheres increasingly resemble figure (\ref{fig21}), except with a velocity $\beta = 0.7$, as $\alpha$ approaches 0. The sphere approaches the form given in equation (\ref{27}). It is also reasonable to claim that the proper acceleration $\alpha $ of spheres doesn{'}t
affect the shape of the sphere as much as the initial velocity does when \(v_i\) is close to 1. This is intuitive as well$-$since the speed of point
can never exceed the speed of light, the world-line of a point in a uniformly accelerating reference frame with a high initial velocity resembles
almost a straight line.
\section{Concluding Remarks}
This project provides multiple new insights about Special Relativity by bridging the gap between equations and visualisation. The Poincar\'e transformation allowed the simulation of complex relativistic transformations for every possible initial configuration. Incorporating time delay and transforming arbitrary shapes paved the way to create some beautiful visualisations. As mentioned in section (\ref{section2}), however, the shape of an object seen by the observer can be determined by positioning a camera at the origin of $S$. For example, below is a series of snapshots of a sphere of radius 2 travelling with a speed of $0.9$ in the positive $x$ direction (with $\Vec{a} = (0,10,0)$) observed at time $x_{\text{obs}}^0 = -3$, $x_{\text{obs}}^0 = 0$, $x_{\text{obs}}^0 = 1$, $x_{\text{obs}}^0 = 10$, and $x_{\text{obs}}^0 = 15$: 
\begin{figure}[h]
    \centering
    \includegraphics[scale = 0.38]{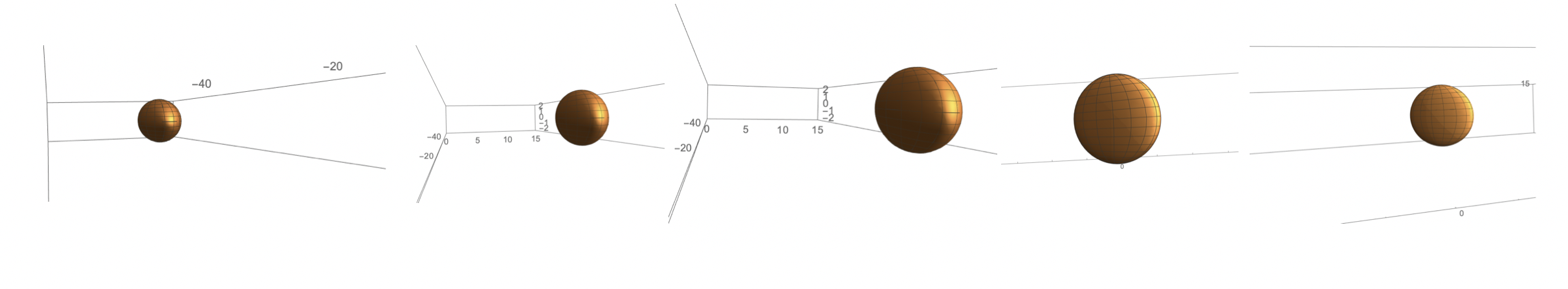}
    \caption{The appearance of a sphere from a camera positioned at the origin. As shown in section (\ref{spheres}), the silhouette appears circular. The above is also a demonstration of the Penrose-Terrell effect, in which the sphere appears `rotated' as it moves across.}
    \label{camera}
\end{figure}
\noindent
By analysing the mathematical properties of various apparent shapes, we were able to develop the intuition required for section (\ref{section3}). The results displayed in section (\ref{section3}) reduce to the ordinary equations in section (\ref{section2}) when acceleration becomes negligible, which can be used to verifying
the formulas presented in the former. We have provided numerous graphical observations as well, and have showed visual appearance
of many objects by placing an observer at the center. The apparent shapes of objects moving with constant velocity were computed using the function \href{https://resources.wolframcloud.com/FunctionRepository/resources/RelativisticInertialDeformedRegion}{\textbf{RelativisticInertialDeformedRegion}} published on the \textit{Wolfram Function Repository} by the author \cite{RelativisticInertial}. This may further be used by teachers to compare the ordinary Lorentz contraction with the results presented here, which can mark the starting point for many more visualisation projects pertaining to Einstein's theory of Relativity. 
\section{Future Work}
An obvious yet interesting future goal is to incorporate optical effects into special relativistic visualisation, as detailed in \cite{illumination} using ray-tracing
methods. The Doppler effect of light should be considered in the presence of an external white light source. In \cite{illumination}, illumination in special
relativity has been explored, and we would like to implement their results in our research as well. Furthermore, it would be worth exploring specular reflection and Lambertian reflectance in the context of special relativity and finding the visual appearance of fast-moving objects. We also see scope for improvement in the aforementioned results. So far, we have taken an algebraic approach. However, a purely geometric approach can be considered using the light cone formalism established in section (\ref{ff}), where the apparent shape of
a 2 dimensional object was found by tracing back the worldlines of each point till they intersect with the past light cone of the observer, followed by projecting the outline onto the \(x-y\) plane (refer to figure (\ref{fig4}) and figure (\ref{fig33})). By analysing the mathematics of such projections,
we can compute the apparent shape of an object in a geometric, coordinate-independent way. Although this may not enable us to perform numerical calculations,
it can give us insight into various geometric features of objects that could not have been gleaned by parametric surfaces. For example, we may compare our results with those in \cite{penrose} by Roger Penrose, who used two-Spinor calculus and relativistic aberration (as explained in equation (\ref{doppler9})) to prove that a fast moving sphere always maintains a circular silhouette. Geometrical methods like these can help us generalise our results further, possibly even to the case of curved spacetime.
\par\noindent
There are also many real-world applications of this project. For example, we can compute the actual speed of an astrophysical jet given its apparent speed observed from telescopes on Earth, which would involve taking into account the time delay of light and the projection of the jet’s $3-D$ trajectory on the $2-D$ sky plane. The formalism established in this paper can also be used in deciphering the actual shape of celestial bodies given their distorted shape observed from Earth, such as the actual shape of a fast moving asteroid. We will model the object as a collection of points, where each point moves in a timelike geodesic. In other words, the object can split into `dust' particles. However, each dust particle will move in a geodesic if and only if there are no extra forces/interactions between them. In that case, the extended object will \emph{actually} distort because the geodesics will deviate from each other. In fact, this deviation can be calculated using the \emph{geodesic deviation} equation\cite{Thanu}:
\begin{equation}\label{deviation1}
    \frac{D^2\xi^{\alpha}}{D\tau^2} = R^{\alpha}_{\delta\beta\gamma}u^\delta u^\beta \xi^\gamma
\end{equation}
where:
\begin{equation}\label{deviation2}
    u^\alpha = \frac{\partial}{\partial \tau}x^\alpha(\tau,\sigma), \xi^\alpha = \frac{\partial}{\partial \sigma}x^\alpha(\tau,\sigma)
\end{equation}
Therefore, we have a `bundle of geodesics' (also known as a \emph{congruence} of geodesics), where each geodesic $x^\alpha$ is parametrized by proper time $\tau$ (so that $u^{\alpha}u_{\alpha} = -1$) and is labelled by a continuous parameter $\sigma$. The vector $\xi^{\alpha}$ denotes the `separation vector', with its tail at a point on one geodesic and its head on a nearby geodesic. Since we want $\xi^{\alpha}$ to measure the `separation' between geodesics, we can initially declare $u_{\alpha}\xi^{\alpha} = 0$. It can be easily derived that $\frac{d}{d\tau}(u_{\alpha}\xi^{\alpha}) = 0$, which implies that $\xi^{\alpha}$ always lies in the orthonormal space of $u_{\alpha}$. 
\par\noindent
From equation (\ref{deviation1}), we can see that the deviation of the geodesics from each other is directly proportional to the components of the Riemann curvature tensor. So, equation (\ref{deviation1}) explains the \emph{tidal deformations} in an extended object due to the curvature of spacetime. For example, the components of the Riemann tensor near a black hole are very large, causing the extended object to `Spaghettify'. When trying to find the visual appearance of an object moving near a black hole, we must take tidal effects into account. However, since we are only concerned with light propagation effects, we may choose scenarios where there are no tidal deformations. 
\par\noindent
As explained in section (\ref{section3}), we must consider motion that satisfies the \emph{Born-Rigidity conditions} i.e. the freely-falling motion of a rigid body. In this particular case, the internal stresses inside the object will keep the `dust particles' from moving in geodesics. As stated in \cite{Thanu}, only the center of mass of the object may be assumed to move in a geodesic, and the remaining points must move in a way as to keep the proper-space distance within the congruence constant. The proper space distance squared is given by:
\begin{equation}
    l^2 := g_{\alpha\beta}\xi^\alpha\xi^\beta = h_{\alpha\beta}\xi^\alpha\xi^\beta
\end{equation}
where $h_{\alpha\beta} = g_{\alpha\beta} + u_{\alpha}u_{\beta}$. The Born Rigidity conditions are satisfied if:
\begin{equation}
\frac{D}{D \tau}(l) = u^\alpha \nabla_{\alpha} l = 0
\end{equation}
There are many other challenges of simulating the appearance of an extended object moving in curved spacetime. We need to find the equation of a geodesic given the initial and final spacetime coordinates, namely the event of emission from the point and the event of the light ray hitting the observer{'}s eye. Essentially, we have to solve the null geodesic equation as a boundary value problem, which is very difficult to do in arbitrary
curved spacetimes. Completing the above can act as a stepping stone to another interesting application: simulating the effects of the expansion of the universe on the apparent shapes of large bodies, such the distortion of galaxies receding away from Earth. This would make use of stationary geodesic congruences as world lines of equally spaced points (stars) across a galaxy.
\par\noindent
Finding the visual appearance of extended objects in special relativity has many potential applications to the \textit{Wolfram Physics Project} \cite{WPP}. Specifically, simulating the visual appearance of an extended object when travelling near the speed of light may reveal interesting facts about the workings of light cones and special relativity in the Wolfram model. In the Wolfram model, space is represented by a hypergraph\textemdash essentially a collection of nodes (`atoms of space') whose only intrinsic property is the causal relationships with other nodes \cite{class}. In this paper, we have assumed that the stationary observer $S$ is isolated from everything, the equivalent of which is the `cosmological rest frame' described in \cite{finallymf}. Then, the moving observer's frame can be described by a specific `foliation' of the original causal graph by spacelike hypersurfaces \cite{class}. We also must have a way to define distances and time in the Wolfram Model. An elementary unit of time can be chosen, which would represent the time it takes it takes to get from one spacelike hypersurface on the causal graph to the next. Once we have the notion of time, we can calculate the time `steps' it takes for a light ray to reach a certain point on the hypergraph. For example, the image below shows  how a light ray might spread across a hypergraph. 
\begin{figure}[H]
    \centering
    \includegraphics[scale  = 0.4]{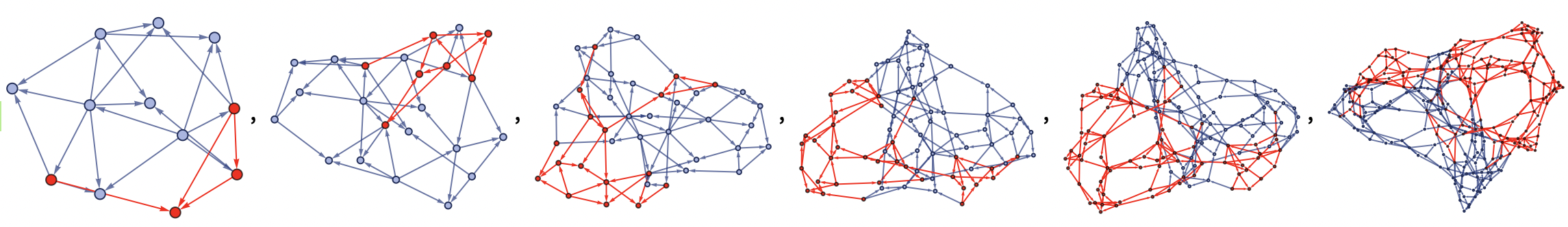}
    \caption{A light ray spreading across a hypergraph. Image taken from \cite{faster}. Since time is merely the progression of rules applied to hypergraph, space itself changes with every time step.}
    \label{WPP_fig_1}
\end{figure}
\noindent
After defining time, we still need a notion of distance to calculated how far away the object is from the stationary observer. As stated in \cite{faster}, there are many ways to define distance on a hypergraph. However, the most appropriate one for this project may use the notion of `a spatial reconstruction graph' \cite{faster}, which reconstructs space from the causal graph of a particular order of evolution. Although notions of distance of time in the Wolfram model are not straightforward, applying them to extended objects once figured out completely may not be as challenging. This is because we can convert arbitrary Riemannian manifolds, such as spheres and even polytopes, into hypergraph approximations. Finding the emission time for each node can then enable us to reconstruct the transformed apparent shape of objects. As a consequence of discreteness underneath, we can compare the transformed shapes to those obtained within the continuum limit in this paper. We expect that close to $\beta =1$, we may obtain asymmetrical transformed shapes from simple initial conditions. This will allow us to probe the inhomogeneity and isotropy of discrete spacetime. As stated in \cite{faster}, there is no fundamental reason for why the speed of light is the same in all directions when considering discrete spacetime.
\section{Acknowledgements}
I express my gratitude towards Dr. Stephen Wolfram for suggesting this project in the first place and for providing me countless support and tips.
I would especially like to thank my mentor, Dr. Matthew Szudzik, for his continuous support and guidance during the entire length of the project.
I express my gratitude towards everyone in the Wolfram Physics Project for helping me with both mathematical and computational problems, especially
Dr. Jonathan Gorard, Mr. Hatem Elshatlawy, and Dr. Xerxes Arsiwalla. Furthermore, I would like to thank Mr. Hatem Elshatlawy and Dr. Xerxes Arsiwalla for proofreading my paper. I would especially like to thank Mr. Nikolay Murzin for helping me with various computational and theoretical problems along the way. I am grateful towards Mr. José Manuel Rodríguez Caballero for acting as a mentor and giving me countless suggestions regarding my paper. All calculations, graphs, shapes, and other graphical elements were made using Mathematica 12.1.

\bibliographystyle{siam}
\raggedright
\bibliography{reference.bib}
\end{document}